\documentclass[12pt]{article}
\usepackage{amssymb}
\begin{document}

\newcommand{\beqn}{\begin{eqnarray}}
\newcommand{\eeqn}{\end{eqnarray}}
\newcommand{\be}{\begin{equation}}
\newcommand{\ee}{\end{equation}}
\newcommand{\ba}{\begin{array}}
\newcommand{\ea}{\end{array}}
\newcommand{\pa}{\partial}
\newcommand{\re}{\ref}
\newcommand{\ci}{\cite}
\newcommand{\la}{\label}
\newcommand{\fr}{\frac}
\newcommand{\ve}{\varepsilon}
\newcommand{\de}{\delta}
\newcommand{\al}{\alpha}
\newcommand{\ga}{\gamma}
\newcommand{\Ga}{\Gamma}
\newcommand{\si}{\sigma}
\newcommand{\ds}{\displaystyle}
\newcommand{\pr}{\prime}
\newcommand{\La}{\Lambda}
\newcommand{\De}{\Delta}
\newcommand{\Si}{\Sigma}
\newcommand{\ti}{\tilde}
\newcommand{\Om}{\Omega}
\newcommand{\om}{\omega}

\newcommand {\PV}{{\rm PV}}

\newcommand{\T}{\mathbb{T}}
\newcommand{\R}{\mathbb{R}}
\newcommand{\Z}{\mathbb{Z}}
\newcommand{\N}{\mathbb{N}}
\newcommand{\C}{\mathbb{C}}
\newcommand{\E}{\mathbb{E}}

\newcommand{\const}{\mathop{\rm const}\nolimits}
\newcommand{\tr}{\mathop{\rm tr}\nolimits}
\newcommand{\tg}{\mathop{\rm tg}\nolimits}
\newcommand{\supp}{\mathop{\rm supp}\nolimits}
\newcommand{\sign}{\mathop{\rm sign}\nolimits}
\newcommand{\dist}{\mathop{\rm dist}\nolimits}

\renewcommand{\theequation}{\thesection.\arabic{equation}}
\newtheorem{theorem}{Theorem}[section]
\renewcommand{\thetheorem}{\arabic{section}.\arabic{theorem}}
\newtheorem{definition}[theorem]{Definition}
\newtheorem{lemma}[theorem]{Lemma}
\newtheorem{example}[theorem]{Example}
\newtheorem{remark}[theorem]{Remark}
\newtheorem{remarks}[theorem]{Remarks}
\newtheorem{cor}[theorem]{Corollary}
\newtheorem{pro}[theorem]{Proposition}

\newcommand{\bo}{{\hfill\loota}}
\newcommand{\loota}{\hbox{\enspace{\vrule height 7pt depth 0pt width 7pt}}}

\begin{titlepage}
\begin{center}
{\Large\bf Caricature of Hydrodynamics for Lattice Dynamics}\\
\vspace{2cm}
{\large T.V.~Dudnikova}
\footnote{Supported partly by
research grant of RFBR (09-01-00288)}\\
{\it Elektrostal Polytechnical Institute\\
 Elektrostal 144000, Russia}\\ 
e-mail:~tdudnikov@mail.ru
\end{center}
 \vspace{1cm}
 \begin{abstract}
The lattice dynamics in $\Z^d$, $d\ge1$, is considered. The initial
data are supposed to be random function.
We introduce the family of initial measures $\{\mu_0^\ve,\ve>0\}$
depending on a small scaling parameter $\ve$.
We assume that the measures $\mu_0^\ve$ are locally homogeneous
for space translations of order much less than
$\ve^{-1}$ and nonhomogeneous for translations of order $\ve^{-1}$.
Moreover, the covariance of $\mu_0^\ve$ decreases with distance
uniformly in $\ve$.
Given $\tau\in\R\setminus 0$, $r\in\R^d$, and $\kappa>0$,
we consider the distributions of random solution in the time moments
$t=\tau/\varepsilon^\kappa$ and at lattice points close to
$[r/\varepsilon]\in\Z^d$. 
The main goil is to study the asymptotics of these distributions
as $\varepsilon\to0$ and derive the limit hydrodynamic equations
of the Euler or Navier-Stokes type.
The similar results are obtained for lattice dynamics 
in the half-space $\Z^d_+$.
 \end{abstract}
\end{titlepage}

 \section{Introduction}

One of the central problems in nonequilibrium statistical physics is
the derivation of hydrodynamic equations of fluid
from the microscopic Hamilton dynamics. 
The main fluid equations are the Euler and Navier--Stokes equations.
The idea that the Euler equation of fluid dynamics 
could be derived from microscopic dynamics goes back to Morrey \cite{Mo}.
A systematic explanation of some basic ideas and first results
are presented in the survey papers by De Masi, Ianiro, Pellegrinotti, 
and Presutti \cite{DeMasi}, by Dobrushin, Sinai and Sukhov \cite{DSS},
and by Spohn \cite{Sp91}. 

One approach is to derive the Euler and Navier--Stokes equations
from the Boltzmann equation (see, for example, \cite{DEL}).
However, the Boltzmann equation is not a microscopic model,
it should itself be derived as a scaling limit of a more basic model.
An alternate approach is to study the hydrodynamic behaviour 
of some simplified or {\it idealized} models of interacting particles. 
For models of {\it stochastic} dynamics, the results were obtained by 
Yau {\it et al.} (see, for example, \cite{EMY, OVY93, QY98},
 the survey paper by Fritz \cite{F} and the bibliography there).
{\it Deterministic} models, where only initial realizations 
can be random,  are more difficult to study in a hydrodynamic framework.
The first results were obtained by Dobrushin {\it et al.} for 
the one-dimensional hard rods \cite{BDS}, and for
the one-dimensional oscillators on the lattice \cite{DPST}--\cite{DPS90}.
The main purpose of these models is to show how hydrodynamic behavior
arises.

The present work continues the papers \cite{DS, D09'},
where as the model the harmonic crystals in $\Z^d$ are considered.
In the harmonic approximation, the crystal is characterized
by the displacements $u(z,t)\in\R^n$, $z\in\Z^d$, of the crystal atoms 
from their equilibrium positions. 
The field  $u(z,t)$ is governed by a discrete wave equation.
The harmonic crystals can be considered as an extension of the model
of the infinite chain of one-dimensional harmonic oscillators 
to the many dimensional case.

The derivation of hydrodynamic equations is connected
with the problem of convergence to an equilibrium measure. 
For harmonic crystals, such convergence was proved in \cite{DKS1, DKM, D08}.  
We outline this result.
The initial data are assumed to be random function with a distribution
$\mu_0$. We suppose that the measure $\mu_0$ 
has zero mean value, a finite mean energy density
and satisfies the mixing condition.
The distribution $\mu_t$ of the random solution $u(\cdot,t)$
in the time moments $t\in\R$ is studied. Then the limit
\be\la{0.1}
\lim_{t\to\infty}\mu_t=\mu_\infty
\ee
is established, where $\mu_\infty$ is an equilibrium Gaussian measure.
For one-dimensional chain of harmonic oscillators,
 this result has been proved by Boldrighini {\it et al.} \cite{BPT}. 
The convergence to equilibrium distribution was proved
also for systems described by partial differential equations
\cite{DKS, DK2}, for the crystal coupled to the scalar field
\cite{DK05}, and for the Klein-Gordon equation coupled to a particle
\cite{D10}. 

To derive the hydrodynamic equations
we apply the special so-called {\it hydrodynamic limit procedure}
in which the notions of hydrodynamic limit and
hydrodynamic space--time rescalings play a central role.
Namely, we introduce a small scale parameter $\ve>0$ giving the relation 
between the microscopic and macroscopic space-time scales 
and consider the family of the initial measures
$\{\mu_0^\varepsilon,\varepsilon>0\}$
 which satisfies some  conditions
(see conditions \textbf{V1} and \textbf{V2} in Section \ref{sec1.2} below).
In particular, we assume that
(i) the measures $\mu_0^\varepsilon$ are  locally homogeneous
for space translations of order much less than
$\ve^{-1}$ and nonhomogeneous for translations of order $\ve^{-1}$;
(ii) the covariance of  $\mu_0^\varepsilon$
 vanishes with distance enough quickly and uniformly in $\varepsilon$.

To deduce the Euler equation we use (see \cite{DS}) 
 a {\it hyperbolic} (or {\it Euler}) {\it scaling},
i.e., the {\it microscopic} time and space variables are
$t=\tau/\ve$ and $z=r/\ve$, where
 the {\it macroscopic} time and space variables are denoted
by $\tau$ and $r$, respectively.
Given nonzero $\tau\in\R$ and $r\in\R^d$, we study the distribution
$\mu_{\tau/\varepsilon,r/\ve}^\varepsilon$ of the random solution $u(z,t)$
at the space points close to $[r/\varepsilon]\in\Z^d$ and 
in time moments $\tau/\varepsilon$. Then the limit is established,
\be\label{0.2}    
\lim_{\varepsilon\to0}\mu_{\tau/\varepsilon,r/\ve}^\varepsilon=\mu^G_{\tau,r},
\ee
where $\mu^G_{\tau,r}$ is a Gaussian measure (see Theorem 3.3 in \cite{DS} 
for harmonic crystals in the whole space $\Z^d$
and Theorem 2.15 in \cite{D09'} 
for harmonic crystals in the half-space $\Z^d_+$).
In particular, we derive the explicit formulas 
for covariance matrix $q_{\tau,r}(z-z')$
of the limit measure $\mu^G_{\tau,r}$. These formulas allow us 
to conclude that in Fourier transform
the matrix function $\hat q_{\tau,r}(\theta)$, $\theta\in\T^d$,  
evolves according to the following equation:
\beqn\label{0.3}
\partial_\tau f(\tau,r;\theta)=i\,C(\theta)
\nabla\Omega(\theta)
\cdot \nabla_r f(\tau,r;\theta),\quad r\in\R^d,\quad \tau>0,
\eeqn 
where $C(\theta)=\left(\ba{cc}
0&\Omega^{-1}(\theta)\\
-\Omega(\theta)&0 \ea\right)$, and, roughly, $\Omega(\theta)$ is 
the dispersion relation of the harmonic
crystal (for details, see Section \ref{sec2.1} below). 
Here and below $\partial_\tau$ denotes 
partial differetiation with respect to a time $\tau$,
$\nabla_r f$ is the gradient of $f$ with respect to $r\in\R^d$,
"$\cdot$" stands for the standard Euclidean scalar product in $\R^d$
(or in $\R^n$).
The equation (\ref{0.3}) should be considered 
as the analog of the Euler equation for our model.
In \cite{DPST}, the similar equation was deduced in the case $d=n=1$.    
These results were extended to 
the harmonic crystals in the entire space $\Z^d$, $d\ge1$, (see \cite{DS})
and in the half-space $\Z^d_+=\{z\in\Z^d:z_1>0\}$, see \cite{D09'}. 

The main result of the given paper is the derivation of 
the equation for the "next approximation" to the Euler equation
(\ref{0.3}).
To obtain the additional term of order $\varepsilon$ in (\ref{0.3}), 
we use a {\it diffusive} (or {\it parabolic}) scaling, that is
 we study the distributions of solution $u(z,t)$ in time moments
of order $\tau/\varepsilon^2$. After the appropriate
change of variables we derive the equation of the Navier--Stokes type 
\be\label{0.4}
\partial_\tau \hat f(\tau,r;\theta)
=i C(\theta)\left(\nabla\omega(\theta)
\cdot \nabla_r \hat f(\tau,r;\theta)
+\frac{i\varepsilon}{2}\,{\rm tr}\left[\nabla^2 \Omega(\theta)
\cdot \nabla^2_r f(\tau,r;\theta)\right]\right).
\ee
Here and below $\nabla^2_r f$ stands for the matrix of second partial
derivatives of $f$ with respect to $r$, "tr" stands for trace.
The precise statement of the result see in Theorem \ref{the2'}
and Corollary \ref{cor2.7} below.
In the case $d=n=1$, these results have been obtained in the work
of Dobrushin {\it et al} \cite{DPST2}. Therefore,
in the proof (see Section 6) we use an approach of \cite{DPST2}
and tools of \cite{DKS1, DS} developed for harmonic crystals
in any dimension.

In Section \ref{sec3.3} we use a scaling of the form
$t=\tau/\ve^k$, with $k\ge2$, $z=r/\ve$, 
and derive the "corrections" of the higher order (i.e. of the order
 $\ve^k$ with $k\ge2$) to equation (\ref{0.3}).
\medskip

Second part of the paper is devoted to the study
of the harmonic crystals in the half-space $\Z^d_+=\{z\in\Z^d:z_1>0\}$, 
with zero boundary condition. 
For such model, we also derive the limiting "hydrodynamic" equations
(of the Euler and Navier--Stokes types). 

The paper is organized as follows.
In Section~2 we introduce the model, impose the main conditions 
on harmonic potentials and initial measures $\mu_0^\ve$ 
and give the examples of the potentials and measures $\mu_0^\ve$
satisfying our conditions. 
In Section~3 the main results are stated. 
Sections~4 and 5 are devoted to
the harmonic crystals in the half-space $\Z^d_+$. 
Sections~6 and 7 contain the main steps of the proof of results.
The technical details of the proof are given in Appendices~A--C.  
Appendix~D is devoted to locally conserved quantities. 

\setcounter{equation}{0}
 \section{Model I: Harmonic crystals in $\Z^d$}
We study the dynamics of the harmonic crystals 
in $\Z^d$, $d\ge 1$, 
\beqn\la{CP1''}
\left\{\ba{l}
\ddot v(z,t)=-\sum\limits_{z'\in \Z^d}V(z-z')v(z',t),\quad z\in\Z^d,
\quad t\in\R,\\
v(z,0)=v_0(z),\quad \dot v(z,0)=v_1(z),\quad z\in\Z^d.
\ea\right.
\eeqn
Here $v(z,t)=(v_1(z,t),\dots,v_n(z,t))$,
$v_0(z)=(v_{01}(z),\dots,v_{0n}(z))\in\R^n$, and
correspondingly for $v_1(z)$,
$V(z)$ is the interaction (or force) matrix, $\left(V_{kl}(z)\right)$,
$k,l=1,\dots,n$. 

Write $X(t)=(X^0(t),X^1(t))\equiv (v(\cdot,t),\dot v(\cdot,t))$ and
$X_0=(X_0^0,X_0^1)\equiv (v_0(\cdot),v_1(\cdot))$. Then
(\ref{CP1''}) becomes
\be\la{CP1'}
\dot X(t)={\cal A}X(t),\quad t\in\R,\quad X(0)=X_0.
\ee
Here ${\cal A}=\left(\ba{cc}0&1\\-{\cal V}&0\ea\right)$,
where ${\cal V}$ is a convolution operator with the matrix kernel $V$,
${\cal V}v=\sum\limits_{z'\in\Z^d} V(z-z')v(z')$.
Formally, (\ref{CP1'}) is a linear Hamiltonian system
with the Hamiltonian functional
 \beqn\label{H}
 H(X)= \frac{1}{2} \sum_{z\in\Z^d}|v_1(z)|^2
+\frac{1}{2} \sum_{z,z'\in\Z^d} v_0(z)\cdot V(z-z')v_0(z'),
 \quad X=(v_0,v_1), 
\eeqn
where the kinetic energy is given by the first term, and the potential
energy by the second term.

  Assume that the initial data $X_0$ for (\ref{CP1'})
belong to the phase space ${\cal H}_{\al}$, $\al\in\R$, defined below.
 \begin{definition}                 \la{d1.1'}
 $ {\cal H}_{\al}$ is the Hilbert space
of $\R^n\times\R^n$-valued functions of $z\in\Z^d$ endowed  with  the norm
$ \Vert X\Vert^2_{\al}
 = \sum_{z\in\Z^d}\vert X(z)\vert^2(1+|z|^2)^{\al} <\infty.
$  
\end{definition}
\subsection{Conditions on the harmonic potentials}
We impose the following conditions {\bf E1}--{\bf E6} on the matrix $V$.
\medskip\\
{\bf E1}. There are positive constants $C$ and $\gamma$ such that
$\|V(z)\|\le C e^{-\gamma|z|}$ for $z\in \Z^d$, where
$\|V(z)\|$ stands for the matrix norm. \smallskip\\
{\bf E2}. 
The matrix $V(z)$ is real and symmetric, 
i.e., $V_{lk}(-z)=V_{kl}(z)\in \R$, $k,l=1,\dots,n$, $z\in \Z^d$.
\smallskip

Let $\hat V(\theta)$ be the Fourier transform of $V(z)$ with the
convention
 $ \hat V(\theta)=
\sum\limits_{z\in\Z^d}V(z)e^{iz\cdot\theta}$, $\theta \in \T^d$,
where  $\T^d$ stands for the $d$-torus $\R^d/(2\pi \Z)^d$.
\medskip

Conditions {\bf E1} and {\bf E2} imply that $\hat V(\theta)$ 
is a real-analytic
 Hermitian matrix-valued function of $\theta\in \T^d\!$.
\smallskip\\
{\bf E3}. 
The matrix $\hat V(\theta)$ is  non-negative definite for 
every $\theta \in \T^d$. \medskip

Let us define the Hermitian  non-negative definite matrix,
 \be\label{Omega}
 \Omega(\theta)=\big(\hat V(\theta )\big)^{1/2}\ge 0.
 \ee
The matrix $\Omega(\theta)$  has the eigenvalues 
$0\leq\omega_1(\theta)<\omega_2(\theta)< \ldots <\omega_s(\theta)$, 
$s\leq n$, and the
corresponding spectral projections $\Pi_\sigma(\theta)$ with
multiplicity $r_\sigma=\tr\Pi_\sigma(\theta)$. 
The following lemma holds.
\begin{lemma}\label{lc*} (see \cite[Lemma 2.2]{DKS1}).
Let conditions {\bf E1} and {\bf E2} hold. 
Then there exists a closed subset ${\cal C}_*\subset \T^d$
such that the following assertions hold.\\
(i) The Lebesgue measure of ${\cal C}_*$ is zero.\\
(ii) The eigenvalue $\omega_\sigma(\theta)$, $\sigma=1,\dots,s$,
 has constant multiplicity in $\T^d\setminus{\cal C}_*$.\\
(iii) The following spectral decomposition holds: 
$
\Omega(\theta)=\sum_{\sigma=1}^s \omega_\sigma
(\theta)\Pi_\sigma(\theta)$,
$\theta\in \T^d\setminus{\cal C}_*$,
 where 
$\Pi_\sigma(\theta)$ is a real-analytic function on
$\T^d\setminus{\cal C}_*$.
\end{lemma}

For $\theta\in \T^d\setminus{\cal C}_*$, denote by
$\nabla^2\omega_\sigma(\theta)$ the matrix of second partial derivatives. 
The next condition on $V$ is as follows.
\smallskip\\
{\bf E4}. 
The functions  $D_\sigma(\theta)
:=\det\big(\nabla^2\omega_\sigma(\theta)\big)$ 
do not vanish identically on
$\T^d\setminus{\cal C}_*$, $\sigma=1,\ldots,s$.
\medskip

Let us write
 \be\label{c0ck}
{\cal C}_0=\{\theta\in \T^d:\det \hat
 V(\theta)=0\},\,\,\,\,
{\cal C}_\sigma=\{\theta\in \T^d\setminus {\cal
C}_*:\,D_\sigma(\theta)=0\},\,\,\, \sigma=1,\dots,s.
 \ee
Then the Lebesgue measure of ${\cal C}_\sigma$ vanishes, $\sigma=0,1,...,s$
(see \cite[Lemma 2.3]{DKS1}).
Usually, the dispersion relations $\omega_k(\theta)$ satisfying
the condition $\omega_k(0)=0$ are called {\it acoustic}.
\smallskip\\
{\bf E5}.  For each $\sigma\ne \sigma'$, 
the identities $\omega_\sigma(\theta)
\pm\omega_{\sigma'}(\theta)\equiv\const_\pm$ for 
$\theta\in \T^d\setminus {\cal C}_*$, do not hold  with $\const_\pm\ne 0$.
\medskip

This condition holds trivially  for $n=1$.
\medskip\\
{\bf E6}. $\Vert \hat V^{-1}(\theta)\Vert\in L^1(\T^d)$.
\medskip

If ${\cal C}_0=\emptyset$, then $\|\hat{V}^{-1}(\theta)\|$ is
bounded, and {\bf E6} evidently holds.
\begin{remark}
{\rm 
(i) Instead of condition {\bf E1} we may assume that
$|V(z)|\le C(1+|z|)^{-N}$ with an $N>0$.
However, in this case, we should assume in addition 
that there exists a set ${\cal K}\subseteq \T^d$ such that
mes$(\T^d\setminus {\cal K})=0$ and for $\theta\in{\cal K}$,
$\omega_\sigma\in C^3({\cal K})$, $\Pi_\sigma\in C({\cal K})$,
$\omega_\sigma(\theta)\not=0$, det$(\nabla^2\omega_\sigma(\theta))\not=0$,
$\nabla\omega_\sigma(\theta)\not=0$, $\sigma=1,\dots,s$.
Note that if condition {\bf E1} holds, 
${\cal K}=\T^d\setminus {\cal C}$ with ${\cal C}$ defined in (\ref{calC}).
\medskip

(ii) Conditions {\bf E1}--{\bf E6} are satisfied, in particular,
in the case of the {\it nearest neighbor crystal} 
(see \cite{DKS1})
in which 
the interaction matrix $V(z)=(V_{kl}(z))_{k,l=1}^n$ is of the form
$$
V_{kl}(z)=0\,\,\mbox{ for }\,k\not=l,\,\,\,\,
V_{kk}(z)=\left\{\ba{ll}-\gamma_k&\mbox{for }\, |z|=1,\\
2d\gamma_k+m_k^2& \mbox{for }\, z=0,\\
0&\mbox{for }\, |z|\ge2,\ea\right.  \quad k=1,\dots,n,
$$ 
with $\gamma_k>0$ and $m_k\ge0$.
In this case, the Hamiltonian functional has a form
$$
 H(v_0, v_1)= \frac{1}{2} \sum\limits_{z\in\Z^d}
\sum\limits_{k=1}^n \Big(|v_{1k}(z)|^2 +m_k^2|v_{0k}(z)|^2+
\sum\limits_{j=1}^d \gamma_k|v_{0k}(z+e_j)-v_{0k}(z)|^2\Big),
$$
$e_j=(\delta_{1j},\dots,\delta_{d\,j})$, and equation (\ref{CP1''}) becomes
$$
\ddot v_k(z,t)=(\gamma_k\Delta_L-m_k^2)v_k(z,t),\quad k=1,\dots,n,
$$
where $\Delta_L$ stands for the discrete
Laplace operator on the lattice $\Z^d$,
$$
\Delta_L v(z):=
\sum\limits_{j=1}^d (v(z+e_j)-2v(z)+v(z-e_j)).
$$
Therefore, the eigenvalues of  $\Omega(\theta)$ are
 \be\la{omega}
 \tilde{\omega}_k(\theta)=
\sqrt{\, 2 \gamma_k(1-\cos\theta_1)+...+2 \gamma_k
(1-\cos\theta_d)+m_k^2}\,,\quad k=1,\dots,n.
 \ee
These eigenvalues still have to be labelled according to magnitude
and degeneracy as in Lemma \ref{lc*}.
Clearly, conditions {\bf {E1}}--{\bf {E5}} hold, and ${\cal C}_*=\emptyset$.
If $m_k>0$ for any $k$, then the set ${\cal C}_0$ is empty and
condition {\bf E6} holds automatically. Otherwise, if $m_k=0$ for
some $k$, then ${\cal C}_0=\{0\}$. In this case,
{\bf E6} is equivalent to the condition $\om_k^{-2}(\theta)\in
L^1(\T^d)$. Therefore, 
conditions {\bf E1}--{\bf E6} hold  if either (i) $d\ge 3$
or (ii) $d=1,2$ and $m_k >0$ for any $k$.
}
\end{remark}
\begin{lemma}\label{lemma}(see \cite[Proposition 2.5]{DKS1})
Let conditions {\bf E1} and {\bf E2} hold, and $\al\in\R$. Then\\
(i) for any  $X_0 \in {\cal H}_{\al}$, there exists  a unique solution
$X(t)\in C(\R, {\cal H}_{\al})$  to the Cauchy problem (\re{CP1'});\\
(ii) for any $t\in\R$,
the operator  $U(t):X_0\mapsto X(t)$ is continuous
on ${\cal H}_{\al}$.
\end{lemma}

The proof of Lemma \ref{lemma} is based on the following formula for 
the solution  $X(t)$ of problem (\ref{CP1'}):  
\be\la{solGr}
X(t)=\sum\limits_{z'\in\Z^d}{\cal G}_t(z-z')X_0(z'),
\ee
where the function ${\cal G}_t(z)$ has the Fourier representation   
\be\la{Grcs}  
{\cal G}_t(z):= F^{-1}_{\theta\to z}[  
\exp\big(\hat{\cal A}(\theta)t\big)]  
=(2\pi)^{-d}\int\limits_{\T^d}e^{-iz\cdot \theta}  
\exp\big(\hat{\cal A}(\theta)t\big)\,d\theta  
\ee  
with 
\be\la{hA}   
\hat{\cal A}(\theta)=\left( \begin{array}{cc}   
0 & 1\\   
-\hat V(\theta) & 0   
\end{array}\right),\quad \theta\in \T^d.    
\ee

\subsection{The family of initial measures}\label{sec1.2}
Let $\ve>0$ be a small scale parameter,
$\{\mu^\varepsilon_0,\varepsilon>0\}$ be a family of initial
measures. 
To formulate the main conditions {\bf V1} and {\bf V2} 
on the covariance of $\mu^\varepsilon_0$,
let us introduce  the complex $2n\times 2n$
matrix-valued function $ {\bf R}_0(r,z)=({\bf R}_0^{ij}(r,z))^1_{i,j=0}$, 
$r\in\R^d$, $z\in\Z^d$, with the following properties.
\medskip\\
{\bf I1}. For every fixed $r\in\R^d$ and $i,j=0,1$, 
the bound holds,
\be\la{3.2} 
|{\bf R}_0^{ij}(r,z)|\le C(1+|z|)^{-\gamma},\quad z\in\Z^d,
 \ee
where $C$ is some  positive constant, $\gamma >d$.\medskip\\
{\bf I2}. For every fixed $r\in\R^d$, 
$\hat{\bf R}_0(r,\theta)$ satisfies
$$
 \hat {\bf R}_0^{00}(r,\theta)\ge0,\,\,
\hat {\bf R}_0^{11}(r,\theta)\ge0,\,\,
\hat {\bf R}_0^{01}(r,\theta)=\hat {\bf R}_0^{10}(r,\theta)^*,
\quad \theta\in \T^d.
$$
{\bf I3}. For every fixed $r\in\R^d$ and $\theta\in \T^d$,
the matrix $\hat{\bf R}_0(r,\theta)$ is non-negative definite.\medskip\\
{\bf I4}. For every $\theta\in\T^d$,
 $\hat {\bf R}^{ij}_0(\cdot,\theta)$  are $C^d$ functions, 
 and the function
$$
r\to \sup_{\theta\in \T^d}\max_{i,j=0,1}
\max_{\alpha_k=0,1,\,k=1,\dots,d}
\Big|\frac{\partial^{\alpha_1+\dots+\alpha_d}}{\partial r_1^{\alpha_1}
\dots\partial r_d^{\alpha_d}}
\hat {\bf R}_0^{ij}(r,\theta)\Big|
$$
is bounded uniformly on bounded sets.
\medskip

To derive the equation of the Navier--Stokes type
we need an additional condition {\bf I4'}.\\
{\bf I4'}.
$\hat {\bf R}_0(\cdot,\theta)\in L^1(\R^d)$, 
 $\forall\theta\in\T^d$, and
 there exist constants $C>0$ and $N>d+3$ such that
\be\label{tildeR}
\sup\limits_{\theta\in\T^d}|\tilde {\bf R}_0(s,\theta)|
\le C(1+|s|)^{-N}, \quad s\in\R^d.
\ee 
Here by $\tilde {\bf R}_0(s,\theta)$ 
we denote the Fourier transform of $\hat {\bf R}_0(r,\theta)$
with respect to $r$:
\be\label{Ftr}
\tilde {\bf R}_0(s,\theta)=F_{r\to s}[\hat {\bf R}_0(r,\theta)]
=\int\limits_{\R^d} 
e^{is\cdot r}\hat {\bf R}_0(r,\theta)\,dr,
\quad s\in\R^d,\quad \theta\in\T^d.
\ee

This condition could be weakened
(see  condition D' in \cite{DPST2} for the case $d=n=1$). 
Instead of {\bf I4'} we may assume that for each
 $\theta\in\T^d$ 
the function $\hat {\bf R}_0(\cdot,\theta)$ admits the representation
\be\la{2.3'}
\hat {\bf R}_0(r,\theta)=(2\pi)^{-d}\int\limits_{\R^d} 
e^{-is\cdot r}\mu(\theta,ds), 
\ee
where $\mu(\theta,ds)$ is some Borel (complex-valued)
measure on $\R^d$, depending on the parameter $\theta\in\T^d$. 
Let $|\mu|(\theta,\cdot)$, $\theta\in\T^d$, 
be the {\it total variation} of the 
measure $\mu(\theta,\cdot)$. This means (see \cite[\S 29]{Halmos}) that
for any measurable set $E\subset\R^d$,
$|\mu|(\theta,E)=\sup|\int\limits_{E} f(s)\mu(\theta,ds)|$,
where supremum is taken over all measurable functions $f$ such that
$|f|\le1$.
We assume that 
there exist constants $C>0$ and $\delta>d^2/2+2d+1$ such that
for any $n\in\N$, 
\be\label{Varm}
\sup_{\theta\in\T^d}|\mu|(\theta,K_n)\le C (1+n)^{-\delta},
\quad \mbox{where }\, K_n=\{s\in\R^d:n\le|s|<n+1\}.
\ee

Denote by  $\E^\ve_0$ expectation with respect to the measure
$\mu_0^\ve$, and by 
$ Q^{ij}_{\ve}(z,z')= \E^\ve_0\big(X^i(z)\otimes
X^j(z')\big)$, $z,z'\in \Z^d$, $i,j=0,1$, 
the correlation functions of $\mu_0^\ve$. 
Assume that $\E^\varepsilon_0\big(X(z)\big)=0$
and functions
$Q^{ij}_{\ve}(z,z')$ satisfy the following conditions
{\bf V1} and {\bf V2}.
\smallskip\\
{\bf V1}. For any $\ve >0$, $z,z'\in\Z^d$,
\beqn\la{2.4}
\left|Q^{ij}_{\ve}(z,z')- {\bf R}_0^{ij}(\ve z,z-z')\right|\le
C\min\Big[(1+|z-z'|)^{-\gamma}, \ve |z-z'|\Big],
\eeqn
with the constants $C$, $\gamma$ as in (\ref{3.2}).
\medskip\\
{\bf V2}. For any $\ve>0$ and all $z,z'\in\Z^d$, $i,j=0,1$, 
$
|Q^{ij}_{\ve}(z,z')|\le C(1+|z-z'|)^{-\gamma}
$
with the constants $C$ and $\gamma$ as in (\ref{3.2}).
\medskip\\
{\bf Remarks}
(i) Condition {\bf V1} can be formulated in the another form 
(see \cite{DPST, DS}). Namely, 
for any $\ve >0$ there exists an  even integer $N_\ve$
such that\\
a) for all $M\in \R^d$ and $z,z'\in I_{M}$,
$$
\left|Q^{ij}_{\ve}(z,z')- {\bf R}_0^{ij}(\ve M,z-z')\right|\le
C\min[(1+|z-z'|)^{-\gamma}, \ve N_\ve],
$$
where $C$, $\gamma$ are the constants from (\ref{3.2}), 
and $I_{M}$ is the cube
centered at the point $M$ with edge length $N_\ve$,
$
I_{M} =\{z=(z_1,\dots,z_d)\in\Z^d:\, |z_j-M_j|\le
N_\ve/2,\,M=(M_1,\dots,M_d)\};
$\\
b) $N_\ve\sim \ve^{-\beta}$ as $\ve\to 0$, with some
$\beta\in(1/2,1)$.
\medskip

The formulation of {\bf V1} in the form (\ref{2.4})
is more natural and convenient for our proof. 
\medskip\\
(ii) By conditions {\bf V1} and {\bf V2},
$\sum\limits_{p\in\Z^d} e^{i\theta\cdot p}
Q^{ij}_\ve([r/\ve+p/2],[r/\ve-p/2])\to 
\hat{\bf R}^{ij}_0(r,\theta)$ as $\ve\to0$
uniformly in $r\in\R^d$ and $\theta\in\T^d$.
Here and below $[x]=([x_1],\dots,[x_d])$ for $x\in\R^d$
and
$[x_i]$ stands for the integer part of $x_i\in\R^1$, $i=1,\dots,d$.

\subsection{Example of initial measures $\mu_0^\ve$}
We construct Gaussian initial measures $\mu_0^\ve$
satisfying conditions {\bf V1} and {\bf V2}.
At first,  we introduce matrix-valued functions 
$q_0^{ij}(z)$, $z\in\Z^d$, such that $q_0^{ij}(z)=0$ for $i\not=j$, and
$$
\hat q_0^{ii}(\theta)= F_{z\to\theta}[q_0^{ii}(z)]\in L^1(\T^d),
\quad \hat q_0^{ii}(\theta)\ge0,\quad i=0,1.
$$
Next, we set 
\be\label{Rex}
{\bf R}_0^{ij}(r,z)=T(r)q_0^{ij}(z),\quad r\in\R^d,\quad z\in\Z^d,
\ee 
where $T\in C^d(\R^d)$, $T(r)\ge0$,
$\sup\limits_{r\in\R^d}\sup\limits_{|\al|\le d}|D^\al T(r)|\le C<\infty$,
$|F_{r\to s}[T(r)]|\le C(1+|s|)^{-N}$ with an $N>d$.
Finally, we put
\be\label{Qexam}
Q_\ve^{ij}(z,z')=\sqrt{T(\ve z)T(\ve z')}q_0^{ij}(z-z'),
\quad z,z'\in\Z^d,\quad i,j=0,1.
\ee
By the Minlos theorem, for any $\ve>0$, there exists a  
Borel Gaussian measure $\mu_0^\ve$ on ${\cal H}_\al$, $\al<-d/2$,
with the correlation functions $Q_\ve^{ij}(z,z')$, because
\beqn
\E_0^\ve(\Vert X\Vert^2_\al)&=&\sum\limits_{z\in\Z^d}
(1+|z|^2)^{\al}\tr[Q_\ve^{00}(z,z)+Q_\ve^{11}(z,z)]
\nonumber\\
&=&\sum\limits_{z\in\Z^d}
(1+|z|^2)^{\al} T(\ve z)\tr[q_0^{00}(0)+q_0^{11}(0)]
\nonumber\\
&\le& C(\al,d) 
\tr\int\limits_{\T^d}(\hat q_0^{00}(\theta)+\hat q_0^{11}(\theta))\,d\theta
\le C_1<\infty. \nonumber
\eeqn
Let us assume that there exist constants $C>0$ and $\gamma>d$ such that 
\be\label{qex}
|q_0^{ij}(z)|\le C (1+|z|)^{-\gamma}, \quad z\in\Z^d.
\ee
Then $Q_\ve^{ij}(z,z')$ satisfy the conditions {\bf V1} and {\bf V2}. 
\begin{definition}\label{def2.5}
Formally, Gibbs measure $g$ is
$
g(dX)= \frac{1}{Z}e^{-\frac{\beta}{2}\sum_{z} H(X)}\prod_{z\in\Z^d}dX(z),
$
where $H(X)$ is defined in (\ref{H}),
$\beta= T^{-1}$, $T\ge0$ is the corresponding absolute temperature.
We define the Gibbs measure $g$ on ${\cal H}_\al$, $\al<-d/2$,
as the Gaussian measure with the correlation matrices 
defined by their Fourier transform as
$\hat q^{00}(\theta)= T\hat V^{-1}(\theta)$,
$\hat q^{11}(\theta)= T I$,
$\hat q^{01}(\theta)= \hat q^{10}(\theta)= 0$,
where $I$ stands for the unit matrix in $\R^n$.
\end{definition}

 Let $\hat q_0^{00}(\theta)=\hat V^{-1}(\theta)$, 
$\hat q^{11}_0(\theta)=I$, $\theta\in\T^d$. Then the functions
${\bf R}^{ij}_0(r,z)$
defined in (\ref{Rex}) are correlation matrices of the Gibbs measures
$g_r$, $r\in\R^d$, with $\beta=1/T(r)$.
If we assume, in addition, that ${\cal C}_0=\emptyset$,
i.e. $\det \hat V(\theta)\not=0$, $\forall\theta\in \T^d$,
then the bound (\ref{qex}) holds, and 
$Q_\ve^{ij}(z,z')$ defined in (\ref{Qexam}) 
satisfy the conditions {\bf V1} and {\bf V2}. 

\setcounter{equation}{0}
\section{Main results}

\begin{definition}\la{d2.9}
(i) $\mu^\ve_t$ is a Borel probability measure on ${\cal H}_{\al}$
which gives the distribution of the random solution $X(t)$,
$\mu^\ve_t(B) = \mu_0^\ve(U(-t)B)$, where 
$B\in {\cal B}({\cal H}_{\al})$ and  $t\in \R$.
\medskip\\ 
(ii) The correlation functions of the  measure $\mu^\ve_t$ are  defined by
$$
Q_{\ve,t}^{ij}(z,z')= \E^{\ve}_t \left(X^i(z)\otimes X^j(z')\right)
= \E^{\ve}_0\big(X^i(z,t)\otimes  X^j(z',t)\big), 
\quad i,j= 0,1,\quad z,z'\in\Z^d,
$$ 
where $\E^{\ve}_t$ stands for expectation with 
respect to the measure $\mu_t^\ve$, 
and $X^i(z,t)$ are  the components of the random solution 
$X(t)=(X^0(\cdot,t),X^1(\cdot,t))$ to problem (\ref{CP1'}).
\medskip\\
(iii) Let $T_h$, $h\in \Z^d$, be the group of space
translations: $T_hX(z)=X(z-h)$, $z\in\Z^d$.  
For $\tau\not=0$, $r\in\R^d$, and $\kappa>0$,
the measures $\mu^\varepsilon_{\tau/\varepsilon^\kappa,r/\ve}$ 
are defined by the rule
$$
\mu^\varepsilon_{\tau/\varepsilon^\kappa,r/\ve}(B)=
 \mu^\varepsilon_{\tau/\varepsilon^\kappa}(T_{[r/\varepsilon]}B),
\quad\mbox{where }\,B\in {\cal B}({\cal H}_{\al}).
$$ 
\end{definition}
{\bf Remarks}
(i) In addition to conditions {\bf V1} and {\bf V2}, let us assume
that the measures $\mu_0^\ve$ satisfy the {\it mixing condition}. 
To formulate this condition, denote by $\sigma ({\cal A})$,
${\cal A}\subset \Z^d$, the $\sigma $-algebra on
${\cal H}_{\alpha}$ generated by $X_0(z)$ with $z\in{\cal A}$.
Define the Ibragimov mixing coefficients of the probability  measure
$\mu^\ve_0$ on ${\cal H}_{\alpha}$ by the rule 
$$ 
\varphi_\ve(r)= \sup_{\scriptsize{\ba{cc} {\cal A},{\cal B}\subset \Z^d\\
\dist({\cal A},\,{\cal B})\geq r \ea}}
\sup_{\scriptsize{
\ba{cc} A\in\sigma({\cal A}),B\in\sigma({\cal B})\\ \mu^\ve_0(B)>0\ea}}
\frac{|\mu^\ve_0(A\cap B) - \mu^\ve_0(A)\mu^\ve_0(B)|}{ \mu^\ve_0(B)}.
$$
 A measure $\mu^\ve_0$ is said to satisfy 
the {\it strong uniform Ibragimov mixing condition} if
$\varphi_\ve(r)\to 0$ as $r\to\infty$.
Moreover, we assume that
$\forall r\in\R^d$,
$\sup_{\ve>0}\varphi_\varepsilon(r)\le C(1+r)^{-2\gamma}$,
with the constant $\gamma$ as in (\ref{3.2}). 
Note that the last bound on $\varphi_\varepsilon(r)$ implies
condition {\bf V2}.

Then  for $\tau\neq 0$, $r\in\R^d$,  
the measures $\mu^\varepsilon_{\tau/\varepsilon,r/\ve}$
converge weakly to a limit measure $\mu^{G}_{\tau,r}$
on the space ${\cal H}_\alpha$, $\alpha<-d/2$.
By definition, this means that
$$
\lim_{\ve\to0}
\int f(X)\mu_{\tau/\varepsilon,r/\varepsilon}^\varepsilon(dX)
=\int f(X)\mu^{G}_{\tau,r}(dX)
$$
for any bounded continuous functional $f$ on ${\cal H}_\alpha$. 
Moreover, the limit measure 
$\mu^{G}_{\tau,r}$ is a Gaussian measure on ${\cal H}_\al$
(see Theorem 3.3 in \cite{DS}). 
\smallskip\\
(ii) 
Let $\kappa<1$. Then for $\tau\neq 0$, $r\in\R^d$, 
the following limit holds,
 $\mu_{\tau/\varepsilon^\kappa,r/\varepsilon}^\varepsilon 
\rightharpoondown\mu_r$ as $\varepsilon\to0$
in the sense of weak convergence on ${\cal H}_\alpha$, $\al<-d/2$.
Moreover, the limit measure $\mu_r$ is Gaussian,
its correlation matrix does not depend on $\tau$ and has a form
(in Forier transform)
$$
\frac{1}2\sum_{\sigma=1}^{s}\Pi_\sigma(\theta)
\Big( \hat {\bf R}_0(r,\theta)+C_\sigma(\theta)
 \hat {\bf R}_0(r,\theta)C_\sigma^*(\theta)\Big)
\Pi_\sigma(\theta),\quad \theta\in\T^d\setminus{\cal C}_*,
$$
where $C_\sigma(\theta)$ is defined in (\ref{Csigma}).

\subsection{Equation of Euler type}\label{sec2.1}
In this subsection we put $\kappa=1$.
Let us introduce the matrix $q_{\tau,r}(z)$, $z\in\Z^d$, $\tau\in\R$,
$r\in\R^d$.
In Fourier space,
\beqn
\hat q_{\tau,r}(\theta)&=&\frac14 \sum_{\sigma=1}^{s}
\Pi_\sigma(\theta)\Big[\sum\limits_\pm
(I\pm iC_\sigma(\theta)) 
\hat{\bf R}_0(r\pm\nabla\om_\sigma(\theta)\tau,\theta) 
(I\mp iC^*_\sigma(\theta)) \Big]
\Pi_\sigma(\theta)\label{qt}\\
&=&\sum_{\si=1}^{s}
\Pi_\si(\theta)\big({\bf M}_+^{\sigma}(\tau,r;\theta)+
i\,{\bf M}_-^{\sigma}(\tau,r;\theta)\big)\Pi_\si(\theta),
\quad \theta\in \T^d\setminus{\cal C}_*\,,\la{qtaur}
\eeqn
where $\Pi_\sigma(\theta)$ is the spectral projection introduced 
in Lemma \re{lc*} (iii), 
\beqn\la{Pi}
\ba{lll}
 {\bf M}^\sigma_{+}(\tau,r;\theta)
&=&\frac{1}2\big( R^\sigma_+(\tau,r;\theta)+C_\sigma(\theta)
 R^\sigma_+(\tau,r;\theta)C_\sigma^*(\theta)\big),\medskip\\
{\bf M}_{-}^\sigma(\tau,r;\theta)&
=&\frac{1}2\big(C_\sigma(\theta) R_{-}^\sigma(\tau,r;\theta)
- R_{-}^{\sigma}(\tau,r;\theta)C^*_\sigma(\theta)\big),
\ea
\eeqn
\be\label{Csigma}
C_\sigma(\theta)=\left(\ba{cc}0&\omega^{-1}_\sigma(\theta)\\
-\omega_\sigma(\theta)&0\ea\right),\quad \sigma=1,\dots,s,
\ee
\beqn\la{gclimcor2}
 R^\sigma_\pm(\tau,r;\theta)&=&\frac{1}2
\Big(\hat {\bf R}_0(r+\nabla\om_\sigma(\theta)\tau,\theta)
\pm\hat {\bf R}_0(r-\nabla\om_\sigma(\theta)\tau,\theta)\Big).
\eeqn
\begin{theorem}\la{t2.4}  (see \cite[Theorem 4.1]{DS})
 Let the conditions {\bf V1}, {\bf V2} and {\bf E1}--{\bf E6} hold.
Then for any $r\in \R^d$, $z,z'\in \Z^d$, $\tau\neq 0$,
the correlation functions of measures 
$\mu^\ve_{\tau/\ve,r/\ve}$ converge to a limit,
\beqn\nonumber
 \lim_{\ve\to 0}Q_{\ve,\tau/\ve}([r/\ve]+z,[r/\ve]+z')
=q_{\tau,r}(z-z').
\eeqn
\end{theorem}
\begin{cor}\la{cor2.5}
Write $f_\sigma(\tau,r;\theta)=
\Pi_\sigma(\theta)\hat q_{\tau,r}(\theta)\Pi_\sigma(\theta)$,
$\sigma=1,\dots,s$.
Then the function $f_\sigma(\tau,r;\theta)$
satisfies the "hydrodynamic" Euler type equation:
\beqn
\partial_\tau f_\sigma(\tau,r;\theta)&=&iC_\sigma(\theta)
\nabla\omega_\sigma(\theta)
\cdot \nabla_r f_\sigma(\tau,r;\theta),\quad r\in\R^d,\quad \tau>0,
\label{Euler}\\
f_\sigma(\tau,r;\theta)|_{\tau=0}&=&
\ds\frac12\Pi_\sigma(\theta)
\left(\hat {\bf R}_0(r,\theta)+C_\sigma(\theta)\hat {\bf R}_0(r,\theta)
C^*_\sigma(\theta)\right)\Pi_\sigma(\theta).\label{initial}
\eeqn
\end{cor}
\begin{remarks}\label{Re3.4} 
{\rm
(i) Note
that $\hat q_{\tau,r}(\theta)=(\hat q^{ij}_{\tau,r}(\theta))_{i,j=0}^1$ 
satisfies the {\it equilibrium condition}, 
i.e., $\hat q^{11}_{\tau,r}(\theta)=
\Omega^2(\theta)\hat q^{00}_{\tau,r}(\theta)$,
$\hat q^{01}_{\tau,r}(\theta)=-\hat q^{10}_{\tau,r}(\theta)$.
Moreover, $\hat q^{ii}_{\tau,r}(\theta)^*=
\hat q^{ii}_{\tau,r}(\theta)\ge0$,
$\hat q^{01}_{\tau,r}(\theta)^*=\hat q^{10}_{\tau,r}(\theta)$.
\medskip\\
(ii)  In the case when $\kappa\in[1,2)$, the correlation matrices of
measures $\mu^\ve_{\tau/\ve^\kappa,r/\ve}$ converge as $\ve\to0$
to the same matrices $q_{\tau,r}(z-z')$ as in Theorem \ref{t2.4}.
This result can be proved by the similar way as Theorem \ref{t2.4}.
}
\end{remarks}

Theorem \ref{t2.4} and Corollary \ref{cor2.5} can be rewritten
in the terms of the Wigner matrices.
Write
$({\cal V}^{k} v)(x)=F^{-1}_{\theta\to x}[\hat V^k(\theta)\hat v(\theta)]$,
$k\in\R$,
and introduce the complex-valued field
 \beqn\la{2.1}
 a(x)= \frac{1}{\sqrt{2}}\Big({\cal V}^{1/4}
 v_0(x)+i{\cal V}^{-1/4}v_1(x)\Big)\in \C^n\,,
\quad x\in \Z^d\,,
 \eeqn
with complex conjugate field $a(x)^*=(1/\sqrt{2})\left({\cal V}^{1/4}
 v_0(x)-i{\cal V}^{-1/4}v_1(x)\right)$.
Define the scaled $n\times n$ Wigner
matrix as
 \beqn\la{3.15}
W^\varepsilon(\tau,r;\theta)=
\sum_{y\in \Z^d}e^{i\theta\cdot y}\,
\E^\varepsilon_{\tau/\varepsilon}
\big(a^\ast([r/\ve+y/2])
\otimes a([r/\ve-y/2])\big)\,.
 \eeqn
By properties of $\mu^\varepsilon_0$, the following limit
exists
 \beqn\la{3.12'}
 \lim_{\varepsilon\to 0}W^\varepsilon(0,r;\theta)&=&
 \frac{1}{2}\Big(\Omega^{1/2}\hat {\bf R}_0^{00}(r,\theta)\Omega^{1/2}+
 \Omega^{-1/2}\hat {\bf R}_0^{11}(r,\theta)\Omega^{-1/2}\nonumber\\
 &&+i\Omega^{1/2}\hat {\bf R}_0^{01}(r,\theta)\Omega^{-1/2}-i
 \Omega^{-1/2}\hat {\bf R}_0^{10}(r,\theta)\Omega^{1/2}\Big)
\equiv W(0,r;\theta)\,.
 \eeqn
\begin{theorem} \la{the2} (see \cite[Theorem 3.2]{DS})
Let the conditions {\bf{V1}}, {\bf{V2}} and {\bf{E1}}--{\bf{E6}}
hold. Then for any $r\in\R^d$ and
$\tau\neq 0$ the following limit exists
 \beqn\nonumber
 \lim_{\varepsilon\to 0} W^\varepsilon(\tau,r;\theta)=
 W^{\rm p}(\tau,r;\theta)\quad \mbox{(in the sense of distributions)},
 \eeqn
where
 \beqn\la{3.16a}
 W^{\rm p}(\tau,r;\theta)=\sum^s_{\sigma=1}
 \Pi_\sigma(\theta)W(0,r-\tau\nabla\omega_\sigma(\theta);\theta)
 \Pi_\sigma(\theta)\,.
 \eeqn
\end{theorem}
{\bf Remarks}
(i) 
The limit correlation matrices 
$\hat{q}_{\tau,r}(\theta)=(\hat{q}^{ij}_{\tau,r}(\theta))_{i,j=0}^1$  
(see (\ref{qtaur}))
are expressed by $W^{p}(\tau,r;\theta)$ as
  \beqn
\ba{rcl}
  \Omega(\theta)\hat{q}^{00}_{\tau,r}(\theta)&=&
  \Omega(\theta)^{-1}\hat{q}^{11}_{\tau,r}(\theta)=
  \frac{1}{2}\big(W^{p}(\tau,r;\theta)+
  W^{p}(\tau,r;-\theta)^*\big),\nonumber\\  
\hat{q}^{01}_{\tau,r}(\theta)&=&
  -\hat{q}^{10}_{\tau,r}(\theta)=
  -\frac{i}{2}\big(W^{p}(\tau,r;\theta)-
  W^p(\tau,r;-\theta)^*\big)\,.  \nonumber
 \ea
\eeqn
(ii) The matrix 
$f_\sigma(\tau,r;\theta)
\equiv\Pi_\sigma(\theta)W^p(\tau,r;\theta)\Pi_\sigma(\theta)$,
 $\sigma=1,\dots,s$, satisfies the {\it energy transport} equation
(see \cite[p.656]{DS})
 \beqn\nonumber
\partial_\tau f_\sigma(\tau,r;\theta)
+\nabla\omega_\sigma(\theta)\cdot \nabla_r
 f_\sigma(\tau,r;\theta)=0,\,\,\,\,\tau>0,\,\,\,r\in\R^d,
\eeqn
with the initial condition
$f_\sigma(\tau,r;\theta)|_{\tau=0}=
\Pi_\sigma(\theta)W(0,r;\theta)\Pi_\sigma(\theta)$, $r\in\R^d$.

\subsection{Navier--Stokes equation}\label{sec2.2}
In this subsection we study the behaviour (as $\ve\to0$)
of the correlation functions of $\mu^\ve_{\tau/\ve^\kappa,r/\ve}$
for $\kappa=2$,
and obtain the next term of the decomposition in $\ve$ 
to equation (\ref{Euler}).
Let us introduce the matrix
 $q^\ve_{\tau,r}(z)$, $z\in\Z^d$, $r\in\R^d$, $\tau\not=0$, $\ve>0$,
which has the following form (in the Fourier transform)
\beqn\label{ftaur}
\hat q^\ve_{\tau,r}(\theta)=\frac14 \sum_{\sigma=1}^{s}
\Pi_\sigma(\theta)\Big[\sum\limits_\pm
(I\pm iC_\sigma(\theta)) 
A^{\pm}_{\ve,\sigma}(\tau,r;\theta) 
(I\mp iC^*_\sigma(\theta)) \Big]
\Pi_\sigma(\theta),\,\,\, \theta\in \T^d\setminus{\cal C}_*,
\eeqn
where 
matrices $C_\sigma(\theta)$ are defined in (\ref{Csigma}),
\beqn\label{Apm}
A^{\pm}_{\ve,\sigma}(\tau,r;\theta)
=\int\limits_{\R^d}
\hat {\bf R}_0(r\pm\nabla\omega_\sigma(\theta)\tau/\ve -x,\theta)
K^{\pm}_{\sigma}(\tau,x,\theta)\,dx,
\eeqn
\be\label{K}
K^{\pm}_{\sigma}(\tau,x,\theta):=F^{-1}_{y\to x}
[e^{\mp i(\tau/2) y\cdot (\nabla^2\omega_\sigma(\theta))y}],
\quad x\in\R^d
\ee
(see also formula (\ref{K2})).
\begin{theorem}\label{the2'}
 Let conditions {\bf I1}--{\bf I4'},
{\bf V1}, {\bf V2} and {\bf E1}--{\bf E6} hold.
Then for any $\tau\not=0$, $r\in\R^d$, $z,z'\in\Z^d$, 
the correlation functions of measures $\mu^\ve_{\tau/\ve^2,r/\ve}$
have the following asymptotics
\beqn\label{asymp}
\lim_{\ve\to0}\left(Q_{\ve,\tau/\ve^2}([r/\ve]+z,[r/\ve]+z')-
q^{\ve}_{\tau,r}(z-z')\right)=0,
\eeqn
where the matrix 
$q^{\ve}_{\tau,r}(z)=F^{-1}_{\theta\to z}[\hat q^{\ve}_{\tau,r}(\theta)]$ 
is defined by (\ref{ftaur}).
\end{theorem}

We omit the proof of this theorem since 
it can be proved by the similar technique as Theorem \ref{the2'+}, below.
\medskip\\
{\bf Remark}.
Note that $\hat q^\ve_{\tau,r}(\theta)$ 
satisfies the equilibrium condition 
(see Remarks \ref{Re3.4} (i)).
\medskip

Set $\tau=\ve t$. 
It follows from formulas (\ref{ftaur})--(\ref{K})
that 
$$
A^\pm_{\ve,\sigma}(\ve t,r;\theta)|_{\ve=0}
= \hat {\bf R}_0(r\pm\nabla\omega_\sigma(\theta)t,\theta).
$$
Hence, $\hat q^{\ve}_{\ve t,r}(\theta)|_{\ve=0}= \hat q_{t,r}(\theta)$,
where $\hat q_{t,r}(\theta)$ is defined in (\ref{qt}).
Denote 
\be\label{nabla-k}
\nabla^k\omega_\sigma(\theta)
\cdot \nabla^k_r f(r):=\sum\limits_{i_1,\dots,i_k=1}^d
\frac{\partial^k \omega_\sigma(\theta)}
{\partial \theta_{i_1}\dots\partial \theta_{i_k}}\,
\frac{\partial^k f}
{\partial r_{i_1}\dots\partial r_{i_k}},\quad k\in\N.
\ee
\begin{cor}\label{cor2.7}
Let  $r\in\R^d$, $t\in\R$, $\theta\in\T^d\setminus {\cal C}_*$.
 It follows from formulas (\ref{ftaur})--(\ref{K}) that 
for each $\sigma=1,\dots,s$,  the matrix-valued function 
$f^\ve_\sigma(t,r;\theta) \equiv
\Pi_\sigma(\theta)\hat q^{\ve}_{\ve t,r}(\theta)\Pi_\sigma(\theta)$ 
evolves according to the following (Navier--Stokes type) equation
\beqn\label{N-S}
\partial_t f^\ve_\sigma(t,r;\theta)=iC_\sigma(\theta)
\Big(\nabla\omega_\sigma(\theta)
\cdot \nabla_r 
+\frac{i\ve}{2} \nabla^2\omega_\sigma(\theta)
\cdot \nabla^2_r \Big)f^\ve_\sigma(t,r;\theta),\quad t\in\R,\quad r\in\R^d,
\eeqn
with the initial condition (\ref{initial}).
\end{cor}

\subsection{Corrections of the higher order}\label{sec3.3}

To obtain the "corrections" of order $\ve^2$, $\ve^3$,\dots 
to equation (\ref{N-S}),
it is neccesary to study the behaviour (as $\ve\to0$) 
of the correlation functions of measures 
$\mu^\ve_{\tau/\ve^\kappa,r/\ve}$ with $\kappa>2$.
\begin{theorem}\label{t-app}
Let $\kappa\ge 2$, $r\in\R^d$, $\tau>0$. Then $\forall z,z'\in\Z^d$,
$$
Q_{\ve,\tau/\ve^\kappa}([r/\ve]+z,[r/\ve]+z')-
q^{\ve,[\kappa]}_{\tau,r}(z-z')\to0,\quad \ve\to0.
$$
The matrix $q^{\ve,k}_{\tau,r}(z)$ ($k=2,3,\dots$)
has the following form (in the Fourier transform)
\beqn\nonumber
\hat q^{\ve,k}_{\tau,r}(\theta)=
\frac14\sum_{\sigma=1}^{s}
\Pi_\sigma(\theta)\Big[\sum\limits_\pm
(I\pm iC_\sigma(\theta)) 
A^{\pm,k}_{\ve,\sigma}(\tau,r;\theta)
(I\mp iC^*_\sigma(\theta))\Big]\Pi_\sigma(\theta),
\eeqn
where
$$
A^{\pm,k}_{\ve,\sigma}(\tau,r;\theta)
=\int\limits_{\R^d} 
\hat {\bf R}_0(r\pm\nabla\omega_\sigma(\theta)\tau/\ve^{k-1}-x,\theta)
K^{\pm,k}_{\ve,\sigma}(\tau,x,\theta)\,dx,\quad
 \theta\in \T^d\setminus{\cal C},
$$
$$
K^{\pm,k}_{\ve,\sigma}(\tau,x,\theta)=F^{-1}_{y\to x}\Big[
\exp\Big\{\mp i\frac{\tau}{\ve^{k-2}}
(\frac{\nabla^2\omega_\sigma(\theta)\cdot y^2}{2!}
+\dots+ \frac{\nabla^k\omega_\sigma(\theta)\cdot y^k}{k!})\Big\}\Big],
\,\,\,x,y\in\R^d.
$$
Here, by definition,
$$
\nabla^k\omega_\sigma(\theta)\cdot y^k:=
\sum\limits_{i_1,\dots,i_k=1}^d
\frac{\partial^k\omega_{\sigma}(\theta)}
{\partial\theta_{i_1}\dots\partial\theta_{i_k}}\,y_{i_1}\dots y_{i_k},\quad
y=(y_1,\dots,y_d)\in\R^d.
$$
\end{theorem}

Note that the matrix $\hat q^{\ve,2}_{\tau,r}(\theta)$
 coincides with $\hat q^\ve_{\tau,r}(\theta)$ from formula (\ref{ftaur}).
The proof of Theorem \ref{t-app} are similar 
to the proof of Theorem \ref{the2'}.
\medskip

Set  $\tau=\ve^{k-1} t$, $k\ge2$. 
Then 
$
\partial_t A^{\pm,k}_{\ve,\sigma}(\ve^{k-1} t,r;\theta)=\pm 
P^k_\ve(\theta,\partial_r)A^{\pm,k}_{\ve,\sigma}(\ve^{k-1} t,r;\theta),
$
where
$$
P^k_\ve(\theta,\partial_r):=
\sum_{p=1}^k \frac{(i\ve)^{p-1}}{p!}\nabla^p\omega_\sigma(\theta)
\cdot\nabla^p_r
$$ 
(see notation (\ref{nabla-k})).
Therefore, the matrix
$\Pi_\sigma(\theta)\hat q^{\ve,k}_{\ve^{k-1}t,r}(\theta)
\Pi_\sigma(\theta)$
(denote its by $f^\ve_\sigma(t,r;\theta)$)
is the solution of the following equation
\beqn\label{app-1}
\partial_t f^\ve_\sigma(t,r;\theta)=iC_\sigma(\theta)
P^k_\ve(\theta,\partial_r)f^\ve_\sigma(t,r;\theta),\quad t>0,\quad r\in\R^d,
\eeqn
with the initial condition (\ref{initial}).

Denote by
$f_\sigma^{ij}$, $i,j=0,1$, the elements of the matrix
 $f^\ve_\sigma(t,r;\theta)$. 
Then $f^{11}_\sigma=\omega^2_\sigma(\theta)f^{00}_\sigma$,
$f^{10}_\sigma=-f^{01}_\sigma$, and the equations 
 (\ref{app-1}) can be rewritten in the form 
\beqn
\partial_t f^{00}_\sigma=(-i/\omega_\sigma(\theta))
P^k_\ve(\theta,\partial_r)f^{01}_\sigma,\quad
\partial_t f^{01}_\sigma=i\omega_\sigma(\theta)
P^k_\ve(\theta,\partial_r)f^{00}_\sigma.\nonumber
\eeqn

\setcounter{equation}{0}
 \section{Model II: Harmonic crystals in the half-space $\Z^d_+$}

We study the dynamics of the harmonic crystals 
in $\Z^d_+$, $d\ge 1$, 
\beqn\la{1+}
\ddot u(z,t)=-\sum\limits_{z'\in \Z^d_+}\left(V(z-z')-V(z-\tilde z')\right)
u(z',t),\,\,\,\,z\in\Z^d_+,\,\,\,\,t\in\R,
\eeqn
with zero boundary condition,
\be\la{2+}
u(z,t)|_{z_1=0}=0,
\ee
and with the initial data
\be\la{3+}
u(z,0)=u_0(z),\quad \dot u(z,0)=u_1(z),\quad z\in\Z^d_+.
\ee
Here $\Z^d_+=\{z\in \Z^d:\,z_1>0\}$, $\tilde z=(-z_1,z_2,\dots, z_d)$.
For convenience, we assume
that  $u_0(z)=u_1(z)=0$ for $z_1=0$.

Write $Y(t)=(Y^0(t),Y^1(t))\equiv (u(\cdot,t),\dot u(\cdot,t))$ and
$Y_0=(Y_0^0,Y_0^1)\equiv (u_0(\cdot),u_1(\cdot))$. Then
(\ref{1+})--(\ref{3+}) becomes the evolution equation
\be\label{CP1}
\dot Y(t)={\cal A}_+Y(t),\quad t\in\R,\,\,z\in\Z^d_+,
\quad Y^0(t)|_{z_1=0}=0,\quad Y(0)=Y_0.
\ee
Here ${\cal A}_+=\left(\ba{cc}0&1\\-{\cal V}_+&0\ea\right)$
with 
${\cal V}_+u(z):= \sum\limits_{z'\in\Z^d_+}(V(z-z')-V(z-\tilde z'))u(z')$.

Let us assume that
\be\label{condE0}
V(z)=V(\tilde z),\quad \mbox{where }\,\tilde z=(-z_1,\bar z),
\quad\bar z=(z_2,\dots,z_d)\in \Z^{d-1}.
\ee
Then the solution to problem  (\ref{CP1}) can be represented as the 
restriction of the solution to the Cauchy problem (\ref{CP1''})
with  odd initial data on the half-space. 
More exactly, 
assume that the initial data $X_0(z)$ form an odd function with respect to
 $z_1\in\Z^1$, i.e., let $X_0(z)=-X_0(\tilde z)$.
Then the solution $v(z,t)$ of (\ref{CP1''})
is also an odd function with respect to $z_1\in\Z^1$.
Restrict the solution $v(z,t)$ 
to the domain $\Z^d_+$ and set $u(z,t)=v(z,t)|_{z_1\ge0}$.
Then $u(z,t)$ is the solution to problem (\ref{1+}) with
the initial data $Y_0(z)=X_0(z)|_{z_1\ge0}$. 
\smallskip

  Assume that the initial data $Y_0$ for (\ref{CP1})
belong to the phase space ${\cal H}_{\al,+}$, $\al\in\R$, defined below.
 \begin{definition}                 \la{d1.1}
 $ {\cal H}_{\al,+}$ is the Hilbert space
of $\R^n\times\R^n$-valued functions  of $z\in\Z^d_+$
 endowed  with  the norm
$ \Vert Y\Vert^2_{\al,+}
 = \sum_{z\in\Z^d_+}\vert Y(z)\vert^2(1+|z|^2)^{\al} <\infty.
$  
\end{definition}

In addition, it is assumed that the initial data vanish ($Y_0=0$)
 at $z_1=0$.
\begin{lemma}\label{c1}(see \cite[Corollary 2.4]{D08})
Let conditions {\bf E1} and {\bf E2} hold. 
Choose some $\al\in\R$. Then
(i) for any  $Y_0 \in {\cal H}_{\al,+}$, there exists  a unique solution
$Y(t)\in C(\R, {\cal H}_{\al,+})$  to the mixed problem (\re{CP1});\\
(ii) the operator  $U_+(t):Y_0\mapsto Y(t)$ is continuous
on ${\cal H}_{\al,+}$.
\end{lemma}
Lemma \ref{c1} follows from Lemma \ref{lemma}
since 
the solution  $Y(t)$ of (\ref{CP1}) admits the representation   
\beqn\la{sol}
Y(t)=\sum\limits_{z'\in\Z^d_+} {\cal G}_{t,+}(z,z')
Y_0(z'),\quad z\in\Z^d_+,\\
\mbox{where }\,\,\,
{\cal G}_{t,+}(z,z'):={\cal G}_t(z-z')-{\cal G}_t(z-\tilde z'), \la{sol1}
\eeqn
and ${\cal G}_t(z)$ is  defined in (\ref{Grcs}).

\subsection{The family of the initial measures}

Let us introduce  the complex $2n\times 2n$
matrix-valued function $R(r,x,y)=(R^{ij}(r,x,y))^1_{i,j=0}$, $r\in\R^d$,
 $x,y\in\Z^d_+$, with the following properties {\bf(a)}--{\bf (d)}.
\medskip\\
{\bf (a)} $R(r,x,y)=0$ for $x_1=0$ or $y_1=0$.  
The $n\times n$ matrix-valued functions $R^{ij}(r,x,y)$
have the form
$$
R^{ij}(r,x,y)={\bf R}^{ij}(r,x_1,y_1,\bar x-\bar y),\quad\mbox{where }\,
x=(x_1,\bar x),\quad y=(y_1,\bar y),\quad i,j=0,1.
$$
Moreover, uniformly in $r\in\R^d$,
\be\la{3.14}
\lim_{y_1\to+\infty} {\bf R}^{ij}(r,y_1+z_1,y_1,\bar z)={\bf R}_0^{ij}(r,z), 
\quad z=(z_1,\bar z)\in\Z^d,\quad i,j=0,1,
\ee
where the matrix ${\bf R}_0(r,z)=({\bf R}_0^{ij}(r,z))_{i,j=0}^1$
satisfies conditions {\bf I1}--{\bf I4} (see Section \ref{sec1.2}).
\medskip\\
{\bf (b)} For every fixed $r\in\R^d$ and $i,j=0,1$, 
the bound holds,
\be\la{3.2'} 
|R^{ij}(r,x,y)|\le C(1+|x-y|)^{-\gamma},\quad x,y\in\Z^d_+,
 \ee
with the same constants $C$ and $\gamma$ as in (\ref{3.2}).
\medskip\\
{\bf (c)} For every fixed $r\in\R^d$, the matrix-valued function
$R(r,x,y)$ satisfies
 \beqn\nonumber
 R^{ii}(r,\cdot,\cdot)\geq 0,\quad
R^{ij}(r,x,y)=(R^{ji}(r,y,x))^T,\quad x,y\in\Z^d_+.
\eeqn
{\bf (d)} 
For every $x,y\in\Z^d_+$, $R^{ij}(\cdot,x,y)$,
$i,j=0,1$, are $C^1$ functions. 

To derive the equation of the Navier-Stokes type 
we need the additional condition {\bf (d')} in the case when $d>1$.\\
{\bf (d')} Let $d>1$. For every  fixed $x,y\in\Z^d_+$,
\be\label{d'1}
\sup_{\bar r=(r_2,\dots,r_d)\in\R^{d-1}}
|R(r_1,\bar r,x,y)-R(0,\bar r,x,y)|\to0\quad \mbox{as }\,r_1\to0.
\ee
Moreover, we assume that for every  fixed $x,y\in\Z^d_+$,
\be\la{d'2}
|\tilde R(0,\bar s,x,y)|\le C(1+|\bar s|)^{-N},\quad \bar s\in\R^{d-1},
\ee
with $N>d+2$. Here by $\tilde R(0,\bar s,x,y)$ we denote the Fourier transform
of $R(0,\bar r,x,y)$ w.r.t. $\bar r\in\R^{d-1}$,
$$
\tilde R(0,\bar s,x,y)=
\int\limits_{\R^{d-1}} e^{i\bar s\bar r}R(0,\bar r,x,y)\,d\bar r,
\quad \bar s\in\R^{d-1}.
$$

Let $\{\mu^\varepsilon_0,\varepsilon>0\}$ be a family of initial
measures on ${\cal H}_{\al,+}$ and 
$\E^\ve_0$ stand for expectation with respect to the measure
$\mu_0^\ve$. Assume that $\E^\varepsilon_0\big(Y(x)\big)=0$
and define the covariance
$ Q^{ij}_{\ve}(x,x')= \E^\ve_0\big(Y^i(x)\otimes
Y^j(x')\big)$, $x,x'\in \Z^d_+$, $i,j=0,1$.
\medskip

The family of measures  $\{\mu_0^\ve,\ve>0\}$
  satisfies the following conditions
{\bf V1'} and {\bf V2'}.\smallskip\\
{\bf V1'}. For any $\ve >0$, $x,x'\in\Z^d_+$,
\beqn\la{2.4+}
\left|Q^{ij}_{\ve}(x,x')- R^{ij}(\ve x,x,x')\right|\le
C\min[(1+|x-x'|)^{-\gamma}, \ve |x-y|],
\eeqn
with the constants $C$ and $\gamma$ as in (\ref{3.2'}).\medskip\\
{\bf V2'}. For any $\ve>0$ and all $x,x'\in\Z^d_+$, $i,j=0,1$, 
$
|Q^{ij}_{\ve}(x,x')|\le C(1+|x-x'|)^{-\gamma}
$
with the constants $C$, $\gamma$ as in (\ref{3.2'}).

\setcounter{equation}{0}
\section{Main results in the half-space }
\begin{definition}\label{def-mut}
(i) $\mu^\ve_t$ is a Borel probability measure on ${\cal H}_{\al,+}$
which gives the distribution of $Y(t)$,
$\mu^\ve_t(B) = \mu_0^\ve(U_+(-t)B)$, where 
$B\in {\cal B}({\cal H}_{\al,+})$ and $t\in \R$.\\ 
(ii) The correlation functions of the  measure $\mu^\ve_t$ are  defined by
$$
Q_{\ve,t}^{ij}(x,y)=\int \left(Y^i(x)\otimes Y^j(y)\right)
\mu^\ve_{t}(dY)= \E^{\ve}_0\big(Y^i(x,t)\otimes  Y^j(y,t)\big), 
\quad i,j= 0,1,\quad x,y\in\Z^d_+. 
$$ 
Here $Y^i(x,t)$ are  the components of the random solution 
$Y(t)=(Y^0(\cdot,t),Y^1(\cdot,t))$ to the problem (\ref{CP1}).
\end{definition}

\subsection{Euler limit}
At first, we introduce the matrix $g_{\tau,r}(z)$, $z\in\Z^d$, 
$r\in\R^d$, $\tau\not=0$, by the  Fourier transform,
$$
\hat g_{\tau,r}(\theta)=\frac14 \sum_{\sigma=1}^{s}
\Pi_\sigma(\theta)\Big[\sum\limits_\pm
(I\pm iC_\sigma(\theta)) 
\hat{\bf R}_0(r\pm\nabla\om_\sigma(\theta)\tau,\theta) 
\chi^\pm_{\tau,r_1}(\theta)
(I\mp iC^*_\sigma(\theta)) \Big]\Pi_\sigma(\theta),
$$
(cf (\ref{qt})), where $\theta\in \T^d\setminus{\cal C}_*$,
\be\label{chipm} 
 \chi^\pm_{\tau,r_1}(\theta)=
\big(1+\sign(r_1\pm\partial_1\omega_\sigma(\theta)\tau)\big)/2.
\ee
\begin{theorem}\label{t4.3} (see Theorem 2.10 in \cite{D09'})
 Let conditions {\bf (a)}--{\bf (d)},
{\bf V1'}, {\bf V2'} and {\bf E1}--{\bf E6} hold.
Then for any $\tau\not=0$, $r\in\R^d$ with $r_1\ge0$,
 the correlation functions of $\mu_{\tau/\ve,r/\ve}$
converge to a limit,
\beqn\label{gclimcor0_+}
 \lim_{\varepsilon\to 0}Q_{\varepsilon,\tau/\varepsilon}
([r/\varepsilon]+z,[r/\varepsilon]+z')=Q_{\tau,r}(z,z').
\eeqn
Here $z,z'\in\Z^d$ if $r_1>0$, $z,z'\in\Z^d_+$ if $r_1=0$,
\be\label{Qtaur}
Q_{\tau,r}(z,z')=\left\{\ba{lr}
{\bf q}_{\tau,r}(z-z')=g_{\tau,r}(z-z')+g_{\tau,\tilde r}(\tilde z-\tilde z'),
& \mbox{if }r_1>0,\\
g_{\tau,r}(z\!-\!z')-g_{\tau,r}(z\!-\!\tilde z')-
g_{\tau,r}(\tilde z\!-\!z')+g_{\tau,r}(\tilde z\!-\!\tilde z'),
& \mbox{if }r_1=0, \ea\right.
\ee
where $\tilde r:=(-r_1,\bar r)$, $\bar r=(r_2,\dots,r_d)$,
 and the matrix $g_{\tau,r}(z)$ is defined above.
\end{theorem}
\begin{cor}
Let  $r\in \R^d_+\equiv\{r\in\R^d:r_1>0\}$ and $\tau>0$.
For each $\sigma=1,\dots,s$, 
the $\sigma$-band of $\hat {\bf q}_{\tau,r}(\theta)$,
$\theta\in \T^d\setminus{\cal C}_*$, 
satisfies the following "hydrodynamic" (Euler type) equation:
$$
\partial_\tau f_\sigma(\tau,r;\theta)=iC_\sigma(\theta)
\nabla\omega_\sigma(\theta)
\cdot \nabla_r f_\sigma(\tau,r;\theta),\quad r\in\R^d_+,\quad \tau>0,
$$
with the initial condition (\ref{initial}) (if $\tau=0$)
and with the boundary condition 
(if $r_1=0$) expressed by $\hat{\bf R}_0$. In particular, if
$\hat{\bf R}_0(r,\tilde\theta)=\hat{\bf R}_0(r,\theta)$, 
the boundary condition has a form
$$
f_\sigma|_{r_1=0}=
\Pi_\sigma(\theta)\frac12
\Big(P^\sigma_++C_\sigma(\theta) P^\sigma_+C^*_\sigma(\theta)
+iC_\sigma(\theta) P^\sigma_--i P^\sigma_-C^*_\sigma(\theta)\Big)
\Pi_\sigma(\theta),\,\,\, \bar r\in\R^{d-1},\,\,\,\tau>0,
$$
where $P^\sigma_\pm$ stands for the $2n\times 2n$ matrix-valued function,
$$
P^\sigma_\pm=\frac12
\Big(\hat {\bf R}_0\left(|\partial_1\omega_\sigma(\theta)|\tau,
\bar r+\nabla_{\bar\theta}\omega_\sigma(\theta)\tau,\theta\right)\pm
\hat {\bf R}_0\left(|\partial_1\omega_\sigma(\theta)|\tau,
\bar r-\nabla_{\bar\theta}\omega_\sigma(\theta)\tau,\theta\right)\Big).
$$
\end{cor}

Introduce the complex-valued field (cf. (\ref{2.1}))
 \beqn\label{2.1+}
 a_+(x)= \frac{1}{\sqrt{2}}\Big(({\cal V}_+^{1/4} u_0)(x)
+i({\cal V}_+^{-1/4}u_1)(x)\Big)\in \C^n\,,
\quad x\in \Z^d\,,
 \eeqn
where 
$$
({\cal V}_+^{k}u)(x)
:=\sum\limits_{z\in\Z^d_+}(V^k(x-z)-V^k(x-\tilde z))u(z),\quad
\mbox{with }\,V^k(z):=F^{-1}_{\theta\to z}\left(\hat V^{k}(\theta)\right).
$$ 

Let us introduce the scaled $n\times n$ Wigner
matrix (cf. (\ref{3.15}))
 \beqn\nonumber
W_+^\varepsilon(\tau,r;\theta)=
\sum_{y\in \Z^d}e^{i\theta\cdot y}\,
\E^\varepsilon_{\tau/\varepsilon}
\Big(a^\ast_+([r/\varepsilon+y/2])
\otimes a_+([r/\varepsilon-y/2])\Big),\quad r\in \R^d_+,
 \eeqn
where $a_+(x)$ is given in (\ref{2.1+}).
By conditions {\bf V1'} and {\bf V2'}, for fixed $r\in\R^d_+$,
$$ 
\lim\limits_{\varepsilon\to 0}W_+^\varepsilon(0,r;\theta)= 
W(0,r;\theta),
$$ 
uniformly on $\theta\in\T^d\setminus{\cal C}_*$,
where $W(0,r;\theta)$ is defined in (\ref{3.12'}).

We also define the limit  Wigner matrix as follows (cf (\ref{3.16a}))
 \beqn\label{3.16+}
 W_+^{p}(\tau,r;\theta)\!=\!
\left\{ \ba{ll}
\sum\limits^s_{\sigma=1} \Pi_\sigma(\theta)
W(0,r\!-\!\tau\nabla\omega_\sigma(\theta);\theta)
\Pi_\sigma(\theta),\!\!&\!\mbox{if }\,
r_1\!>\!\tau\partial_1\omega_\sigma(\theta),\\
\sum\limits^s_{\sigma=1} \Pi_\sigma(\theta)
W(0,-r_1\!+\!\tau\partial_1\omega_\sigma(\theta),
\bar r\!-\!\tau\nabla_{\bar\theta}\,\omega_\sigma(\theta);\tilde\theta)
\Pi_\sigma(\theta),\!\!&\!\mbox{if }
\,r_1\!<\!\tau\partial_1\omega_\sigma(\theta),
\ea
\right.
 \eeqn
where $\tilde\theta=(-\theta_1,\bar\theta)$, 
$\bar \theta=(\theta_2,\dots,\theta_d)$, 
$\nabla_{\bar\theta}\,\omega_\sigma(\theta)=(\partial_2\omega_\sigma(\theta),
\dots,\partial_d\omega_\sigma(\theta))$,
$\partial_k=\partial/\partial \theta_k$.
\begin{theorem} \label{t4.5} (see Theorem 2.12 in \cite{D09'})
Let conditions {\bf V1'}, {\bf V2'} 
and {\bf E1}--{\bf E6} hold. Then  for any $r\in\R^d_+$ and
$\tau>0$, the following limit exists in the sense of distributions,
 \be\label{3.17}
 \lim_{\varepsilon\to 0} W_+^\varepsilon(\tau,r;\theta)=
 W_+^{p}(\tau,r;\theta)\,.
 \ee
\end{theorem}
\begin{cor}\label{cor2.13} (see Corollary 2.13 in \cite{D09'})
Denote by 
$W_{\sigma}^{p}(\tau,r;\theta)$,
$\sigma=1,\dots,s$, the $\sigma$-th band of the Wigner function
$W_+^{p}(\tau,r;\theta)$.
Then, for any fixed $\sigma=1,\dots,s$,
$W_{\sigma}^{p}$ is a solution of the "energy transport" equation 
$$
\partial_\tau W_{\sigma}^{p}(\tau,r;\theta)
+\nabla\omega_\sigma(\theta)\cdot\nabla_rW_{\sigma}^{p}(\tau,r;\theta)
=0,\quad \tau>0,\quad r\in\R^d_+,
$$
with the initial and boundary conditions
\beqn
W_{\sigma}^{p}(\tau,r;\theta)|_{\tau=0}&=&
 W_\sigma(0,r;\theta),\quad r\in\R^d_+,
\nonumber\\
 W_{\sigma}^{p}(\tau,r;\theta)|_{r_1=0}&=&b_\sigma(\tau,\bar r;\theta),
\quad \bar r\in\R^{d-1}, \quad \tau>0.\label{2.28'}
\eeqn
Here $W_\sigma(0,r;\theta)$ is the $\sigma$-th band of 
the initial Wigner matrix $W(0,r;\theta)$ (see (\ref{3.12'})),
$$
b_\sigma(\tau,\bar r;\theta):=\left\{\ba{cc}
W_\sigma(0,-\tau\partial_1\omega_\sigma(\theta),
\bar r-\tau\nabla_{\bar\theta}\,\omega_\sigma(\theta);\theta),&
\mbox{if }\,\partial_1\omega_\sigma(\theta)<0,\\
W_\sigma(0,\tau\partial_1\omega_\sigma(\theta),
\bar r-\tau\nabla_{\bar\theta}\,\omega_\sigma(\theta);\tilde\theta),&
\mbox{if }\,\partial_1\omega_\sigma(\theta)>0.
\ea\right.
$$
In particular, if 
we assume that $\hat {\bf R}_0(r,\tilde\theta)=\hat{\bf R}_0(r,\theta)$
for $r\in\R^d$, $\theta\in\T^d$,
then the boundary condition (\ref{2.28'}) can be rewritten in the form
$$
W_{\sigma}^{p}(\tau,r;\theta)|_{r_1=0}=
W_\sigma(0,\tau|\partial_1\omega_\sigma(\theta)|, 
\bar r-\tau\nabla_{\bar\theta}\,\omega_\sigma(\theta);\theta),
\quad \bar r\in\R^{d-1}, \quad \tau>0.
$$
\end{cor}
{\bf Remark} (see \cite[Theorem 2.15]{D09'})\, 
Let the measures $\mu^\varepsilon_0$
satisfy the mixing condition (of the Rosenblatt or Ibragimov type).
Then for $\tau\neq 0$, $r\in\R^d$ with $r_1\ge0$,
in the sense of weak convergence on ${\cal H}_{\alpha,+}$,
 $ \lim\limits_{\varepsilon\to 0}\mu^\varepsilon_{\tau/\varepsilon,r/\ve}
=\mu^{G}_{\tau,r}$.
The measure  $\mu^{G}_{\tau,r}$ is a Gaussian measure 
on ${\cal H}_{\alpha,+}$, which is invariant under the time
translation $U_+(t)$. $\mu^{G}_{\tau,r}$ has mean zero and covariance
$Q_{\tau,r}(z,z')$ defined by (\ref{Qtaur}).

\subsection{Second approximation}\label{sec4.3}
In this subsection we treat the main result of this paper.
Let us introduce the matrix
 $ g^\ve_{\tau,r}(z)$, $z\in\Z^d$, $r\in\R^d$, $\tau\not=0$, $\ve>0$,
which has the following form in the Fourier transform (cf (\ref{ftaur}))
\be\label{gve}
\hat g^\ve_{\tau,r}(\theta)=\frac14 \sum_{\sigma=1}^{s}
\Pi_\sigma(\theta)\Big[\sum\limits_\pm
(I\pm iC_\sigma(\theta)) 
{\bf A}^{\pm}_{\ve,\sigma}(\tau,r;\theta) 
(I\mp iC^*_\sigma(\theta)) \Big]
\Pi_\sigma(\theta),\quad \theta\in \T^d\setminus{\cal C}_*,
\ee
where the matrices $C_\sigma(\theta)$ are defined in (\ref{Csigma}),
$$
{\bf A}^{\pm}_{\ve,\sigma}(\tau,r;\theta)
=A^\pm_{\ve,\sigma}(\tau,r;\theta)
\chi_{\tau/\ve,r_1}^\pm (\theta),
$$
with the matrix-valued functions 
$A^\pm_{\ve,\sigma}(\tau,r;\theta)$ from (\ref{Apm})
and $\chi_{\tau,r_1}^\pm$ from (\ref{chipm}).
\begin{theorem}\label{the2'+}
 Let conditions {\bf (a)}--{\bf (d')},
{\bf V1'}, {\bf V2'} and {\bf E1}--{\bf E6} hold.
Then for any $\tau\not=0$ and $r\in\R^d$ with $r_1\ge0$, 
the correlation functions of measures $\mu_{\tau/\ve^2,r/\ve}$
have the following asymptotics
\beqn\label{4.13}
\lim_{\ve\to0}\Big(Q_{\ve,\tau/\ve^2}([r/\ve]+z,[r/\ve]+z')-
Q^\ve_{\tau,r}(z,z')\Big)=0,
\eeqn
where $z,z'\in\Z^d$ if $r_1>0$, $z,z'\in\Z_+^d$ if $r_1=0$,
$$
Q^\ve_{\tau,r}(z,z')=\left\{\ba{lr} 
{\bf q}^{\ve}_{\tau,r}(z-z')=g^\ve_{\tau,r}(z-z')
+ g^\ve_{\tau,\tilde r}(\ti z-\tilde z'),
&\mbox{if }\, r_1>0,\\
 g^\ve_{\tau,r}(z-z')- g^\ve_{\tau,r}(z-\ti z')
- g^\ve_{\tau,r}(\ti z-z')
+ g^\ve_{\tau, r}(\ti z-\tilde z'),&\mbox{if }\, r_1=0,
\ea\right.
$$
 $g^{\ve}_{\tau,r}(z)
=F^{-1}_{\theta\to z}[\hat g^{\ve}_{\tau,r}(\theta)]$
and $\hat g^{\ve}_{\tau,r}(\theta)$ defined by (\ref{gve}).
\end{theorem}
{\bf Remark}.
The matrices $\hat {\bf q}_{\tau,r}(\theta)$ and
 $\hat {\bf q}^{\ve}_{\tau,r}(\theta)$
satisfy the equilibrium condition (see Remarks \ref{Re3.4} (i)). 
\medskip

Set $\tau=\ve t$. In this case,
 ${\bf A}^{\pm}_{\ve,\sigma}(\ve t,r;\theta)|_{\ve=0}
=\hat {\bf R} _0(r\pm\nabla\omega_\sigma(\theta) t,\theta)
\chi_{t,r_1}^\pm (\theta)$,
and, then, $\hat g^{\ve}_{\ve t,r}(\theta)|_{\ve=0}= \hat g_{t,r}(\theta)$.
Therefore, for $r_1>0$, 
$\hat {\bf q}^{\ve}_{\ve t,r}(\theta)|_{\ve=0}= \hat {\bf q}_{t,r}(\theta)$,
where $\hat {\bf q}_{t,r}(\theta)=F_{z\to\theta}[{\bf q}_{t,r}(z)]$ 
and ${\bf q}_{t,r}(z)$ from (\ref{Qtaur}). 
\begin{cor}
Let  $r_1>0$ and $t>0$. Then for each $\sigma=1,\dots,s$,
the matrix-valued function 
 $F_\sigma\equiv F_\sigma^\ve(t,r;\theta)=
\Pi_\sigma(\theta)\hat{\bf q}^{\ve}_{\ve t,r}(\theta)\Pi_\sigma(\theta)$,
$\theta\in\T^d\setminus{\cal C}_*$, 
evolves according to the following mixing problem
\beqn\nonumber
\partial_t F_\sigma&=& iC_\sigma(\theta)
\left(\nabla\omega_\sigma(\theta)
\cdot \nabla_r F_\sigma
+\ds\frac{i\ve}{2}\nabla^2\omega_\sigma(\theta)
\cdot\nabla^2_r F_\sigma\right),\quad r_1>0,\quad t>0,\nonumber\\
F_\sigma|_{t=0}&=&
\frac12\Pi_\sigma(\theta)
\left(\hat{\bf R}_0(r,\theta)+C_\sigma(\theta)\hat{\bf R}_0(r,\theta)
C^*_\sigma(\theta)\right)\Pi_\sigma(\theta),\quad r_1>0,\nonumber\\
F_\sigma|_{r_1=0}&=&\frac14\sum\limits_\pm
\Pi_\sigma(\theta)(I\pm iC_\sigma(\theta)) 
{\bf A}^\pm_{\ve,\sigma}(\ve t,r;\theta)|_{r_1=0}
(I\mp iC^*_\sigma(\theta))\Pi_\sigma(\theta),\quad t>0. \nonumber
\eeqn
\end{cor}

\setcounter{equation}{0}
\section{Convergence of correlation functions}\label{sec.5}
\subsection{Bounds for initial covariance}
\begin{definition}
By $\ell^p\equiv \ell^p(\Z^d)\otimes \R^n$ 
(by $\ell^p_+\equiv \ell^p(\Z^d_+)\otimes \R^n)$, where $p\ge 1$ and
 $n\ge 1$, denote the space of sequences
$f(z)=(f_1(z),\dots,f_n(z))$ endowed with norm
$\Vert f\Vert_{\ell^p}=\Big(\sum_{z\in\Z^d}|f(z)|^p\Big)^{1/p}$,
respectively,
$\Vert f\Vert_{\ell^p_+}=\Big(\sum_{z\in\Z^d_+}|f(z)|^p\Big)^{1/p}$.
\end{definition}

The following lemma follows from condition {\bf V2}.
\begin{lemma} \label{l4.1}
Let  condition {\bf V2} hold. Then, 
for $i,j=0,1$, the following bounds hold:
\beqn
\sum\limits_{z'\in\Z^d_+} |Q^{ij}_\varepsilon(z,z')|
&\le& C<\infty\,\,\,\mbox{ for all }\,z\in\Z^d_+,
\nonumber\\
\sum\limits_{z\in\Z^d_+} |Q^{ij}_\varepsilon(z,z')|
&\le& C<\infty\,\,\,\mbox{ for all }\,z'\in\Z^d_+.\label{pr2}
\nonumber
\eeqn
Here the constant $C$ does not depend on $z,z'\in \Z^d_+$
and $\varepsilon>0$.
\end{lemma}
\begin{cor}\label{c4.10}
By the Shur lemma, it follows from Lemma \ref{l4.1} that  
\beqn\nonumber
|\langle Q_\varepsilon(z,z'),\Phi(z)\otimes\Psi(z')\rangle_+|\le
C\Vert\Phi\Vert_{\ell^2_+} \Vert\Psi\Vert_{\ell^2_+},\,\,\,\,
\mbox{for any }\,\,\Phi,\Psi\in \ell^2_+,
\eeqn
where the constant $C$ does not depend on  $\varepsilon>0$.
\end{cor}

\subsection{Stationary phase method}

By (\ref{Grcs}) and (\ref{hA}) we see that $\hat{\cal G}_t( \theta)$ is
of the form 
\beqn\label{4.5}
\hat{\cal G}_t( \theta)=
\left( \begin{array}{cc}
 \cos\Omega t &~ \sin \Omega t~\Omega^{-1}  \\
 -\sin\Omega t~\Omega
&  \cos\Omega t\end{array}\right),
\eeqn
where $\Omega=\Omega(\theta)$ is the
Hermitian matrix defined by (\ref{Omega}).
Hence, by Lemma \ref{lc*} (iii),
by formulas $\cos\omega_\sigma(\theta)t
=(e^{i\omega_\sigma t}+e^{-i\omega_\sigma t})/2$, 
$\sin\omega_\sigma(\theta)t=
(e^{i\omega_\sigma t}-e^{-i\omega_\sigma t})/(2i)$ and by (\ref{Csigma}),
the matrix ${\cal G}_t(x)$ can be rewritten in the form
\be\label{3.4'}
{\cal G}_t(x)=
\sum\limits_{\pm,\sigma=1}^s \int\limits_{\T^d}
e^{-ix\cdot \theta}e^{\pm i\omega_\sigma(\theta)\,t}
c^\mp_\sigma(\theta)\,d\theta,
\ee
where 
$c^\mp_\sigma(\theta):=\Pi_\sigma(\theta)
(I\mp i C_\sigma(\theta))/2$.
We are going to apply the stationary phase arguments
to the integral (\ref{3.4'}) which require a smoothness in $\theta$. 
Then we have to choose certain smooth branches of the functions 
$c^\pm_\sigma(\theta)$ and $\omega_\sigma(\theta)$ 
and cut off all singularities.
First, introduce the {\it critical set} as
\beqn\label{calC}
{\cal C} ={\cal C}_0\bigcup
{\cal C}_*\bigcup_{\sigma=1}^{s}\Big({\cal C}_\sigma
 \bigcup\limits_{i=1}^{d}\,
\Big\{\theta\in \T^d\setminus {\cal C}_*:\,
 \frac{\partial^2\omega_\sigma(\theta)}{\partial \theta_i^2}=0\Big \}
\bigcup\Big\{\theta\in \T^d\setminus {\cal C}_*:\,
\frac{\partial\, \omega_\sigma(\theta)}{\partial\theta_1}=0\Big\}\Big),
\eeqn
with ${\cal C}_*$ as in Lemma \ref{lc*} and sets ${\cal C}_0$
and ${\cal C}_\sigma$ defined by (\ref{c0ck}).
Obviously, ${\rm mes}\,{\cal C}=0$ (see \cite[Lemmas 2.2, 2.3]{DKS1}).
  Secondly, fix an $\delta>0$ and choose a finite partition of unity,
\beqn\label{part}
f(\theta)+ g(\theta)=1,\,\,\,\,
g(\theta)=\sum_{k=1}^K g_k(\theta),
\,\,\,\,\theta\in \T^d,
\eeqn
where $f,g_k$ are non-negative functions in $C_0^\infty(\T^d)$, and
\be\label{fge}
\supp f\subset \{\theta\in \T^d:\,
{\rm dist}(\theta,{\cal C})<\delta\},\,\,\,
\supp g_k\subset \{\theta\in \T^d:\,
{\rm dist}(\theta,{\cal C})\ge\delta/2\}.
\ee
Then we represent ${\cal G}_t(x)$ in the form
${\cal G}_t(x)={\cal G}^f_t(x)+{\cal G}^g_t(x)$,
where
\beqn\label{frepecut'}
{\cal G}^f_t(x)=F^{-1}_{\theta\to x}[f(\theta)\,\hat{\cal G}_t(\theta)],
\quad
{\cal G}^g_t(x)=F^{-1}_{\theta\to x}[g(\theta)\,\hat{\cal G}_t(\theta)].
\eeqn
By Lemma \ref{lc*} and the compactness arguments,
we can choose the supports of $g_k$ so small that the eigenvalues 
$\omega_\sigma(\theta)$ and the amplitudes $c^\pm_\sigma(\theta)$
are real-analytic functions inside
the $\supp g_k$ for every $k$. (We do not label the
functions by the index $k$ to simplify the notation.)
The Parseval identity, (\ref{4.5}), and condition {\bf E6} imply
\beqn
\ba{rcl}\label{3.12}
\Vert {\cal G}^f_t(\cdot)\Vert^2_{\ell^2}&=&C
\ds\int\limits_{\T^d} |\hat{\cal G}_t(\theta)|^2|f(\theta)|^2\,d\theta
\le C\int\limits_{{\rm dist}(\theta,{\cal C})<\delta}
|\hat{\cal G}_t(\theta)|^2\,d\theta \to 0\,\,\,\mbox{as } \,\,\delta\to 0,\\
\Vert {\cal G}^g_t(\cdot)\Vert^2_{\ell^2}&\le&
\ds C\int\limits_{\T^d} |\hat{\cal G}_t(\theta)|^2\,d\theta
\le C_1<\infty,
\ea
\eeqn
uniformly in $t\in\R$.
For the function ${\cal G}^g_t(x)$, the following lemma holds.
\begin{lemma}\label{l5.3} (see \cite[Lemma 4.2]{DS})
Let conditions {\bf E1}--{\bf  E4} and {\bf E6} hold. Then
the following bounds hold.

(i) $\sup_{x\in\Z^d}|{\cal G}^g_t(x)| \le  C~t^{-d/2}$.

(ii) For  any $p>0$, there exist $C_p,\gamma_g>0$ such that
$|{\cal G}^g_t(x)|\le C_p(|t|+|x|+1)^{-p}$ for
$|x|\ge \gamma_g t$.
\end{lemma}

\subsection{Proof of Theorem 5.6}
 The representation (\ref{sol}) yields
$$   
Q_{\varepsilon,t}(z,z')=\E^\varepsilon_0\big(Y(z,t)\otimes Y(z',t)\big)
=\sum\limits_{x,y\in \Z^d_+}
{\cal G}_{t,+}(z,x)Q_\varepsilon(x,y){\cal G}_{t,+}(z',y)^T,
\,\,\, z,z'\in\Z^d_+,
$$
for any $t\in\R$.
It follows from condition (\ref{condE0}) and from formulas
(\ref{Grcs}) and (\ref{hA}) that ${\cal G}_t(z)={\cal G}_t(\tilde z)$
with $\tilde z=(-z_1,z_2,\dots,z_d)$. In this case, by (\ref{sol1}),
the covariance $Q_{\varepsilon,t}(z,z')$ can be decomposed into the
 sum of fourth terms,
$$
Q_{\varepsilon,t}(z,z')=S_{\varepsilon,t}(z,z')
-S_{\varepsilon,t}(\tilde z,z')-S_{\varepsilon,t}(z,\tilde z')
+S_{\varepsilon,t}(\tilde z,\tilde z'),\quad z,z'\in\Z^d_+,
$$
where
$$
S_{\varepsilon,t}(z,z'):=\sum\limits_{x,y\in \Z^d_+}
{\cal G}_{t}(z-x)Q_\varepsilon(x,y){\cal G}_{t}(z'-y)^T.
$$
\begin{pro}\label{p4}
Let $r\in\R^d$ and $z,z'\in\Z^d$. Then
\be\label{pro4}
\lim_{\varepsilon\to+0}
(S_{\varepsilon,\tau/\varepsilon^2}([r/\varepsilon]+z,[r/\varepsilon]+z')
- g^{\varepsilon}_{\tau,r}(z-z'))=0,
\ee
where $ g^\ve_{\tau,r}(z)$  is defined by (\ref{gve}). 
\end{pro}

This proposition implies Theorem \ref{the2'+}. Indeed, 
let $r_1=0$. Then $\tilde r/\varepsilon=r/\varepsilon$, 
and for $z,z'\in\Z^d_+$,
\beqn
Q_{\varepsilon,t}([r/\varepsilon]+z,[r/\varepsilon]+z')&=&
S_{\varepsilon,t}([r/\varepsilon]+z,[r/\varepsilon]+z')
-S_{\varepsilon,t}([r/\varepsilon]+\tilde z,[r/\varepsilon]+z')
\nonumber\\
&&-S_{\varepsilon,t}([r/\varepsilon]+z,
[r/\varepsilon]+\tilde z')
+S_{\varepsilon,t}([r/\varepsilon]+\tilde z,
[r/\varepsilon]+\tilde z').\nonumber
\eeqn
Therefore, convergence (\ref{pro4}) implies (\ref{4.13}).

Let $r_1>0$. In this case,
the matrix-valued functions 
$S_{\varepsilon,\tau/\varepsilon^2}([\tilde r/\varepsilon]+\tilde z,
[r/\varepsilon]+z')$ and 
$S_{\varepsilon,\tau/\varepsilon^2}([r/\varepsilon]+z,
[\tilde r/\varepsilon]+\tilde z')$ 
vanish as $\varepsilon\to+0$, and 
$$
S_{\varepsilon,\tau/\varepsilon^2}([\tilde r/\varepsilon]+\tilde z,
[\tilde r/\varepsilon]+\tilde z')-
g^{\varepsilon}_{\tau,\tilde r}(\tilde z-\tilde z')
\to0,\quad \varepsilon\to+0.
$$
It can be proved similarly to Proposition \ref{p4}.
\medskip\\
{\bf Proof of Proposition \ref{p4}}.
Let us denote
$$
\bar Q_\varepsilon(x,y)=\left\{
\ba{cl} Q_\varepsilon(x,y)&\mbox{for }\, x,y\in\Z^d_+,\\
0& \mbox{otherwise}
\ea\right.
$$
The partition (\ref{part}), 
Corollary \ref{c4.10} and the bound (\ref{3.12})  yield
$$
S_{\varepsilon,t}(z,z')
=\sum\limits_{x,y\in \Z^d}{\cal G}^g_t(z-x)\bar Q_\varepsilon(x,y)
{\cal G}^{g}_t(z'-y)^T+o(1),
$$
where ${\cal G}^g_t$ is defined in (\ref{frepecut'}),
$o(1)\to 0$ as  $\delta\to 0$ uniformly in $t\in \R$ and $z,z'\in\Z^d$.
In particular, setting $t=\tau/\varepsilon^2$,
$z=[r/\varepsilon]+l$ and $z'=[r/\varepsilon]+p$ we obtain
\beqn\label{10.1}
S_{\varepsilon,\tau/\varepsilon^2}
([r/\varepsilon]\!+\!l,[r/\varepsilon]\!+\!p)\!\!&=&\!\!\!\!
\sum\limits_{x,y\in \Z^d}
{\cal G}^g_{\tau/\varepsilon^2}([r/\varepsilon]+l-x)
\bar Q_\varepsilon(x,y){\cal G}^{g}_{\tau/\varepsilon^2}([r/\varepsilon]+p-y)^T
+o(1)\nonumber\\
\!\!&=&\!\!\!\!\sum\limits_{x,y\in \Z^d}
{\cal G}^g_{\tau/\varepsilon^2}(l+x)
\bar Q_\varepsilon([r/\varepsilon]\!-\!x,[r/\varepsilon]\!-\!y)
{\cal G}^{g}_{\tau/\varepsilon^2}(p+y)^T\!+\!o(1).\,\,\,\,\,\,\,
\eeqn
Let $c =\gamma_g+\max(|l|,|p|)$ with $\gamma_g$ from Lemma \ref{l5.3}.
Then Lemma \ref{l5.3} (ii) and condition {\bf V2'}  imply that 
the series in (\ref{10.1})
can be taken over $x,y\in [-c\tau/\ve^2,c\tau/\ve^2]^d$.
\medskip

By definition, the function $\bar R$ is equal to
$$
\bar R(r,x,y)=\left\{
\ba{cl} R(r,x,y)&\mbox{if }\, x,y\in\Z^d_+,\\
0& \mbox{otherwise} \ea\right.
$$
The asymptotics of $S_{\varepsilon,\tau/\varepsilon^2}$
is not changed when we replace 
$\bar Q_\varepsilon([r/\varepsilon]\!-\!x,[r/\varepsilon]\!-\!y)$
in the r.h.s. of (\ref{10.1}) by 
$\bar R(\dots)\equiv\bar R(\varepsilon[r/\varepsilon]-\varepsilon x, 
[r/\varepsilon]-x,[r/\varepsilon]-y)$, i.e.,
\beqn\label{6.10}
S_{\varepsilon,\tau/\varepsilon^2}([r/\varepsilon]\!+\!l,
[r/\varepsilon]\!+\!p)=\!\!
\sum\limits_{x,y\in [-c\tau/\ve^2,c\tau/\ve^2]^d}\!
{\cal G}^g_{\tau/\varepsilon^2}(l+x)
\bar R(\dots){\cal G}^{g}_{\tau/\varepsilon^2}(p+y)^T+o(1).
\eeqn
 Indeed, by Lemma~\ref{l5.3} (i) and condition (\ref{2.4+}),
\beqn\nonumber
&&\Big|\sum\limits_{x,y\in[-c\tau/\ve^2,c\tau/\ve^2]^d}
{\cal G}^g_{\tau/\varepsilon^2}(l+x)
\Big(\bar Q_\varepsilon([r/\varepsilon]\!-\!x,[r/\varepsilon]\!-\!y)
-\bar R(\dots)\Big){\cal G}^{g}_{\tau/\varepsilon^2}(p+y)^T\Big|\nonumber\\
&\le& C\sum\limits_{y\in\Z^d}\min[(1+|x-y|)^{-\gamma},\varepsilon b|x-y|],
\nonumber
\eeqn
where the last series is order of $\varepsilon^{(\gamma-d)/(\gamma+1)}$.
This goes to zero as $\ve\to0$,  since $\gamma>d$.
Using Lemma \ref{l5.3} and properties of $R$, we can take the series
in (\ref{6.10}) over $x,y\in\Z^d$. Hence,
\beqn\label{3.24}
S_{\varepsilon,\tau/\varepsilon^2}([r/\varepsilon]+l,
[r/\varepsilon]+p)=\sum\limits_{x,y\in \Z^d}
{\cal G}^g_{\tau/\varepsilon^2}(l+x)
\bar R(\dots){\cal G}^{g}_{\tau/\varepsilon^2}(p+y)^T+o(1),
\,\,\, \varepsilon\to0.
\eeqn
Let us split the function $\bar R$ into the following three matrix functions:
\beqn
R^+(r,x,y)&:=&\frac12 {\bf R}_0(r,x-y),\label{d1'}\\
R^-(r,x,y)&:=&\frac12 {\bf R}_0(r,x-y)\sign (y_1),\label{d1''}\\
R^0(r,x,y)&:=&\bar R(r,x,y)-R^+(r,x,y)-R^-(r,x,y).\label{d1'''}
\eeqn
Next, introduce the  matrices
\beqn\label{Qta}
S^a_{\varepsilon,\tau/\varepsilon^2}
&\equiv& S^a_{\varepsilon,\tau/\varepsilon^2} 
([r/\varepsilon]\!+\!l,[r/\varepsilon]\!+\!p)\nonumber\\
&=&\sum\limits_{x,y\in\Z^d}
{\cal G}^g_{\tau/\varepsilon^2}(l+\!x)\bar R^a
(\varepsilon[r/\varepsilon]-\varepsilon x,[r/\varepsilon]-x,[r/\varepsilon]-y)
{\cal G}^{g}_{\tau/\varepsilon^2}(p+y)^T,
\eeqn
for each $a=\{+,-,0\}$ and split 
$S_{\varepsilon,\tau/\varepsilon^2}$ into three terms,
$S_{\varepsilon,\tau/\varepsilon^2}=S^+_{\varepsilon,\tau/\varepsilon^2}
+S^-_{\varepsilon,\tau/\varepsilon^2}+S^0_{\varepsilon,\tau/\varepsilon^2}$.
The convergence (\ref{pro4}) results now from the following three Lemmas
\ref{Qt1}--\ref{Qt3}, since (see (\ref{gve}))
$ g^\ve_{\tau,r}(z)=
 (1/2)q^\ve_{\tau,r}(z)+ (1/2) f^\ve_{\tau,r}(z)$,
where $q^\ve_{\tau,r}(z)$ is defined by (\ref{ftaur})
and $f^\ve_{\tau,r}(z)$ is defined in Lemma \ref{Qt2}.
\begin{lemma}\label{Qt1}
 $\lim\limits_{\varepsilon\to0} 
(S^+_{\varepsilon,\tau/\varepsilon^2}([r/\varepsilon]+l,[r/\varepsilon]+p)
- (1/2)q^\ve_{\tau,r}(l-p))=0$, $l,p\in\Z^d$,
 where $q^\ve_{\tau,r}(l)$ is defined by (\ref{ftaur}).
\end{lemma}
\begin{lemma}\label{Qt2}
 $\lim\limits_{\varepsilon\to0}
(S^-_{\varepsilon,\tau/\varepsilon^2}([r/\varepsilon]+l,[r/\varepsilon]+p)
-(1/2) f^{\ve}_{\tau,r}(l-p))=0 $, $l,p\in\Z^d$.\\
Here $f^\ve_{\tau,r}(l)$ stands for the matrix which is defined 
similarly to $q^\ve_{\tau,r}(l)$ but with the matrix
 $
A^\pm_{\ve,\sigma}(\tau,r;\theta)
\sign(r_1\pm\partial_1\omega_\sigma(\theta)\tau/\ve)
$ 
instead of $A^\pm_{\ve,\sigma}(\tau,r;\theta)$ in the r.h.s. of (\ref{ftaur}).
\end{lemma}
\begin{lemma}\label{Qt3}
 $\lim\limits_{\varepsilon\to0} 
S^0_{\varepsilon,\tau/\varepsilon^2}
([r/\varepsilon]\!+\!l,[r/\varepsilon]\!+\!p)=0$, $l,p\in\Z^d$. 
\end{lemma}

The proofs of Lemmas \ref{Qt2} and \ref{Qt3} see in Appendices~A and B, resp.
\bigskip\\
{\bf Proof of Lemma \ref{Qt1}}. 
{\it Step (i)}: By (\ref{d1'}) and (\ref{Qta}), 
the function $S^+_{\varepsilon,\tau/\varepsilon^2}$
can be represented as
$$
S^+_{\varepsilon,\tau/\varepsilon^2}=
\frac12\sum\limits_{x\in \Z^d}
{\cal G}^g_{\tau/\varepsilon^2}(l+x)\sum\limits_{y\in \Z^d}
 {\bf R}_0(\varepsilon[r/\varepsilon]-\varepsilon x, y-x)
{\cal G}^g_{\tau/\varepsilon^2}(p+y)^T.
$$
Using the Fourier transform and the Parseval identity
we rewrite $S^+_{\varepsilon,\tau/\varepsilon^2}$ as
\beqn\label{4.9}
S^+_{\varepsilon,\tau/\varepsilon^2}=(2\pi)^{-2d}\frac12
 \sum\limits_{x\in \Z^d}\int\limits_{\T^{2d}}
e^{-il\cdot\theta'+ip\cdot\theta+ix\cdot(\theta-\theta')}
\hat{\cal G}^g_{\tau/\varepsilon^2}(\theta')
\hat {\bf R}_0(\varepsilon[r/\varepsilon]-\varepsilon x,\theta)
\hat{\cal G}^g_{\tau/\varepsilon^2}(\theta)^*\,d\theta d\theta'.
\eeqn
Change variables in (\ref{4.9}): 
$(\theta,\theta')\to(\theta,\varphi)$, $\varphi=\theta-\theta'$ and rewrite 
$S^+_{\varepsilon,\tau/\varepsilon^2}$   in the form
\beqn\label{4.9'}
S^+_{\varepsilon,\tau/\varepsilon^2}=(2\pi)^{-d}\frac12
\int\limits_{\T^{d}} e^{-i(l-p)\cdot\theta}
I_\varepsilon(\theta)
\hat{\cal G}^g_{\tau/\varepsilon^2}(\theta)^*\,d\theta,
\eeqn
where $I_\varepsilon(\theta)$ stands for the matrix-valued function,
\beqn\label{I}
I_\varepsilon(\theta)=
\sum\limits_{x\in \Z^d}(2\pi)^{-d}
\int\limits_{\T^{d}} e^{i(l+x)\cdot\varphi}
\hat{\cal G}^g_{\tau/\varepsilon^2}(\theta-\varphi)\,d\varphi\,
\hat {\bf R}_0(\varepsilon[r/\varepsilon]-\varepsilon x,\theta).
\eeqn
{\it Step (ii)}:
For simplicity of the proof, 
let us assume that $\hat {\bf R}_0(r,\theta)$
satisfies condition (\ref{tildeR}).
Under the more weakened condition (\ref{Varm}) 
the proof is given in Appendix~C.

We apply the Poisson summation formula 
(see, for example, \cite{Tit})
and obtain
\be\label{Psf}
\sum\limits_{x\in \Z^d}e^{i x\cdot\varphi} 
\hat {\bf R}_0(\ve[r/\ve]-\ve x,\theta)
=\ve^{-d}\sum\limits_{n\in \Z^d}  
\tilde {\bf R}_0
(-\varphi/\ve-2\pi n/\ve,\theta)e^{i [r/\ve]\cdot\varphi},
\ee
where $\tilde {\bf R}_0(\cdot,\theta)$
stands for the Fourier transform of $\hat {\bf R}_0(\cdot,\theta)$
(see (\ref{Ftr})).
We substitute the last expression to (\ref{I})
and obtain
\beqn\label{5.20}
I_\varepsilon(\theta)=(2\pi\ve)^{-d}
\int\limits_{\T^{d}} e^{i(l+[r/\ve])\cdot\varphi}
\hat{\cal G}^g_{\tau/\varepsilon^2}(\theta-\varphi)
\Big(\sum\limits_{n\in \Z^d}  
\tilde {\bf R}_0(-\varphi/\ve-2\pi n/\ve,\theta)\Big)
\,d\varphi+o(1).
\eeqn
Condition {\bf I4'} implies that 
\be\label{5.21}
\ve^{-d}\sum\limits_{n\not=0}
\tilde {\bf R}_0(-\varphi/\ve-2\pi n/\ve,\theta)\to0
\quad \mbox{as }\,\ve\to0
\ee
uniformly in $\theta\in\T^d$, $\varphi\in[-\pi,\pi]^d$.
Hence (\ref{5.20}) and (\ref{5.21}) yield
$$
I_\varepsilon(\theta)=
(2\pi\ve)^{-d}\int\limits_{[-\pi,\pi]^{d}} 
e^{i\varphi\cdot(l+[r/\ve])}
\hat{\cal G}^g_{\tau/\varepsilon^2}(\theta-\varphi)
\tilde {\bf R}_0(-\varphi/\ve,\theta)\,d\varphi+o(1).
$$
{\it Step (iii)}:
Change variables $\varphi\to -\ve\varphi$ in the last integral
and obtain
 $$
I_\varepsilon(\theta)=(2\pi)^{-d}
\int\limits_{[-\pi/\ve,\pi/\ve]^d} e^{-i\ve \varphi\cdot(l+[r/\ve])}
\hat{\cal G}^g_{\tau/\varepsilon^2}(\theta+\ve\varphi)
\tilde {\bf R}_0(\varphi,\theta)\,d\varphi+o(1).
$$
Now we use  the following representation for
$\hat{\cal G}^g_{t}(\theta)$ (see (\ref{3.4'})):
\beqn\label{A2}
\hat{\cal G}^g_{t}(\theta)
=\sum\limits_{\sigma=1}^s\sum\limits_{\pm}
e^{\pm i\omega_\sigma(\theta)t}\,h^\mp_\sigma(\theta),
\,\,\,\,\mbox{where }\,
h^\mp_\sigma(\theta)=g(\theta)c^\mp_\sigma(\theta)
=g(\theta)\frac{I\mp i C_\sigma(\theta)}{2}\Pi_\sigma(\theta),
\eeqn
and rewrite $I_\varepsilon(\theta)$ in the form
\beqn\nonumber
I_\varepsilon(\theta)&=&(2\pi)^{-d}
\sum\limits_{\sigma,\pm}
 \int\limits_{[-\pi/\ve,\pi/\ve]^d}e^{-i\ve\varphi\cdot(l+[r/\ve])}
e^{\pm i\omega_\sigma(\theta+\ve\varphi)\tau/\varepsilon^2}
h^\mp_\sigma(\theta+\ve\varphi)\tilde {\bf R}_0(\varphi,\theta)
\,d\varphi+o(1)\nonumber\\
&=&(2\pi)^{-d}
\sum\limits_{\sigma,\pm}h^\mp_\sigma(\theta)
 \int\limits_{[-\pi/\ve,\pi/\ve]^d} e^{-i\varphi\cdot r}
e^{\pm i\omega_\sigma(\theta+\ve\varphi)\tau/\varepsilon^2} 
\tilde {\bf R}_0(\varphi,\theta) \,d\varphi+o(1),
\nonumber
\eeqn
since $e^{-i\ve\varphi\cdot l}-1=O(\ve)$ and 
$e^{-i\varphi\cdot\ve[r/\ve]}-e^{-i\varphi\cdot r}=O(\ve)$
by (\ref{tildeR}).
\medskip\\
{\it Step (iv)}:
Let us take the Taylor sum representation 
for $\omega_\sigma(\theta+\ve\varphi)$:
\be\la{Tsr}
\omega_\sigma(\theta+\ve\varphi)\tau/\varepsilon^2=
\omega_\sigma(\theta)\tau/\varepsilon^2
+\varphi\cdot\nabla\omega_\sigma(\theta)\tau/\varepsilon+
(\tau/2)\varphi\cdot H_\sigma(\theta)\varphi+O(\ve|\varphi|^3),
\ee
where $H_\sigma(\theta)$ denotes the matrix 
$\nabla^2\omega_\sigma(\theta)$.
Hence, by (\ref{tildeR}),
\beqn\label{3.13}
I_\varepsilon(\theta)=
\sum\limits_{\sigma,\pm}h^\mp_\sigma(\theta)
e^{\pm i\omega_\sigma(\theta)\tau/\varepsilon^2} 
A^{\mp}_{\ve,\sigma}(\tau,r;\theta)
+o(1),\,\,\,\,\quad\ve\to 0.
\eeqn
Here (see formulas (\ref{Apm}) and (\ref{K}))
\beqn\nonumber
A^{\mp}_{\ve,\sigma}(\tau,r;\theta)
&=& (2\pi)^{-d}\int\limits_{\R^d}
e^{-i\varphi\cdot(r\mp\nabla\omega_\sigma(\theta)\tau/\ve)}
e^{\pm i(\tau/2)\varphi\cdot H_\sigma(\theta)\varphi}
\tilde {\bf R}_0(\varphi,\theta)\,d\varphi
\nonumber\\
&=&F^{-1}_{\varphi\to(r\mp\nabla\omega_\sigma(\theta)\tau/\ve)}
\Big[\tilde {\bf R}_0(\varphi,\theta)
e^{\pm i(\tau/2)\varphi\cdot H_\sigma(\theta)\varphi}\Big]
\nonumber\\
&=&\int\limits_{\R^d} 
\hat {\bf R}_0(r\mp\nabla\omega_\sigma(\theta)\tau/\ve-x,\theta)
K^{\mp}_{\sigma}(\tau,x,\theta)\,dx,\nonumber
\eeqn
and $K^{\mp}_{\sigma}(\tau,x,\theta)$ ($\tau>0$, $x\in\R^d$,
$\theta\in \T^d\setminus{\cal C}$) stands 
for the following matrix-valued function
\be\label{K2}
K^{\mp}_{\sigma}(\tau,x,\theta)=F^{-1}_{\varphi\to x}
[e^{\pm i(\tau/2) \varphi \cdot H_\sigma(\theta)\varphi}]
=\frac{e^{\pm i\pi s/4}}{(2\pi\,\tau)^{d/2}}
\frac{e^{\mp i/(2\tau) x\cdot H^{-1}_\sigma(\theta)x}}
{\sqrt{|\det H_\sigma(\theta)|}},
\ee
where 
$s$ denotes the signature of the matrix $H_\sigma(\theta)$.
By definition, the signature of the nondegenerate symmetric matrix $A$ 
means the signature of the quadratic form with this matrix (or 
the difference between the sums of positive and negative eigenvalues of $A$).
Formula (\ref{K2}) follows from the following equality
(see, for example, {\it Ramanujan's integrals} in \cite[section 7.5]{Tit}
or \cite[\S3]{Fed}): for any $a>0$,
$$
\int\limits_{-\infty}^{+\infty} 
e^{-ix\varphi\pm ia \varphi^2}\,d\varphi
=\sqrt{\frac{\pi}{a}} e^{\pm i\pi/4}e^{\mp i x^2/(4a)},\quad x\in\R^1.
$$
{\it Step (v)}:
We substitute (\ref{3.13}) in (\ref{4.9'}), 
 apply the decomposition (\ref{A2}) 
to $\hat{\cal G}^g_{\tau/\ve^2}(\theta)^*$ and obtain
\beqn\label{6.22'}
S^+_{\varepsilon,\tau/\varepsilon^2}&=&(2\pi)^{-d}
\frac12
\int\limits_{\T^{d}} e^{-i(l-p)\cdot\theta}
\Big(\sum\limits_{\sigma,\pm}
h^\mp_\sigma(\theta)
e^{\pm i\omega_\sigma(\theta)\tau/\varepsilon^2} 
A^{\mp}_{\ve,\sigma}(\tau,r;\theta)
\Big)\nonumber\\
&&\,\,\,\,\,\,\,\,\,\,\,\,\,\,\,\,\,\,\,\,\,\,\,\,
\times\Big(\sum\limits_{\sigma',\pm}
e^{\pm i\omega_{\sigma'}(\theta)\tau/\varepsilon^2}
h^\mp_{\sigma'}(\theta)^*\Big)\,d\theta+o(1), \quad \ve\to0.
\eeqn
Therefore, to find the asymptotics of $S^+_{\varepsilon,\tau/\varepsilon^2}$ 
it suffices to study the behaviour as $\ve\to0$ of the following integrals
\be\label{4.9''}
I_{\sigma\sigma'}^\pm(\ve)\equiv (2\pi)^{-d}
\frac12\int\limits_{\T^{d}} e^{-i(l-p)\cdot\theta}
e^{ i(\omega_\sigma(\theta)\pm \omega_{\sigma'}(\theta))\tau/\varepsilon^2}
h^-_\sigma(\theta)A^{-}_{\ve,\sigma}(\tau,r;\theta)
h^\mp_{\sigma'}(\theta)^*\,d\theta,
\ee
$\sigma,\sigma'=1,\dots,s$. Note that 
$\sup\limits_{\theta\in\T^d\setminus {\cal C}}
|A^{-}_{\ve,\sigma}(\tau,r;\theta)|\le C<\infty$
by condition {\bf I4'}. Hence, the 
function
$
h^-_\sigma(\theta)A^{-}_{\ve,\sigma}(\tau,r;\theta)
h^\mp_{\sigma'}(\theta)^*\in L^1(\T^d)
$
by condition {\bf E6}. Therefore, the oscillatory integrals with 
$\omega_\sigma(\theta)\pm\omega_{\sigma'}(\theta)\not\equiv{\rm const}_\pm$
vanish as $\varepsilon\to0$ by the Lebesgue--Riemann theorem.
Furthermore, the identities
$\omega_\sigma(\theta)\pm\omega_{\sigma'}(\theta)\equiv{\rm const}_\pm$ 
 in the exponent of (\ref{4.9''})
with ${\rm const}_\pm\ne 0$ are impossible by condition {\bf E5}.
Hence, only the integrals with 
$\omega_\sigma(\theta)-\omega_{\sigma'}(\theta)\equiv 0$
contribute to the integral (\ref{4.9''})
since  $\omega_\sigma(\theta)+\omega_{\sigma'}(\theta)\equiv 0$ would imply
$\omega_\sigma(\theta)\equiv\omega_{\sigma'}(\theta)\equiv 0$ which
is impossible by {\bf E4}.
Therefore, if $\sigma\not=\sigma'$,  
$I_{\sigma\sigma'}^\pm(\varepsilon)=o(1)$ as $\varepsilon\to0$.
If $\sigma=\sigma'$, 
$I_{\sigma\sigma}^+(\varepsilon)=o(1)$,
$$
I_{\sigma\sigma}^-(\varepsilon)=(2\pi)^{-d}\frac12
\int\limits_{\T^{d}} e^{-i(l-p)\cdot\theta}
h^-_{\sigma}(\theta)A^{-}_{\ve,\sigma}(\tau,r;\theta)
h^+_{\sigma}(\theta)^*\,d\theta +o(1),\quad \varepsilon\to0.   
$$ 
Finally, using (\ref{6.22'}) and the equalities
$h^\mp_\sigma(\theta)=g(\theta)\Pi_\sigma(\theta)(I\mp i C_\sigma(\theta))/2$
 and (\ref{ftaur}), we obtain
\beqn\label{6.28}
S^+_{\varepsilon,\tau/\varepsilon^2}\!\!&=&\!\!(2\pi)^{-d}
\frac12 \int\limits_{\T^{d}} e^{-i(l-p)\cdot\theta}
\sum\limits_{\sigma=1}^s\sum\limits_\pm
h^\mp_{\sigma}(\theta)A^{\mp}_{\ve,\sigma}(\tau,r;\theta)
h^\pm_{\sigma}(\theta)^*
\,d\theta+o(1)\\
&=&\frac12 q^\ve_{\tau,r}(l-p)+o(1),\quad \varepsilon\to0.  
\,\,\,\bo  \nonumber
\eeqn
\setcounter{equation}{0}
\section{Convergence of Wigner matrices}\label{sec.4}
Here we prove Theorem \ref{t4.5}.
Theorem \ref{the2'+} implies that for any fixed $r\in\R^d_+$,
$\tau\not=0$, and $y\in (2\Z)^d$, the following convergence holds,
\beqn\label{5.1}
\lim_{\varepsilon\to0}
\E^\varepsilon_{\tau/\varepsilon}
\big(a_+^\ast([r/\varepsilon]+y/2) \otimes a_+([r/\varepsilon]-y/2)\big)
={\cal W}^p_+(\tau,r;y),
\eeqn
where in the Fourier space one has
\beqn\label{5.2}
\hat {\cal W}^p_+(\tau,r;\theta)\!&=&\!
 \frac{1}{2}\Big(\Omega^{1/2}\hat {\bf q}^{00}_{\tau,r}(\theta)\Omega^{1/2}+
 \Omega^{-1/2}\hat {\bf q}^{11}_{\tau,r}(\theta)\Omega^{-1/2}
\nonumber\\
&&\!\!+i\Omega^{1/2}\hat {\bf q}^{01}_{\tau,r}(\theta)\Omega^{-1/2}-i
 \Omega^{-1/2}\hat {\bf q}^{10}_{\tau,r}(\theta)\Omega^{1/2}\Big)
=W^p_+(\tau,r;\theta),\,\,\,r\in\R^d_+,\,\,\,
\eeqn
by formulas (\ref{3.16+}) and (\ref{Qtaur}).
Therefore, convergence (\ref{3.17}) follows from (\ref{5.1}),
(\ref{5.2}) and the following bound:
$$
\sup\limits_{\ve>0}\sup\limits_{t\in\R} \sup\limits_{z,z'\in \Z^d}
\left|\E^\varepsilon_{t}
\big(a_+^\ast(z) \otimes a_+(z')\big)\right|\le C<\infty.
$$
The proof of this bound follows from Lemma \ref{lcom}.
\begin{lemma}\label{lcom}
Let conditions {\bf V2'}  and {\bf E1}--{\bf E3}, {\bf E6} hold
and let $\alpha<-d/2$. Then the following bound holds:
$
\sup\limits_{\ve>0}\sup\limits_{t\in\R} \sup\limits_{z,z'\in \Z^d_+}
\Vert Q_{\varepsilon,t}(z,z')\Vert\le C<\infty.
$
\end{lemma}
{\bf Proof}.
The representation (\ref{sol}) gives
\beqn
Q^{ij}_{\varepsilon,t}(z,z')
&=&\E^\varepsilon_0\Big(Y^i(z,t)\otimes Y^j(z',t)\Big)
= \sum\limits_{y,y'\in \Z^d_+} \sum\limits_{k,l=0,1}
{\cal G}^{ik}_{t,+}(z,y)Q^{kl}_\varepsilon(y,y')
{\cal G}^{jl}_{t,+}(z',y')\nonumber\\
&=& \langle Q_\varepsilon(y,y'), \Phi^i_{z}(y,t)\otimes
\Phi^j_{z'}(y',t)\rangle_+,\nonumber
\eeqn
where $\Phi^i_{z}(y,t)$ is given by
\beqn
\Phi^i_{z}(y,t)&=&\Big(
{\cal G}^{i0}_{t,+}(z,y),{\cal G}^{i1}_{t,+}(z,y)\Big)\nonumber\\
&=&({\cal G}_t^{i0}(z-y)-{\cal G}_t^{i0}(z-\tilde y),
{\cal G}_t^{i1}(z-y)-{\cal G}_t^{i1}(z-\tilde y)),
\,\,\,\,\,i=0,1.\nonumber
\eeqn
The Parseval identity, formula (\ref{4.5}),
and condition {\bf E6} imply that
$$
\Vert\Phi^i_{z}(\cdot,t)\Vert^2_{l^2}= (2\pi)^{-d}
\int\limits_{\T^d} |\hat\Phi^i_{z}(\theta,t)|^2\,d\theta
\le C\int\limits_{\T^d}
\Big( |\hat{\cal G}^{i0}_t(\theta)|^2
+|\hat{\cal G}^{i1}_t(\theta)|^2\Big)
\,d\theta \le C_0<\infty,
$$
where the constant $C_0$ does not depend on
$z\in\Z^d$ and $t\in\R$.
Corollary \ref{c4.10} gives now
$$
|Q^{ij}_{\varepsilon,t}(z,z')|=
|\langle Q_\varepsilon(y,y'), \Phi^i_{z}(y,t)\otimes
\Phi^j_{z'}(y',t)\rangle_+|
\le C\Vert\Phi^i_{z}(\cdot,t)\Vert_{l^2_+}\,
\Vert\Phi^j_{z'}(\cdot,t)\Vert_{l^2_+}\le C_1<\infty,
$$
where the constant $C_1$  does  not depend on
$z,z'\in\Z^d_+$, $t\in\R$, and $\varepsilon>0$.\bo

\appendix
\setcounter{section}{1}
\setcounter{equation}{0}
\section*{\large\bf Appendix A: Proof of Lemma 6.7}

By (\ref{d1''}) and (\ref{Qta}), the function 
$S^-_{\varepsilon,\tau/\varepsilon^2}$
can be represented as 
$$
S^-_{\varepsilon,\tau/\varepsilon^2}
=\frac12\sum\limits_{x\in \Z^d}
{\cal G}^g_{\tau/\varepsilon^2}(l+x)\sum\limits_{y\in \Z^d}
 {\bf R}_0(\kappa_{\varepsilon,x}, y-x)
\sign([r_1/\varepsilon]-y_1)
{\cal G}^g_{\tau/\varepsilon^2}(p+y)^T,
$$
where, by definition,  $\kappa_{\varepsilon,x}
=\varepsilon[r/\varepsilon]-\varepsilon x\in\R^d$. 
The Parseval equality yields
\beqn
&&\sum\limits_{y\in \Z^d}\sign([r_1/\varepsilon]-y_1)
{\bf R}_0(\kappa_{\varepsilon,x}, y-x)
{\cal G}^g_{\tau/\varepsilon^2}(p+y)^T\nonumber\\
&=&(2\pi)^{-d}\int\limits_{\T^{d}}
F_{y\to\theta}\Big[\sign([r_1/\varepsilon]-y_1)
 {\bf R}_0(\kappa_{\varepsilon,x}, y-x)\Big]
\overline{F_{y\to\theta}\Big[{\cal G}^g_{\tau/\varepsilon^2}(p+y)^T\Big]}
\,d\theta.\nonumber
\eeqn
Note that
$$
F_{y\to\theta}\Big[\sign([r_1/\varepsilon]-y_1)
 {\bf R}_0(\kappa_{\varepsilon,x}, y-x)\Big]=(2\pi)^{-d}
F_{y\to\theta}\Big[\sign([r_1/\varepsilon]-y_1)\Big]*
F_{y\to\theta}\Big[ {\bf R}_0(\kappa_{\varepsilon,x}, y-x)\Big],
$$ 
where
$F_{y\to\theta}[{\bf R}_0(\kappa_{\varepsilon,x}, y-x)]
=e^{ix\cdot \theta}\hat {\bf R}_0(\kappa_{\varepsilon,x},\theta)$,
$$
F_{y\to\theta}\Big[\sign([r_1/\varepsilon]-y_1)\Big]=-i\,(2\pi)^{d-1}
\delta(\bar \theta)\PV\left(\frac1{\tg(\theta_1/2)}\right)
e^{i[r_1/\varepsilon]\theta_1},
$$
$\theta=(\theta_1,\bar\theta)$, $\bar\theta=(\theta_2,\dots,\theta_d)$, and
$\PV$ stands for the Cauchy principal part.
Hence,
 \beqn
S^-_{\varepsilon,\tau/\varepsilon^2}&=&
-\frac{i}{2}(2\pi)^{-d-1}
\sum\limits_{x\in \Z^d}{\cal G}^g_{\tau/\varepsilon^2}(l+x)
\int\limits_{\T^{d}}\Big(\PV\int\limits_{\T^1}
\frac{e^{i[r_1/\varepsilon](\theta_1-z)+ix_1z}}{\tg((\theta_1-z)/2)}
\hat {\bf R}_0(\kappa_{\varepsilon,x},z,\bar\theta)dz\times\nonumber\\
&&\quad\quad\quad\quad\quad\quad\quad\quad\quad\quad\quad\quad\quad\quad
\times e^{i\bar x\cdot\bar\theta}
\hat{\cal G}^g_{\tau/\varepsilon^2}(\theta)^*e^{ip\cdot\theta}\Big)\,d\theta
\nonumber\\
&=&-\frac{i}2(2\pi)^{-2d-1}
\sum\limits_{x\in \Z^d}\int\limits_{\T^{2d}}
\Big(e^{-i(l+x)\cdot\theta'} \hat{\cal G}^g_{\tau/\varepsilon^2}(\theta')\times
\nonumber\\
&&\quad\quad\quad\quad\times \PV\int\limits_{\T^1}
\frac{e^{i[r_1/\varepsilon](\theta_1-z)+ix_1z}}{\tg((\theta_1-z)/2)}
\hat {\bf R}_0(\kappa_{\varepsilon,x},z,\bar\theta)\,dz\,
e^{i\bar x\cdot\bar\theta}\hat{\cal G}^g_{\tau/\varepsilon^2}(\theta)^*
e^{ip\cdot\theta}\Big)\,d\theta d\theta'.\nonumber
\eeqn 
We change variables $\theta_1\to\phi=\theta_1-z$,
then denote $z=\theta_1$, and change variables
 $\theta'\to\varphi=\theta-\theta'$. 
Therefore,
\beqn\label{A1}
S^-_{\varepsilon,\tau/\varepsilon^2}=(2\pi)^{-d}\frac1{4\pi i}
\int\limits_{\T^{d}} e^{-i(l-p)\cdot\theta}
I_\varepsilon(\theta)P_{\varepsilon}(\theta)\,d\theta,  
\eeqn
where $I_\varepsilon(\theta)$ is defined by (\ref{I}),
and $P_\varepsilon(\theta)$ stands for the matrix-valued function 
$$
P_{\varepsilon}(\theta)=\PV\int\limits_{\T^1}
\frac{e^{i([r_1/\varepsilon]+p_1)\phi}}{\tg(\phi/2)}
\hat{\cal G}^g_{\tau/\varepsilon^2}(\theta_1+\phi,\bar\theta)^*\,d\phi,
\quad\theta=(\theta_1,\bar\theta).
$$
Applying the partition of unity (\ref{part}), (\ref{fge}), 
and formula (\ref{A2}),
we rewrite $P_\varepsilon(\theta)$ in the form 
$$
P_\varepsilon(\theta)=\sum\limits_{\sigma=1}^s
\sum\limits_{\pm}\PV\int\limits_{\T^1} 
\frac{e^{i([r_1/\varepsilon]+p_1)\phi}
e^{\pm i\omega_\sigma(\theta_1+\phi,\bar \theta)
\tau/\varepsilon^2}}{\tg(\phi/2)}\,
g(\theta_1+\phi,\bar\theta)c^\mp_\sigma(\theta_1+\phi,\bar\theta)^*\,d\phi,
$$
where $c^\mp_\sigma(\theta)^*=\Pi_\sigma(\theta)(I\mp iC^*_\sigma(\theta))/2$.
Let us fix $r_1\in\R$, $\tau\not=0$.
Since $\partial_1\omega_\sigma(\theta)\not=0$ 
for $\theta\in\supp g$,
we can choose $\ve_g\equiv\ve_g(r_1,\tau)>0$ such that for all $\ve<\ve_g$
and $\theta\in\supp g$,
$\partial_1\omega_\sigma(\theta)\not=\pm \ve r_1/\tau$.
\bigskip\\
{\bf Lemma A.1}\, 
{\it Let us fix $r_1\in\R$ and $\tau\not=0$.
Then \\
(i) $\sup\limits_{\theta\in\T^d}\sup\limits_{\varepsilon>0}
|P_{\varepsilon}(\theta)|<\infty$.\\
(ii) Let $r_1\pm\partial_1\omega_\sigma(\theta)\tau/\ve\not=0$
for all $\theta\in\supp g$ and $\ve\in(0,\ve_g)$. Then
$$ 
P_\varepsilon(\theta)-2\pi i\sum\limits_{\sigma=1}^s
\sum\limits_{\pm} e^{\pm i\omega_\sigma(\theta)\tau/\varepsilon^2}
\sign(r_1\pm\partial_1\omega_\sigma(\theta)\tau/\ve)\,
g(\theta)c^\mp_{\sigma}(\theta)^*\to0
\quad\mbox{as }\,\, \varepsilon\to+0.  
$$
}

Lemma~A.1 can be proved by using the technique of \cite[Lemma 8.3]{DKM} or 
of \cite[Proposition A.4 (i), (ii)]{BPT}. 
The proof of this lemma is based on the following well-known assertion:
$$
\lim_{\lambda\to+\infty}
\left(\PV\int\limits_{-\pi}^{\pi}\frac{e^{i\lambda \omega(z)}\chi(z)}{z}\,dz-
\pi ie^{i\lambda \omega(0)}\chi(0)\sign \omega'(0)\right)=0,
$$
where $\chi\in C^1$, $\omega\in C^2$, and $\omega'(0)\not=0$.
\medskip

By Lemma~A.1 we can rewrite the r.h.s. of (\ref{A1}) 
in the form
\beqn \label{A3} 
S^-_{\varepsilon,\tau/\varepsilon^2}
=(2\pi)^{-d}\frac12\sum\limits_{\sigma,\pm}
\int\limits_{\T^{d}} e^{-i(l-p)\cdot\theta}
I_\varepsilon(\theta) 
e^{\pm i\omega_{\sigma}(\theta)\tau/\varepsilon^2}
\sign(r_1\pm\partial_1\omega_{\sigma}(\theta)\tau/\ve) 
h^\mp_{\sigma}(\theta)^*\,d\theta,
\eeqn
where $h^\mp_\sigma(\theta)=g(\theta)c^\mp_\sigma(\theta)$.
Substituting (\ref{3.13}) in (\ref{A3}) we obtain
\beqn\label{A4}
S^-_{\varepsilon,\tau/\varepsilon^2}&=&
(2\pi)^{-d}\frac12
\int\limits_{\T^{d}} e^{-i(l-p)\cdot\theta}
\Big(\sum\limits_{\sigma,\pm}
h^\mp_\sigma(\theta)
e^{\pm i\omega_\sigma(\theta)\tau/\varepsilon^2} 
A^{\mp}_{\ve,\sigma}(\tau,r;\theta)\Big)\times\nonumber\\
&&\times\Big(\sum\limits_{\sigma',\pm}
e^{\pm i\omega_{\sigma'}(\theta)\tau/\varepsilon^2}
\sign(r_1\pm\partial_1\omega_{\sigma'}(\theta)\tau/\ve) 
h^\mp_{\sigma'}(\theta)^*\Big)\,d\theta+o(1), \quad \ve\to0.
\eeqn
Finally, comparing the r.h.s. of (\ref{A4}) and (\ref{6.22'}) we see that
 the problem of evaluating the limit value of (\ref{A4}) is solved 
by the similar way as in the step (v) of the proof of Lemma \ref{Qt1}. 
\bo

\setcounter{section}{2}
\setcounter{equation}{0}
\section*{\large\bf  Appendix B: Proof of Lemma 6.8}

By (\ref{Qta}) we  write
$$
S^0_{\varepsilon,\tau/\varepsilon^2}=
\sum\limits_{x,y\in\Z^d}
{\cal G}^g_{\tau/\varepsilon^2}(l+\!x) R^0(\kappa_{\varepsilon,x},
[r/\varepsilon]\!-\!x,[r/\varepsilon]\!-\!y)
{\cal G}^{g}_{\tau/\varepsilon^2}(p+y)^T,
$$ 
where $\kappa_{\varepsilon,x}
=\varepsilon[r/\varepsilon]-\varepsilon x$. 
Change variables $y\to z=y-x$ and denote the series over $x$ by
$\Phi_\varepsilon(z)$,
\beqn\label{B1}
\Phi_\varepsilon(z)&\equiv& \Phi_\varepsilon(z,\tau,r,l,p)\nonumber\\
&=&
\sum\limits_{x\in \Z^d}
{\cal G}^g_{\tau/\varepsilon^2}(l+\!x) R^0(\kappa_{\varepsilon,x},
[r/\varepsilon]\!-\!x,[r/\varepsilon]\!-\!x-z)
{\cal G}^{g}_{\tau/\varepsilon^2}(p+x+z)^T.
\eeqn 
Therefore,
\be \label{B2}
S^0_{\varepsilon,\tau/\varepsilon^2}
=\sum\limits_{z\in\Z^d}\Phi_\varepsilon(z).  
\ee
The estimate (\ref{3.2'}) for $R$
and the notation (\ref{d1'''}) imply the same estimate for $R^0$,
\be \label{B3}
|R^0(r,x,y)|\le C(1+|x-y|)^{-\gamma},\quad x,y\in\Z^d. 
\ee
Next, the Cauchy--Schwartz inequality yields
\beqn\label{B4}
\sum\limits_{x\in \Z^d}
|{\cal G}^g_{\tau/\varepsilon^2}(l+\!x)| 
|{\cal G}^{g}_{\tau/\varepsilon^2}(p+x+z)^T|
\le \Vert {\cal G}^g_{\tau/\varepsilon^2}(\cdot)\Vert^2_{\ell^2}
\le C (1+\Vert \hat V^{-1}\Vert^2_{L^2(\T^d)}).
\eeqn
Hence, condition {\bf E6} and the estimate (\ref{B3}) imply that
$|\Phi_\varepsilon(z)|\le C(1+|z|)^{-\gamma}$.
Since $\gamma>d$,
$$
\sum\limits_{z\in\Z^d}|\Phi_\varepsilon(z)|\le C<\infty,
$$
and the series in (\ref{B2}) converges uniformly in $\varepsilon$
(and also in $\tau,r,l,p$). Therefore, it suffices to prove that
\be\label{B5}
\lim_{\varepsilon\to0}\Phi_\varepsilon(z)=0\quad\mbox{for each }
\, z\in\Z^d.
\ee
Let us consider the series  in (\ref{B1}). At first, note that
by the property {\bf (a)} for $R$ and the definition (\ref{d1'''}), 
the function $R^0(r,x,y)$ has the form   
$R^0(r,x,y)={\bf R}^0(r,x_1,y_1,\bar x-\bar y)$, and
${\bf R}^0(r,x_1,y_1,\bar z)=0$ for $y_1<0$. Hence,
$$
{\bf R}^0(r,[r_1/\varepsilon] -x_1,[r_1/\varepsilon]-x_1-z_1,\bar z)=0
\quad \mbox{for }\, x_1> [r_1/\varepsilon]-z_1.
$$
Secondly, it follows from condition (\ref{3.14}) that  $\forall \delta>0$
$\exists K_\delta>0$ such that for any $y_1>K_\delta$ and $\forall r\in\R^d$,
$|{\bf R}^0(r,y_1,y_1-z_1,\bar z)|<\delta$.
Hence, $\forall \delta>0$ $\exists M_\delta=\max(K_\delta,z_1)>0$
such that
\beqn
&&\Big|\sum\limits_{x\in\Z^d:\,x_1<[r_1/\varepsilon]-M_\delta}\!\!\!\!\!\!
{\cal G}^g_{\tau/\varepsilon^2}(l+x) {\bf R}^0(\kappa_{\varepsilon,x},
[r_1/\varepsilon]\!-\!x_1,[r_1/\varepsilon]\!-\!x_1-z_1,\bar z)
{\cal G}^{g}_{\tau/\varepsilon^2}(p+x+z)^T\Big|
\nonumber\\
&\le&\delta \sum\limits_{x\in \Z^d}
\Big|{\cal G}^g_{\tau/\varepsilon^2}(l+\!x)
{\cal G}^{g}_{\tau/\varepsilon^2}(p+x+z)^T\Big|
\le C\delta,      \nonumber
\eeqn 
by the bound (\ref{B4}). Let us fix $\delta>0$. 
Therefore, it suffices to prove that for each $l,p,z\in\Z^d$,
\be\label{B6}
\sum\limits_{x_1\in A_\ve}\sum\limits_{\bar x\in \Z^{d-1}}
{\cal G}^g_{\tau/\varepsilon^2}(l+x) {\bf R}^0(\kappa_{\varepsilon,x},
[r_1/\varepsilon]-x_1,[r_1/\varepsilon]-x_1-z_1,\bar z)
{\cal G}^{g}_{\tau/\varepsilon^2}(p+x+z)^T \to0  
\ee
as $\varepsilon\to0$, where $A_\ve:=\{x_1\in\Z^1:\,
x_1\in[[r_1/\varepsilon]-M_\delta,[r_1/\varepsilon]-z_1]\}$.

 At first, note that
if $x_1\in A_\ve$,  $[r_1/\varepsilon]-x_1\in[z_1,M_\delta]$.
Denote $\bar\kappa_{\varepsilon,\bar x}\equiv
\varepsilon[\bar r/\varepsilon]-\varepsilon \bar x$,
$\bar r\in\R^{d-1}$, $\bar x\in\Z^{d-1}$.
Then, by property (\ref{tildeR}) for ${\bf R}_0$ 
and property (\ref{d'1}) for $R$,
$$
\sup_{\bar x\in\Z^{d-1}}|{\bf R}^0(\varepsilon[r_1/\varepsilon]
-\varepsilon x_1,\bar\kappa_{\ve,\bar x},y_1,y'_1,\bar z)
-{\bf R}^0(0,\bar\kappa_{\varepsilon,\bar x},y_1,y'_1,\bar z)|
\to 0,\quad \varepsilon\to0,
$$
for every $x_1\in A_\ve$, $y_1,y'_1\in\Z^1$, and $\bar z\in\Z^{d-1}$.
Hence, by the bound (\ref{B4}), we can replace 
$\kappa_{\varepsilon,x}$ into $(0,\bar\kappa_{\varepsilon,\bar x})$ 
in the series (\ref{B6}).

Further, we change the variable $x_1\to x'_1$:
$x'_1=[r_1/\varepsilon]-x_1$ in (\ref{B6})
and denote $x'_1=x_1$, $x_\varepsilon=([r_1/\varepsilon]-x_1,\bar x)$.
Therefore, to derive (\ref{B6}) it suffices to check that for any  fixed
$l,p,z\in\Z^d$, and $x_1\in[z_1,M_\delta]$, 
$$
D_\varepsilon:=\sum\limits_{\bar x\in \Z^{d-1}}
{\cal G}^{g}_{\tau/\varepsilon^2}(l+x_\varepsilon) 
{\bf R}^{0}(0,\bar\kappa_{\varepsilon,\bar x},x_1,x_1-z_1,\bar z)
{\cal G}^{g}_{\tau/\varepsilon^2}(p+x_\varepsilon+z)^T\to0 
\quad \mbox{as }\,\varepsilon\to0.
$$
Let $d=1$. Then 
$D_\varepsilon={\cal G}^{g}_{\tau/\varepsilon^2}(l_1+[r_1/\ve]-x_1) 
{\bf R}^{0}(0,x_1,x_1-z_1)
{\cal G}^{g}_{\tau/\varepsilon^2}(p_1+[r_1/\ve]-x_1+z_1)^T$
vanishes as $\ve\to0$. Indeed, applying the inverse Fourier transform
and the decomposition (\ref{A2}), we have
\beqn
{\cal G}^{g}_{\tau/\varepsilon^2}(l_1+[r_1/\ve]-x_1)
&=&\frac1{2\pi}
\int_{\T^1} e^{-i(l_1+[r_1/\varepsilon]-x_1)\theta_1}
\hat{\cal G}^g_{\tau/\varepsilon^2}(\theta_1)\,d\theta_1\nonumber\\
&=&
\frac1{2\pi}\sum\limits_{\sigma=1}^s
\sum\limits_{\pm}
\int\limits_{\T^1} e^{-i(l_1+[r_1/\varepsilon]-x_1)\theta_1}
e^{\pm i\omega_\sigma(\theta_1)\tau/\varepsilon^2}
g(\theta_1)c^\mp_\sigma(\theta_1)\,d\theta_1.\nonumber
\eeqn
The last integral vanishes as $\ve\to0$, since
$\omega'_\sigma(\theta_1)\not=0$ for all $\theta_1\in\supp g$.
\medskip

Let $d>1$. In this case, we put, for simplicity, 
${\bf R}^{0}(0,\bar\kappa_{\varepsilon,\bar x},\dots)=
{\bf R}^{0}(0,\bar\kappa_{\varepsilon,\bar x},x_1,x_1-z_1,\bar z)$. 
Using the Fourier transform we rewrite $D_{\varepsilon}$ as
\beqn\nonumber
D_\varepsilon=\sum\limits_{\bar x\in \Z^{d-1}}
(2\pi)^{-2d}\int\limits_{\T^{2d}}
e^{-i(l+x_\varepsilon)\cdot\theta'}
\hat{\cal G}^g_{\tau/\varepsilon^2}(\theta')
{\bf R}^{0}(0,\bar\kappa_{\varepsilon,\bar x},\dots)
e^{i(p+x_\varepsilon+z)\cdot\theta}
\hat{\cal G}^g_{\tau/\varepsilon^2}(\theta)^*\,d\theta\,d\theta'.
\eeqn
Let us change variables $\theta=(\theta_1,\bar\theta)$,
$\theta'=(\theta'_1,\bar\theta')$ as follows: 
$\bar\theta'\to\bar\varphi=\bar\theta-\bar\theta'$,
and denote $\theta_1=\varphi_1$, $\theta'_1=\theta_1$. Hence,
\beqn\label{B.7}
D_\varepsilon=(2\pi)^{-d+1}\int\limits_{\T^{d-1}}
e^{-i(\bar l-\bar p-\bar z)\cdot\bar\theta}
A_\varepsilon(\bar\theta)B_\varepsilon(\bar\theta)\,d\bar\theta,
\eeqn
where
\beqn\label{B.8}
A_\varepsilon(\bar\theta)&=&\frac1{2\pi}
\int_{\T^1} e^{-i(l_1+[r_1/\varepsilon]-x_1)\theta_1}
{\cal I}_\varepsilon(\theta)\,d\theta_1,\\
B_\varepsilon(\bar\theta)&=&\frac1{2\pi}
\int_{\T^1} e^{i(p_1+[r_1/\varepsilon]-x_1+z_1)\varphi_1}
\hat{\cal G}^g_{\tau/\varepsilon^2}(\varphi_1,\bar\theta)^*
\,d\varphi_1,\label{B.9}\\
{\cal I}_\varepsilon(\theta)&=&(2\pi)^{-d+1} 
\int\limits_{\T^{d-1}} e^{i\bar l\cdot\bar\varphi}
\hat{\cal G}^g_{\tau/\varepsilon^2}(\theta_1,\bar\theta-\bar\varphi)\,
\Big(\sum\limits_{\bar x\in \Z^{d-1}}
e^{i \bar x\cdot\bar\varphi}
{\bf R}^0(0,\ve[\bar r/\ve]-\ve\bar x,\dots)
\Big)d\,\bar\varphi\,\,\,\,\,\,\,\,\, \label{B.10}
\eeqn
(cf. formula (\ref{I})).
Below we prove the following bound for ${\cal I}_\varepsilon(\theta)$
(cf. (\ref{3.13})):
\beqn\label{B.11}
{\cal I}_\varepsilon(\theta)=
\sum\limits_{\sigma=1}^s\sum\limits_{\pm}
h^\mp_\sigma(\theta)\,e^{\pm i\omega_\sigma(\theta)\tau/\varepsilon^2}\,
{\cal A}^{\mp}_{\ve,\sigma}(\tau,\bar r;\theta)+o(1),\quad\varepsilon\to0,
\eeqn
where $h^\mp_\sigma(\theta)$ is defined in (\ref{A2}), 
\beqn\label{B.12}
{\cal A}^{\mp}_{\ve,\sigma}(\tau,\bar r;\theta)=\int\limits_{\R^{d-1}} 
{\bf R}^0(0,\bar r\mp\nabla_{\bar\theta}\,
\omega_\sigma(\theta)\tau/\ve\!-\!\bar y,\dots)
{\cal K}^{\mp}_\sigma(\tau,\bar y, \theta)\,d\bar y,
\eeqn
${\cal K}^{\mp}_{\sigma}(\tau,\bar y,\theta)$, $\tau>0$, $\bar y\in\R^{d-1}$,
$\theta\in \T^d\setminus{\cal C}$, stands 
for the following matrix-valued function
$$
{\cal K}^{\mp}_{\sigma}(\tau,\bar y,\theta)=F^{-1}_{\bar\varphi\to \bar y}
[e^{\pm i(\tau/2) \bar\varphi \cdot 
(\nabla^2_{\bar\theta}\omega_\sigma(\theta))\bar\varphi}]
=\frac{e^{\pm i\pi s/4}}{(2\pi\,\tau)^{(d-1)/2}}
\frac{e^{\mp i/(2\tau)\bar y\cdot(\bar\nabla^2\omega_\sigma(\theta))^{-1}
\bar y}}
{\sqrt{|\det (\nabla^2_{\bar\theta}\omega_\sigma(\theta))|}},
$$
and 
$s$ denotes the signature of the matrix 
$(\nabla^2_{\bar\theta}\omega_\sigma(\theta))$,
$\theta=(\theta_1,\bar\theta)\in\T^d\setminus{\cal C}$ (cf. (\ref{K2})).
\medskip

To prove (\ref{B.11}) we first apply the Poisson summation formula 
(see (\ref{Psf})) and obtain
$$
\sum\limits_{\bar x\in \Z^{d-1}}
e^{i \bar x\cdot\bar\varphi} 
{\bf R}^0(0,\ve[\bar r/\ve]-\ve \bar x,\dots)
=\ve^{-d+1}\sum\limits_{\bar n\in \Z^{d-1}}  
\tilde {\bf R}^0
(0,-\bar\varphi/\ve-2\pi \bar n/\ve,\dots)e^{i [\bar r/\ve]\cdot\bar\varphi},
$$
where $\tilde {\bf R}^0(0,\bar\varphi,\dots)$
stands for the Fourier transform 
$\tilde {\bf R}^0(0,\bar\varphi,\dots)=F_{\bar r\to\bar \varphi}
[{\bf R}^0(0,\bar r,\dots)]$, $\bar r,\bar \varphi\in\R^{d-1}$.
Conditions (\ref{tildeR}) and (\ref{d'2}) yield
\be\label{B.13}
\ve^{-d+1}\sum\limits_{\bar n\not=0}
\tilde {\bf R}^0(0,-\bar\varphi/\ve-2\pi \bar n/\ve,\dots)\to0
\quad \mbox{as }\,\ve\to0
\ee
uniformly in $\bar\varphi\in[-\pi,\pi]^{d-1}$.
Hence, by (\ref{B.10}) and (\ref{B.13}),
\beqn
{\cal I}_\varepsilon(\theta)&=&
(2\pi\ve)^{-d+1}\int\limits_{[-\pi,\pi]^{d-1}} 
e^{i\bar\varphi\cdot(\bar l+[\bar r/\ve])}
\hat{\cal G}^g_{\tau/\varepsilon^2}(\theta_1,\bar\theta-\bar\varphi)
\tilde {\bf R}^0(0,-\bar\varphi/\ve,\dots)\,d\bar\varphi+o(1)
\nonumber\\
&=&(2\pi)^{-d+1}
\int\limits_{[-\pi/\ve,\pi/\ve]^{d-1}} 
e^{-i\ve \bar\varphi\cdot(\bar l+[\bar r/\ve])}
\hat{\cal G}^g_{\tau/\varepsilon^2}(\theta_1,\bar\theta+\ve\bar\varphi)
\tilde {\bf R}^0(0,\bar\varphi,\dots)\,d\bar\varphi+o(1).\nonumber
\eeqn
Secondly, using the representation (\ref{A2}) for
$\hat{\cal G}^g_{t}(\theta)$ 
we rewrite ${\cal I}_\varepsilon(\theta)$ in the form
\beqn\nonumber
{\cal I}_\varepsilon(\theta)\!\!&=&\!\!(2\pi)^{-d+1}
\sum\limits_{\sigma,\pm}
 \int\limits_{[-\pi/\ve,\pi/\ve]^{d-1}}\!\!\!\!\!\!\!
e^{-i\ve\bar\varphi\cdot(\bar l+[\bar r/\ve])}
e^{\pm i\omega_\sigma(\theta_1,\bar\theta+\ve\bar\varphi)\tau/\varepsilon^2}
h^\mp_\sigma(\theta_1,\bar\theta+\ve\bar\varphi)
\tilde {\bf R}^0(0,\bar\varphi,\dots)
\,d\bar\varphi\nonumber\\
\!\!&&+o(1)=(2\pi)^{-d+1} 
\sum\limits_{\sigma,\pm}h^\mp_\sigma(\theta)
\int\limits_{\R^{d-1}} e^{-i\bar\varphi\cdot \bar r}
e^{\pm i\omega_\sigma(\theta_1,\bar\theta+\ve\bar\varphi)\tau/\varepsilon^2} 
\tilde {\bf R}^0(0,\bar\varphi,\dots) \,d\bar\varphi+o(1),\nonumber
\eeqn
by conditions (\ref{tildeR}) and (\ref{d'2}).
Finally, we use the Taylor sum representation for 
$\omega_\sigma(\theta_1,\bar\theta+\ve\bar\varphi)$
(cf (\ref{Tsr})) and obtain the bound (\ref{B.11}),
since 
\beqn\nonumber
(2\pi)^{-d+1}\int\limits_{\R^{d-1}}
e^{-i\bar\varphi
\cdot(\bar r\mp\nabla_{\bar\theta}\omega_\sigma(\theta)\tau/\ve)}
e^{\pm i(\tau/2)\bar\varphi\cdot (\nabla^2_{\bar\theta}\omega_\sigma(\theta))
\bar\varphi}
\tilde {\bf R}^0(0,\bar\varphi,\dots)\,d\bar\varphi\nonumber\\
=F^{-1}_{\bar\varphi\to
\bar r\mp\nabla_{\bar\theta}\omega_\sigma(\theta)\tau/\ve}
[\tilde {\bf R}^0(0,\bar\varphi,\dots)
e^{\pm i(\tau/2)\bar\varphi\cdot (\nabla^2_{\bar\theta}\omega_\sigma(\theta))
\bar\varphi}]
={\cal A}^{\mp}_{\ve,\sigma}(\tau,\bar r;\theta),\nonumber
\eeqn
where ${\cal A}^{\mp}_{\ve,\sigma}(\tau,\bar r;\theta)$ is defined in (\ref{B.12}).
The bound (\ref{B.11}) is proved.
\medskip

Now we prove that the r.h.s. of (\ref{B.7})
vanishes as $\varepsilon\to0$, using the Lebesgue dominated convergence
theorem. At first, we show that 
\beqn\nonumber
|A_\varepsilon(\bar\theta)B_\varepsilon(\bar\theta)|\le v(\bar\theta),\quad
\forall\varepsilon>0,\quad
\mbox{where }\,v(\bar\theta)\in L^1(\T^{d-1}).
\eeqn
Indeed, by conditions {\bf I4'} and (\ref{d'2}),
$$
\sup_{\ve>0}\sup_{\theta\in\T^d}
|{\cal A}^{\mp}_{\ve,\sigma}(\tau,\bar r;\theta)|\le C<\infty.
$$ 
Since $h^{\mp}_\sigma(\theta)
=g(\theta)\Pi_\sigma(\theta)(I\mp i C_\sigma(\theta))/2$,
 the bound (\ref{B.11}) yields
$$
|{\cal I}_\varepsilon(\theta)|\le \sum\limits_{\sigma=1}^s
(C_1\omega^{-1}_\sigma(\theta)+C_2\omega_\sigma(\theta))+C_3,\,\,\,\,
\theta\in\T^d\setminus{\cal C}.
$$
Write 
$v(\bar\theta):= 
\int\limits_{\T^1}(1+\Vert \hat V^{-1}(\theta)\Vert)\,d\theta_1$,
$\bar \theta\in\T^{d-1}$.
Therefore, by (\ref{B.8}) and (\ref{B.9}),
\beqn
|A_\varepsilon(\bar\theta)|
&\le& C\int\limits_{\T^1}|{\cal I}_\varepsilon(\theta)|\,d\theta_1
\le C_1(v(\bar\theta))^{1/2},\nonumber\\
|B_\varepsilon(\bar\theta)|
&\le& C\int\limits_{\T^1}|\hat{\cal G}^g_{\tau/\ve^2}(\theta)|\,d\theta_1
\le C_1 (v(\bar\theta))^{1/2},\nonumber
\eeqn
and, evidently, $|A_\varepsilon(\bar\theta)B_\varepsilon(\bar\theta)|
\le C\,v(\bar\theta)$, 
where $v(\bar\theta)\in L^1(\T^{d-1})$ by condition {\bf E6}.
Therefore, to prove that the integral in (\ref{B.7}) tends to zero 
it suffices to check that $B_\varepsilon(\bar\theta)$
vanishes as $\varepsilon\to0$ for a.a. $\bar\theta\in\T^{d-1}$.
By the representation (\ref{A2}),
\beqn
B_\varepsilon(\bar\theta)=\frac1{2\pi}\sum\limits_{\sigma=1}^s
\sum\limits_{\pm}
\int\limits_{\T^1} e^{i(p_1+[r_1/\varepsilon]-x_1+z_1)\varphi_1}
e^{\pm i\omega_\sigma(\varphi_1,\bar\theta)\tau/\varepsilon^2}
h^\mp_\sigma(\varphi_1,\bar\theta)^*\,d\varphi_1. 
\nonumber
\eeqn
For fixed $r_1\in\R^1$, $\tau\not=0$, and $p_1, x_1, z_1\in\Z^1$,
mes$\{\varphi_1\in\T^1:\partial_1\omega_\sigma(\varphi_1,\bar\theta)=0\}=0$
for a.a. fixed $\bar\theta\in\T^{d-1}$.
Hence, $B_\varepsilon(\bar\theta)$ vanishes as $\ve\to0$.
It can be proved similarly to the Lebesgue--Riemann theorem.
\bo

\setcounter{section}{3}
\setcounter{equation}{0}
\section*{\large\bf  Appendix C: Proof of Lemma 6.6 in the general case}
We prove Lemma \ref{Qt1} under the weaker condition (\ref{2.3'})--(\ref{Varm})
on $\hat {\bf R}_0(r,\theta)$ than (\ref{tildeR}).
\medskip

{\it Step (i)} of the proof is not changed and we derive formulas 
(\ref{4.9})--(\ref{I}). 

Next step is to apply the Poisson summation formula (\ref{Psf}).
However, in general, we can not apply (\ref{Psf})
to $\hat {\bf R}_0(r,\theta)$. Therefore, introduce
\be\label{5.19}
\hat {\bf R}_{r,\ve}(y,\theta)=\hat {\bf R}_0(y,\theta)e^{-\ve^\beta(y-r)^2},
\quad y\in\R^d,\quad \theta\in\T^d,
\ee
with  $\beta\in(d+2,2(\delta-1-d)/d)$ and $\delta$ from (\ref{Varm}).
The asymptotics of (\ref{I}) is not changed if we 
replace the function $\hat {\bf R}_0(y,\theta)$
 by $\hat {\bf R}_{r,\ve}(y,\theta)$ in (\ref{I}).
Indeed, consider the following series:
\beqn
&&\sum\limits_{x\in \Z^d}\int\limits_{\T^{d}} 
e^{i(l+x)\cdot\varphi}
\hat{\cal G}^g_{\tau/\varepsilon^2}(\theta-\varphi)\,d\varphi
\Big(\hat {\bf R}_0(\varepsilon[r/\varepsilon]-\varepsilon x,\theta)
-\hat {\bf R}_{r,\ve}(\varepsilon[r/\varepsilon]-\varepsilon x,\theta)\Big)
\nonumber\\
&=&\sum\limits_{|x|\le c\tau/\ve^2}\dots+
\sum\limits_{|x|\ge c\tau/\ve^2}\dots=
A_\ve(\theta)+B_\ve(\theta). \nonumber
\eeqn
and prove that the series $A_\ve(\theta)$ and $B_\ve(\theta)$ vanish 
as $\ve\to0$. 
For the first series $A_\ve(\theta)$, we apply Lemma \ref{l5.3} (i),
definition (\ref{5.19}) and property {\bf I1} and obtain
\beqn
A_\ve(\theta)&\le& 
C \sum\limits_{|x|\le c\tau/\ve^2}\Big|\int\limits_{\T^{d}} 
e^{i(l+x)\cdot\varphi}
\hat{\cal G}^g_{\tau/\varepsilon^2}(\theta-\varphi)\,d\varphi\Big|
\Big|\hat {\bf R}_0(\varepsilon[r/\varepsilon]-\varepsilon x,\theta)
-\hat {\bf R}_{r,\ve}(\varepsilon[r/\varepsilon]-\varepsilon x,\theta)\Big|
\nonumber\\
&\le& C_1
\sum\limits_{|x|\le c\tau/\ve^2} |{\cal G}^g_{\tau/\varepsilon^2}(l+x)|
\ve^{\beta}(\ve|x|+\ve)^2
\le C\ve^{d}\ve^{\beta+2} (\tau/\ve^2)^{2+d}\to0,\quad \ve\to0,\nonumber
\eeqn
uniformly in $\theta$,  since $\beta>2+d$. 
For the second series $B_\ve(\theta)$, Lemma \ref{l5.3} (ii) yields
\beqn
B_\ve(\theta)&\le& 
C \sum\limits_{|x|\ge c\tau/\ve^2}\Big|\int\limits_{\T^{d}} 
e^{i(l+x)\cdot\varphi}
\hat{\cal G}^g_{\tau/\varepsilon^2}(\theta-\varphi)\,d\varphi\Big|
|\hat {\bf R}_0(\varepsilon[r/\varepsilon]-\varepsilon x,\theta)|
\nonumber\\
&\le& C_1 \sum\limits_{|x|\ge c\tau/\ve^2}
\left|{\cal G}^g_{\tau/\varepsilon^2}(l+x)\right|\to0,\quad\ve\to0.\nonumber
\eeqn
{\it Step (ii)}: Now we apply formula (\ref{Psf}) 
to $\sum\limits_{x\in\Z^d} e^{ix\cdot \varphi}
\hat {\bf R}_{r,\ve}(\ve[r/\ve]-\ve x,\theta)$.
By $\tilde {\bf R}_{r,\ve}(s,\theta)$, $s\in\R^d$, $\theta\in\T^d$,
we denote the Fourier transform of $\hat {\bf R}_{r,\ve}(y,\theta)$
w.r.t. $y$ (see (\ref{Ftr})) and obtain
\beqn\label{5.20D}
I_\varepsilon(\theta)=(2\pi\ve)^{-d}
\int\limits_{\T^{d}} e^{i(l+[r/\ve])\cdot\varphi}
\hat{\cal G}^g_{\tau/\varepsilon^2}(\theta-\varphi)
\Big(\sum\limits_{n\in \Z^d}  
\tilde {\bf R}_{r,\ve}(-\varphi/\ve-2\pi n/\ve,\theta)\Big)
\,d\varphi+o(1)
\eeqn
(cf (\ref{5.20})). We check that the contribution 
of the series with $n\not=0$ in (\ref{5.20D}) vanishes,
\be\label{5.21D}
\ve^{-d}\sum\limits_{n\in \Z^d,n\not=0}
\tilde {\bf R}_{r,\ve}(-\varphi/\ve-2\pi n/\ve,\theta)\to0
\quad \mbox{as }\,\ve\to0,
\ee
uniformly in $\theta\in\T^d$, $\varphi\in[-\pi,\pi]^d$
(cf (\ref{5.21})). At first, we derive the formula 
for $\tilde {\bf R}_{r,\ve}(\xi,\theta)$:
\be\label{5.22D}
\tilde {\bf R}_{r,\ve}(\xi,\theta)= e^{ir\cdot\xi}
(4\pi k)^{-d/2}
\int\limits_{\R^d}e^{-is\cdot r}e^{-|s-\xi|^2/(4k)}\mu(\theta,ds),\quad
\xi\in\R^d,\quad \theta\in\T^d,
\ee
where $k:=\ve^\beta$ and $\mu(\theta,ds)$ from (\ref{2.3'}). 
Indeed, by definition (\ref{5.19}),
$$
\tilde {\bf R}_{r,\ve}(\xi,\theta)=\int\limits_{\R^d}
e^{i\xi\cdot y-k(y-r)^2}\hat {\bf R}_0(y,\theta)\,dy
=e^{ir\cdot\xi-|\xi|^2/(4k)}
\int\limits_{\R^d} e^{-k(y-r-i\xi/(2k))^2}\hat {\bf R}_0(y,\theta)\,dy.
$$
On the other hand, by definition (\ref{2.3'}), for any $\al\in\R^d$,
\beqn
\int\limits_{\R^d} e^{-k(y-\al)^2}\hat {\bf R}_0(y,\theta)\,dy
&=&
(2\pi)^{-d}\int\limits_{\R^d}\int\limits_{\R^d} e^{-iy\cdot s}e^{-k(y-\al)^2}
\,\mu(\theta,ds)\,dy\nonumber\\
&=&(2\pi)^{-d}\int\limits_{\R^d}
 e^{-is\cdot \al}\left(\frac{\pi}{k}\right)^{d/2}
e^{-|s|^2/(4k)}\,\mu(\theta,ds).\nonumber
\eeqn
Hence,
$$
\tilde {\bf R}_{r,\ve}(\xi,\theta)=e^{ir\cdot\xi-|\xi|^2/(4k)}
(4\pi k)^{-d/2}\int\limits_{\R^d} e^{-is\cdot (r+i\xi/(2k))}
e^{-|s|^2/(4k)}\,\mu(\theta,ds).
$$
This implies formula (\ref{5.22D}).

Now we prove (\ref{5.21D}). Let us set $\xi_{n,\ve}=-\varphi/\ve-2\pi n/\ve$, 
where $|\varphi_i|\le\pi$, $i=1,\dots,d$,
apply (\ref{5.22D}) and devide the integration into two:
$|s|\le\pi/(2\ve)$ and $|s|\ge\pi/(2\ve)$:
\beqn\nonumber
\ve^{-d}\Big|\sum\limits_{n\not=0}
\tilde {\bf R}_{r,\ve}(\xi_{n,\ve},\theta)\Big|&\le&
\ve^{-d}\sum\limits_{n\not=0}(4\pi k)^{-d/2}
\Big|\int\limits_{\R^d}e^{-is\cdot r}e^{-|s-\xi_{n,\ve}|^2/(4k)}\mu(\theta,ds)
\Big|\nonumber\\
&=&I_1+I_2,
\eeqn
where
\beqn
I_1&=&(\ve\sqrt{4\pi k})^{-d}\sum\limits_{n\not=0}
\Big|\int\limits_{|s|\le\pi/(2\ve)}
e^{-is\cdot r}e^{-|s-\xi_{n,\ve}|^2/(4k)}\mu(\theta,ds)\Big|,\nonumber\\
I_2&=&(\ve\sqrt{4\pi k})^{-d}\sum\limits_{n\not=0}
\Big|\int\limits_{|s|\ge\pi/(2\ve)}
e^{-is\cdot r}e^{-|s-\xi_{n,\ve}|^2/(4k)}\mu(\theta,ds)\Big|. \label{C.6'}
\eeqn
Let $|s|\le\pi/(2\ve)$.  Put $\xi_{n_i,\ve}=-\varphi_i/\ve-2\pi n_i/\ve$.
Then  
$|s_i-\xi_{n_i,\ve}|\ge \left|2\pi|n_i|/\ve-|s_i+\varphi_i/\ve|\right|
\ge(\pi/\ve)(2|n_i|-3/2)$ if $n_i\not=0$.
Therefore, there exist positive constants $a,b$ such that
$$
\sum\limits_{n_i\in\Z^1:n_i\not=0}e^{-(s_i-\xi_{n_i,\ve})^2/(4k)}
\le \sum\limits_{n_i\not=0}e^{-\pi^2(2|n_i|-3/2)^2/(4k\ve^2)}
\le a\,e^{-b/(k\ve^2)}\,\sqrt{k}\ve,\quad i=1,\dots,d.
$$
Hence, there exist positive constants $a_s$ and $b_s$ such that
$$
I_1\sim \sum\limits_{s=0}^{d-1}
a_s\, e^{-b_s/(k\ve^2)}(\sqrt{k}\ve)^{-s} \to0,\quad \ve\to0,
$$
since $k=\ve^\beta$ with $\beta>0$.
Let $|s|\ge\pi/(2\ve)$. Then  
\beqn\label{C.7'}
\sum\limits_{n\in\Z^d}e^{-(s-\xi_{n,\ve})^2/(4k)}
=\sum\limits_{n\in\Z^d}e^{-(\pi^2/(k\ve^2))(n+\varphi/(2\pi)+s\ve/(2\pi))^2}
\le C_1 (\ve \sqrt k)^d+C_2. 
\eeqn
Using (\ref{C.6'}), (\ref{C.7'}), and condition (\ref{Varm}), we obtain 
$$
I_2\sim (C_1+C_2 (\ve\sqrt k)^{-d})
\sup_{\theta\in\T^d} 
\int\limits_{|s|\ge\pi/(2\ve)}|\mu(\theta,ds)|\le
C_3\ve^{\delta-1-(\beta/2+1)d}\to0,\quad \ve\to0,
$$
because $k=\ve^\beta$ and $\delta>1+(\beta/2+1)d$.
This completes the proof of the bound (\ref{5.21D}).
\medskip\\
{\it Step (iii)}: We apply (\ref{5.20D}), (\ref{5.21D}),
change variables $\varphi\to -\ve\varphi$ 
and use formula (\ref{A2}):
\beqn
I_\varepsilon(\theta)\!\!&=&\!\!
(2\pi\ve)^{-d}\int\limits_{[-\pi,\pi]^{d}} 
e^{i\varphi\cdot(l+[r/\ve])}
\hat{\cal G}^g_{\tau/\varepsilon^2}(\theta-\varphi)
\tilde {\bf R}_{r,\ve}(-\varphi/\ve,\theta)\,d\varphi+o(1)\nonumber\\
\!\!&=&\!\!(2\pi)^{-d}
\int\limits_{[-\pi/\ve,\pi/\ve]^d} e^{-i\ve \varphi\cdot(l+[r/\ve])}
\hat{\cal G}^g_{\tau/\varepsilon^2}(\theta+\ve\varphi)
\tilde {\bf R}_{r,\ve}(\varphi,\theta)\,d\varphi+o(1)\nonumber\\
\!\!&=&\!\!
(2\pi)^{-d}\sum\limits_{\sigma,\pm}
 \int\limits_{[-\pi/\ve,\pi/\ve]^d}\!\!\!\!\!\!\!\!
e^{-i\ve\varphi\cdot(l+[r/\ve])}
e^{\pm i\omega_\sigma(\theta+\ve\varphi)\tau/\varepsilon^2}
h^\mp_\sigma(\theta+\ve\varphi)\tilde {\bf R}_{r,\ve}(\varphi,\theta)
\,d\varphi+o(1). \,\,\,\,\label{C.6}
\eeqn  
The asymptotics of (\ref{C.6}) is not changed if we replace 
 $e^{-i\ve\varphi\cdot(l+[r/\ve])}$
by $e^{-i\varphi\cdot r}$ in the integrand. Indeed,
we check that the integral 
$$
D_\ve(\theta)=\int\limits_{[-\pi/\ve,\pi/\ve]^d}
(e^{-i\ve\varphi\cdot(l+[r/\ve])}-e^{-i\varphi\cdot r})
e^{\pm i\omega_\sigma(\theta+\ve\varphi)\tau/\varepsilon^2}
h^\mp_\sigma(\theta+\ve\varphi)\tilde {\bf R}_{r,\ve}(\varphi,\theta)
\,d\varphi
$$
vanishes as $\ve\to0$. We apply formula (\ref{5.22D}) 
and change the order of the integration $\mu(\theta,ds)\,d\varphi\to
d\varphi\,\mu(\theta,ds)$:
\beqn\label{C.7}
D_\varepsilon(\theta)\!&=&\!
 \int\limits_{[-\pi/\ve,\pi/\ve]^d}\!\!\!
(e^{-i\ve\varphi\cdot(l+[r/\ve])}-e^{-i\varphi\cdot r})
e^{\pm i\omega_\sigma(\theta+\ve\varphi)\tau/\varepsilon^2}
h^\mp_\sigma(\theta+\ve\varphi)\times\nonumber\\
\!&&\times
\Big( \frac{e^{ir\cdot\varphi}}{(4\pi k)^{d/2}}
\int\limits_{\R^d}e^{-is\cdot r}e^{-|s-\varphi|^2/(4k)}\mu(\theta,ds)
\Big)\,d\varphi=\int\limits_{\R^d}e^{-is\cdot r}
C_\ve(s,\theta)\,\mu(\theta,ds),\,\,
\eeqn
where $C_\ve(s,\theta)$ stands for the inner integral in $d\varphi$:
$$
C_\ve(s,\theta)=\frac1{(4\pi k)^{d/2}}
 \int\limits_{[-\pi/\ve,\pi/\ve]^d}
e^{-|s-\varphi|^2/(4k)}(e^{i\varphi\cdot(r-\ve[r/\ve]-\ve l)}-1)
e^{\pm i\omega_\sigma(\theta+\ve\varphi)\tau/\varepsilon^2}
h^\mp_\sigma(\theta+\ve\varphi)\,d\varphi.
$$
Note that
$$
|C_\ve(s,\theta)|\le \ve
\sup_{\theta\in\T^d}|h^\mp_\sigma(\theta)|\frac{1}{(4\pi k)^{d/2}}
 \int\limits_{[-\pi/\ve,\pi/\ve]^d}
e^{-|s-\varphi|^2/(4k)}|\varphi|\,d\varphi,
$$ 
where 
$$
\frac{1}{(4\pi k)^{d/2}} \int\limits_{[-\pi/\ve,\pi/\ve]^d}\!\!\!\!\!\!
e^{-|s-\varphi|^2/(4k)}|\varphi|\,d\varphi
\le \frac{1}{(4\pi k)^{d/2}}
\int\limits_{\R^d}e^{-|\varphi|^2/(4k)}(|s|+|\varphi|)\,d\varphi
\le C_1|s|+C_2\sqrt k.
$$
Hence, by (\ref{C.7}) and (\ref{Varm}), 
$\sup_{\theta\in\T^d}|D_\ve(\theta)|\le C\ve$.
Similarly, the asymptotics of (\ref{C.6}) is not changed if we replace 
 $h^\mp_\sigma(\theta+\ve\varphi)$ by $h^\mp_\sigma(\theta)$
in the integrand. Therefore,
\beqn\label{C.8}
I_\ve(\theta)=
\sum\limits_{\sigma,\pm} h^\mp_\sigma(\theta) 
N^{\mp}_{\ve,\sigma} (\tau,r;\theta)+o(1),\quad \ve\to0,
\eeqn
where $N^\mp_{\ve,\sigma}(\tau,r;\theta)$ stands for 
the matrix-valued function,
\beqn\label{C11}
N^{\mp}_{\ve,\sigma} (\tau,r;\theta)=
(2\pi)^{-d} \int\limits_{[-\pi/\ve,\pi/\ve]^d}
e^{-i\varphi\cdot r}
e^{\pm i\omega_\sigma(\theta+\ve\varphi)\tau/\varepsilon^2}
\tilde {\bf R}_{r,\ve}(\varphi,\theta)\,d\varphi.
\eeqn
{\it Step (iv)}: 
Now we apply the Taylor sum representation (cf (\ref{Tsr}))
and replace $\omega_\sigma(\theta+\ve\varphi)$ by
$\omega_\sigma(\theta)+\ve\nabla\omega_\sigma(\theta)\cdot \varphi
+(\ve^2/2)\varphi\cdot H_\sigma(\theta)\varphi$.
To do it we check that 
$$
L_\ve(\theta):=\int\limits_{[-\pi/\ve,\pi/\ve]^d}\!\!\!\!e^{-i\varphi\cdot r}
\Big[e^{\pm i\omega_\sigma(\theta+\ve\varphi)\tau/\varepsilon^2}
-e^{\pm i(\omega_\sigma(\theta)+\ve\nabla\omega_\sigma(\theta)\cdot \varphi
+\frac{\ve^2}{2}\varphi\cdot H_\sigma(\theta)\varphi)\tau/\varepsilon^2}
\Big]\tilde {\bf R}_{r,\ve}(\varphi,\theta)\,d\varphi\to0
$$
as $\ve\to0$.
As in {\it Step (iii)}, we use formula (\ref{5.22D})  and
change the order of the integration $\mu(\theta,ds)\,d\varphi\to
d\varphi\,\mu(\theta,ds)$. Therefore,
$$
L_\ve(\theta)=\int\limits_{\R^d}e^{-is\cdot r}
D_\ve(s,\theta)\,\mu(\theta,ds),
$$
where $D_\ve(s,\theta)$ stands for the inner integral in $d\varphi$:
$$
D_\ve(s,\theta)=\frac1{(4\pi k)^{d/2}}\!
 \int\limits_{[-\pi/\ve,\pi/\ve]^d}\!\!\!\!\!\!\!\!\!
e^{-|s-\varphi|^2/(4k)}
\Big[e^{\pm i\omega_\sigma(\theta+\ve\varphi)\tau/\varepsilon^2}
-e^{\pm i(\omega_\sigma(\theta)+\ve\nabla\omega_\sigma(\theta)\cdot \varphi
+\frac{\ve^2}{2}\varphi\cdot H_\sigma(\theta)\varphi)\tau/\varepsilon^2}
\Big]d\varphi.
$$
Moreover, by condition (\ref{Varm}),
\beqn\nonumber
\sup_{\theta\in\T^d}
\Big|\int\limits_{|s|\ge N}e^{-is\cdot r}
D_\ve(s,\theta)\,\mu(\theta,ds)\Big|\!&\le&\!
\sup_{\theta\in\T^d}\int\limits_{|s|\ge N}
\Big(\frac2{(4\pi k)^{d/2}} \int\limits_{\R^d}
e^{-|s-\varphi|^2/(4k)}d\varphi\Big)|\mu|(\theta,ds)\nonumber\\
\!&\le&\! C(1+N)^{-\delta+1}\to0 \quad \mbox{as }\,N\to\infty,\nonumber
\eeqn
uniformly in $\ve$. Hence, it suffices to prove that for any fixed $N\in\N$,
\beqn\label{C9}
L_\ve^N(\theta):=\int\limits_{|s|\le N}e^{-is\cdot r}
D_\ve(s,\theta)\,\mu(\theta,ds)\to 0 \quad \mbox{as }\,\ve\to0.
\eeqn
To check (\ref{C9}) we estimate the inner integral $D_\ve(s,\theta)$:
\beqn\nonumber
|D_\ve(s,\theta)|&\le& C\ve\tau \frac1{(4\pi k)^{d/2}}
 \int\limits_{[-\pi/\ve,\pi/\ve]^d}
e^{-|s-\varphi|^2/(4k)}|\varphi|^3\,d\varphi\nonumber\\
&\le&
C_1\ve\frac1{(4\pi k)^{d/2}} \Big[
\int\limits_{\R^d} e^{-|\varphi|^2/(4k)}|\varphi|^3\,d\varphi
+|s|^3\int\limits_{\R^d} e^{-|\varphi|^2/(4k)}\,d\varphi\Big]
\le C_2\ve[k^{3/2}+|s|^3].\nonumber
\eeqn
Hence, for fixed $N$,
$$ 
|L_\ve^N(\theta)|\le C\ve
\int\limits_{|s|\le N}(\ve^{3\beta/2}+|s|^3)|\mu|(\theta,ds)
\le C_2(N)\ve,
$$
and (\ref{C9}) follows.
Therefore, $\sup_{\theta\in\T^d}|L_\ve(\theta)|\to0$ as $\ve\to0$, and 
\beqn\label{C.9}
N^{\mp}_{\ve,\sigma}(\tau,r;\theta)=
e^{\pm i\omega_\sigma(\theta)\tau/\varepsilon^2}
B^{\mp}_{\ve,\sigma}(\tau,r;\theta)+o(1),\quad \ve\to0,
\eeqn
where, by definition, 
\beqn\label{C.10}
B^{\mp}_{\ve,\sigma}(\tau,r;\theta)\!&=&\!
(2\pi)^{-d} \int\limits_{\R^d}
e^{-i\varphi\cdot (r\mp\nabla\omega_\sigma(\theta)\tau/\ve)}
e^{\pm i(\tau/2)\varphi\cdot H_\sigma(\theta)\varphi}
\tilde {\bf R}_{r,\ve}(\varphi,\theta)\,d\varphi\nonumber\\
\!&=&\!
\int\limits_{\R^d} 
\hat {\bf R}_{r,\ve}(r\mp\nabla\omega_\sigma(\theta)\tau/\ve-x,\theta)
K^{\mp}_{\sigma}(\tau,x,\theta)\,dx,
\eeqn
with the function $K^{\mp}_{\sigma}(\tau,x,\theta)$ ($\tau>0$, $x\in\R^d$,
$\theta\in \T^d\setminus{\cal C}$) from (\ref{K2}).
\medskip\\
{\bf Remark}. The integral $N^{\mp}_{\ve,\sigma}(\tau,r;\theta)$ 
can be rewritten in the another form. Namely,
\beqn\label{C.11}
N^{\mp}_{\ve,\sigma}(\tau,r;\theta)=
e^{\pm i\omega_\sigma(\theta)\tau/\varepsilon^2}
C^{\mp}_{\ve,\sigma}(\tau,r;\theta)+o(1),\quad \ve\to0,
\eeqn
where, by definition,
\beqn\label{C.12}
C^{\mp}_{\ve,\sigma}(\tau,r;\theta):=
(2\pi)^{-d}\int\limits_{\R^d}
e^{-is\cdot (r\mp\nabla\omega_\sigma(\theta)\tau/\ve)}
e^{\pm i(\tau/2) s\cdot H_\sigma(\theta)s}
\mu(\theta,ds).
\eeqn
{\bf Proof}. We first substitute formula (\ref{5.22D})
in (\ref{C11})  and
change the order of the integration $\mu(\theta,ds)\,d\varphi\to
d\varphi\,\mu(\theta,ds)$. Therefore,
\beqn\label{C.13}
N^{\mp}_{\ve,\sigma}(\tau,r;\theta)=(2\pi)^{-d}
\int\limits_{\R^d}\!e^{-is\cdot r}
\Big(\frac1{(4\pi k)^{d/2}} \int\limits_{[-\pi/\ve,\pi/\ve]^d}
\!\!\!\!\!\!e^{-|s-\varphi|^2/(4k)}
e^{\pm i\omega_\sigma(\theta+\ve\varphi)\tau/\varepsilon^2}
d\varphi\Big)\mu(\theta,ds).\,\,\,
\eeqn
Secondly, we replace $\omega_\sigma(\theta+\ve \varphi)$ 
by $\omega_\sigma(\theta+\ve s)$ in the inner integral.
To do this we devide the integration into two: $|s-\varphi|\ge\ve^{\gamma}$
and $|s-\varphi|\le \ve^{\gamma}$ with a $\gamma$,
$\gamma \in(1,\beta/2)$. For $|s-\varphi|\ge\ve^{\gamma}$,
\beqn\label{C.14}
\frac1{(4\pi k)^{d/2}}
\int\limits_{[-\pi/\ve,\pi/\ve]^d,|s-\varphi|>\ve^\gamma}
\!\!\!\!\!e^{-|s-\varphi|^2/(4k)}\,d\varphi
\le\frac{e^{-\ve^{2\gamma}/(4k)}}{(4\pi k)^{d/2}}\left(\frac\pi\ve\right)^{d}
\le C\,  \frac{e^{-\ve^{(2\gamma-\beta)}/4}}{\ve^{\beta d/2+d}}\to0
\eeqn 
as $\ve\to+0$, since $k=\ve^\beta$ and $\gamma<\beta/2$.
Note that
$$
\frac1{(4\pi k)^{d/2}}
 \int\limits_{[-\pi/\ve,\pi/\ve]^d}
e^{-|s-\varphi|^2/(4k)}\,d\varphi=1+o(1),\quad \ve\to0,
$$
where $k=\ve^\beta$. Therefore, 
for $|s-\varphi|\le\ve^{\gamma}$, we estimate the difference
\beqn\label{C.15}
\frac1{(4\pi k)^{d/2}}
\Big|\int\limits_{[-\pi/\ve,\pi/\ve]^d,|s-\varphi|<\ve^\gamma}
e^{-|s-\varphi|^2/(4k)}
(e^{\pm i\omega_\sigma(\theta+\ve\varphi)\tau/\varepsilon^2}
-e^{\pm i\omega_\sigma(\theta+\ve s)\tau/\varepsilon^2})
\,d\varphi\Big|\nonumber\\
\le \frac{C}{(4\pi k)^{d/2}}
\int\limits_{[-\pi/\ve,\pi/\ve]^d}
e^{-|s-\varphi|^2/(4k)}\ve^{\gamma+1}\tau/\ve^2\,d\varphi
\le C\ve^{\gamma-1}(1+o(1)),
\eeqn
where the last expression vanishes as $\ve\to0$ since
 $\gamma>1$. By (\ref{C.14}) and (\ref{C.15}),
the inner integral in (\ref{C.13}) is 
\beqn\nonumber
&&\frac1{(4\pi k)^{d/2}}
\int\limits_{[-\pi/\ve,\pi/\ve]^d,|s-\varphi|<\ve^\gamma}
e^{-|s-\varphi|^2/(4k)}
e^{\pm i\omega_\sigma(\theta+\ve \varphi)\tau/\varepsilon^2}
\,d\varphi+o(1)\nonumber\\
&=& e^{\pm i\omega_\sigma(\theta+\ve s)\tau/\varepsilon^2}
\frac1{(4\pi k)^{d/2}}\int\limits_{[-\pi/\ve,\pi/\ve]^d,|s-\varphi|<\ve^\gamma}
e^{-|s-\varphi|^2/(4k)}\,d\varphi+o(1)\nonumber\\
&=& e^{\pm i\omega_\sigma(\theta+\ve s)\tau/\varepsilon^2}
\frac1{(4\pi k)^{d/2}}\int\limits_{[-\pi/\ve,\pi/\ve]^d}
e^{-|s-\varphi|^2/(4k)}\,d\varphi+o(1)\nonumber\\
&=&e^{\pm i\omega_\sigma(\theta+\ve s)\tau/\varepsilon^2}(1+o(1)),
\quad \ve\to0,
\nonumber
\eeqn
uniformly in $\theta\in\T^d$.
We substitute the last expression in (\ref{C.13}) and get
$$
N^{\mp}_{\ve,\sigma}(\tau,r;\theta)=(2\pi)^{-d} 
\int\limits_{\R^d}e^{-is\cdot r}
e^{\pm i\omega_\sigma(\theta+\ve s)\tau/\varepsilon^2}
\,\mu(\theta,ds)+o(1),\quad \ve\to0,
$$
uniformly in $\theta\in\T^d$.
Finally, we apply the Taylor sum representation (\ref{Tsr}) 
to $\omega_\sigma(\theta+\ve s)$ and obtain (\ref{C.11})--(\ref{C.12}).
The integral in (\ref{C.12}) can be taken over $|s|\le N$
with some $N\in\N$, by condition (\ref{Varm}).
\medskip\\
{\it Step (v)}: We substitute $I_\ve(\theta)$ of the form
(\ref{C.8}) with $N^{\mp}_{\ve,\sigma}(\tau,r;\theta)$
from (\ref{C.9}) (or (\ref{C.11})) in the r.h.s. of (\ref{4.9'}).
Applying the decomposition (\ref{A2}) 
to $\hat{\cal G}^g_{\tau/\ve^2}(\theta)^*$ we obtain (\ref{6.22'}),
with $B^{\mp}_{\ve,\sigma}$ (or $C^{\mp}_{\ve,\sigma}$, respectively)
instead of $A^{\mp}_{\ve,\sigma}$.
It remains to study the behaviour (as $\ve\to0$) of integrals
of the form (\ref{4.9''}) with $B^{-}_{\ve,\sigma}$
(or $C^{-}_{\ve,\sigma}$, resp.)
instead of $A^{-}_{\ve,\sigma}$.

Let 
$\omega_\sigma(\theta)\pm\omega_{\sigma'}(\theta)\not\equiv{\rm const}_\pm$.
In this case, the oscillatory integrals 
\be\label{C.20}
I_{\sigma\sigma'}^\pm(\ve)\equiv (2\pi)^{-d}
\frac12\int\limits_{\T^{d}} e^{-i(l-p)\cdot\theta}
e^{ i(\omega_\sigma(\theta)\pm \omega_{\sigma'}(\theta))\tau/\varepsilon^2}
h^-_\sigma(\theta)C^{-}_{\ve,\sigma}(\tau,r;\theta)
h^\mp_{\sigma'}(\theta)^*\,d\theta
\ee
vanish as $\ve\to0$, since $\sup\limits_{\ve>0,\theta\in\T^d}
|C^{-}_{\ve,\sigma}(\tau,r;\theta)|\!\le\! C<\infty$,
and hence, 
$h^-_\sigma(\theta)C^{-}_{\ve,\sigma}(\tau,r;\theta)
h^\mp_{\sigma'}(\theta)^*\in L^1(\T^d)$.
The identities
$\omega_\sigma(\theta)\pm\omega_{\sigma'}(\theta)\equiv{\rm const}_\pm$ 
 in the exponent of (\ref{C.20})
with ${\rm const}_\pm\ne 0$ are impossible by condition {\bf E5}.
The identity $\omega_\sigma(\theta)+\omega_{\sigma'}(\theta)\equiv0$
implies
$\omega_\sigma(\theta)\equiv\omega_{\sigma'}(\theta)\equiv 0$ which
is impossible by {\bf E4}. 
 Therefore, if $\sigma\not=\sigma'$,  
$I_{\sigma\sigma'}^\pm(\varepsilon)=o(1)$ as $\varepsilon\to0$.
If $\sigma=\sigma'$, $I_{\sigma\sigma}^+(\varepsilon)=o(1)$,
and only $I_{\sigma\sigma}^-(\varepsilon)$ contrubutes in the limit.
Hence, (cf (\ref{6.28}))
\be\label{C.16}
S^+_{\varepsilon,\tau/\varepsilon^2}=(2\pi)^{-d}
\frac12 \int\limits_{\T^{d}} e^{-i(l-p)\cdot\theta}
\sum\limits_{\sigma,\pm}
h^\mp_{\sigma}(\theta)B^{\mp}_{\ve,\sigma}(\tau,r;\theta)
h^\pm_{\sigma}(\theta)^*\,d\theta+o(1),\quad \ve\to0.
\ee
{\it Step (vi)}: To obtain (\ref{6.28}), 
it remains to replace $B^{\mp}_{\ve,\sigma}$ by $A^{\mp}_{\ve,\sigma}$
in (\ref{C.16}) 
(or $\hat {\bf R}_{r,\ve}$ by $\hat {\bf R}_0$ in (\ref{C.10})).
We check that the difference
$B^{-}_{\ve,\sigma}-A^{-}_{\ve,\sigma}$ vanishes as $\ve\to0$
(for the difference $B^{+}_{\ve,\sigma}-A^{+}_{\ve,\sigma}$
 the proof is similar).
Indeed, by (\ref{Apm}), (\ref{5.19}), (\ref{C.10}), and (\ref{K2}),
$$
A^{-}_{\ve,\sigma}-B^{-}_{\ve,\sigma}=
C(\theta)\int\limits_{\R^d} 
\hat {\bf R}_{0}(r-\nabla\omega_\sigma(\theta)\tau/\ve-x,\theta)
\Big(1-e^{-\ve^\beta(\nabla\omega_\sigma(\theta)\tau/\ve+x)^2}\Big)
 e^{-i/(2\tau)x\cdot H^{-1}_\sigma(\theta) x}dx,
$$
with $C(\theta)=e^{i\pi s/4}(2\pi\tau)^{-d/2}|\det H_\sigma(\theta)|^{-1/2}$.
Write
$c=2|\tau|\max\limits_{\theta\in\T^d,\sigma}|\nabla\omega_\sigma(\theta)|$.
We devide the integration into two:
$|x+\nabla\omega_\sigma(\theta)\tau/\ve|\le c/\ve$
and $|x+\nabla\omega_\sigma(\theta)\tau/\ve|\ge c/\ve$.
For $|x+\nabla\omega_\sigma(\theta)\tau/\ve|\le c/\ve$,
\beqn
&&\Big|\int\limits_{|x+\nabla\omega_\sigma(\theta)\tau/\ve|\le c/\ve} 
\hat {\bf R}_{0}(r-\nabla\omega_\sigma(\theta)\tau/\ve-x,\theta)
\Big(1-e^{-\ve^\beta(\nabla\omega_\sigma(\theta)\tau/\ve+x)^2}\Big)
 e^{-i/(2\tau)x\cdot H^{-1}_\sigma(\theta) x}\,dx\Big|\nonumber\\
&\le& C \int\limits_{|x+\nabla\omega_\sigma(\theta)\tau/\ve|\le c/\ve} 
\ve^\beta|x+\nabla\omega_\sigma(\theta)\tau/\ve|^2\,dx
\le C_1\ve^{\beta-2-d}\to0,\,\,\,\,\ve\to0,\nonumber
\eeqn
since $\beta>d+2$.
For $|x+\nabla\omega_\sigma(\theta)\tau/\ve|\ge c/\ve$,
we apply the integration by parts in every variable $x_i$, $i=1,\dots,d$,
For simplicity, let $d=1$
and $h$ stand for $-1/(2\tau)H^{-1}_{\sigma}(\theta)$. Therefore, 
\beqn
&&\int\limits_{c/\ve-\omega'_\sigma(\theta)\tau/\ve}^{+\infty} 
\hat {\bf R}_{0}(r-\omega'_\sigma(\theta)\tau/\ve-x,\theta)
\Big(1-e^{-\ve^\beta(\omega'_\sigma(\theta)\tau/\ve+x)^2}\Big)
e^{ihx^2}\,dx\nonumber\\
&=&\frac1{2ih}\Big[
-\hat {\bf R}_{0}(r-c/\ve,\theta)
\Big(1-e^{-\ve^\beta(c/\ve)^2}\Big)
\frac{e^{ihx^2}}{x}\Big|_{x=(c-\omega'_\sigma(\theta)\tau)/\ve}
\nonumber\\
&&+
\int\limits_{(c-\omega'_\sigma(\theta)\tau)/\ve}^{+\infty} 
\partial _r\hat {\bf R}_{0}(r-\omega'_\sigma(\theta)\tau/\ve-x,\theta)
\Big(1-e^{-\ve^\beta(\omega'_\sigma(\theta)\tau/\ve+x)^2}\Big)
\frac{e^{ihx^2}}{x}\,dx\nonumber\\
&&-\ve^\beta 2
\int\limits_{(c-\omega'_\sigma(\theta)\tau)/\ve}^{+\infty} 
\hat {\bf R}_{0}(r-\omega'_\sigma(\theta)\tau/\ve\!-\!x,\theta)
(\omega'_\sigma(\theta)\tau/\ve+x)
e^{-\ve^\beta(\omega'_\sigma(\theta)\tau/\ve+x)^2}
\frac{e^{ihx^2}}{x}\,dx\nonumber\\
&&+\int\limits_{(c-\omega'_\sigma(\theta)\tau)/\ve}^{+\infty} 
\hat {\bf R}_{0}(r-\omega'_\sigma(\theta)\tau/\ve-x,\theta)
e^{-\ve^\beta(\omega'_\sigma(\theta)\tau/\ve+x)^2}
\frac{e^{ihx^2}}{x^2}\,dx\Big]\nonumber\\
&=&I_1+I_2+I_3+I_4.\nonumber
\eeqn
The terms $I_1$ and 
$I_4$ tend to zero as $\ve\to0$, since $|\hat {\bf R}_0|\le C<\infty$
by condition {\bf I1}. 
Note that $|x|\ge c/(2\ve)$ if 
$|x+\omega'_\sigma(\theta)\tau/\ve|\ge c/\ve$. Hence,
\beqn
|I_3|&\le& C_1\ve^\beta 
\int\limits_{(c-\omega'_\sigma(\theta)\tau)/\ve}^{+\infty} 
(\omega'_\sigma(\theta)\tau/\ve+x)
e^{-\ve^\beta(\omega'_\sigma(\theta)\tau/\ve+x)^2}
\frac1{|x|}dx\nonumber\\
&\le& C_2\ve^{\beta+1}\int\limits_{c/\ve}^{+\infty}
ye^{-\ve^{\beta}y^2}dy
=C_3\,\ve\, e^{-\ve^\beta(c/\ve)^2}\to0,\quad\ve\to0.
\nonumber
\eeqn
To prove that the contribution of the integral $I_2$ vanishes, 
we repeat the integration by parts and use the bound  
$\sup\limits_{r\in\R^d,\theta\in\T^d}
|\partial^k_r\hat {\bf R}_{0}(r,\theta)|\le C<\infty$,
where $k\in[0,d^2/2+2d]$.
This bound follows from condition (\ref{2.3'})--(\ref{Varm}). 

\setcounter{section}{4}
\setcounter{equation}{0}
\section*{\large\bf Appendix D: Local conservation law} 
Let $v(x,t)$ be a solution of (\ref{CP1''})
with the finite energy.
The local energy in the point $x\in\Z^d$ is defined as 
$$
{\cal E}(x,t):=
\frac12\Big\{|\dot v (x,t)|^2+\sum\limits_{y\in\Z^d}
v(x,t)\cdot V(x-y)v(y,t)\Big\}.
$$
We derive formally the expression for the
energy current of the finite energy solutions $v(x,t)$.
For the half-space $\Omega_k:=\{x\in\Z^d:\,x=(x_1,\dots,x_d),\,x_k\le 0\}$,
$k=1,\dots,d$, we define the energy in the region
$\Omega_k$ as
$$
{\cal E}_{\Omega_k}(t):=\frac12\sum\limits_{x\in\Omega_k}
\Big\{|\dot v(x,t)|^2+\sum\limits_{y\in\Z^d}
v(x,t)\cdot V(x-y)v(y,t)\Big\}.
$$
By formal calculation, using Eqn (\ref{CP1''})  we obtain
$\dot{\cal E}_{\Omega_k}(t)=-\sum\limits_{x'\in\Z^d:\,x'_k=0} j_k(x',t)$,
where  $j_k(x',t)$ stands for the  energy current density
in the direction $e_k=(\de_{1k},\dots,\de_{dk})$, $k=1,\dots,d$, 
$x'\in\Z^d$  with $x'_k=0$,
\beqn\label{D.0}
j_k(x',t)\!&:=&\!-\fr12 \sum\limits_{y'\in\Z^d:\,y'_k=0}
\Big\{\sum\limits_{m\ge1,\,p\le 0}
\dot v(x'+me_k,t)\cdot V(x'+me_k-y'-pe_k)v(y'+pe_k,t)
\nonumber\\
&&\!\!-\sum\limits_{m\le0,\,p\ge 1}\dot v(x'+me_k,t)\cdot
V(x'+me_k-y'-pe_k)v(y'+pe_k,t)\Big\}.
\eeqn

Now let $v(x,t)$ be the random solution to (\ref{CP1''}) with
the initial measure $\mu_0^\ve$ satisfying {\bf V1} and {\bf V2}.
Therefore, for any $x\in\Z^d$, $\tau\in\R\setminus 0$, and $r\in\R^d$,
the {\it average energy} is
\beqn
&&\E_0^\ve [{\cal E}(x+[r/\ve],\tau/\ve)]=
\E_{\tau/\ve}^\ve [{\cal E}(x+[r/\ve],0)]\nonumber\\
&&=\frac12\tr\Big[Q^{11}_{\ve,\tau/\ve}([r/\ve]+x,[r/\ve]+x)
+ \sum\limits_{y\in\Z^d}Q^{00}_{\ve,\tau/\ve}([r/\ve]+x,[r/\ve]+y)\cdot 
V^T(x-y)\Big],\nonumber
\eeqn
by condition {\bf E1} and
the uniform bound for the correlation functions of the measure
$\mu^\ve_{\tau/\ve}$ (see \cite[Lemma 5.1]{DS} or Lemma \ref{lcom}):
\be\label{D.2}
\sup_{\ve>0}\sup_{i,j=0,1}\sup_{z,z'\in\Z^d}
\Vert Q^{ij}_{\ve,\tau/\ve}(z,z')\Vert\le C<\infty.
\ee
It follows from Theorem \ref{t2.4} that for any $\tau\not=0$, $r\in\R^d$,  
$\E_0^\ve [{\cal E}(x+[r/\ve],\tau/\ve)]
\to  {\bf e}(\tau,r)$ as $\ve\to0$, where
\beqn\label{etaur}
 {\bf e}(\tau,r)\!&=&\!\fr{1}{2}
\tr\Big[ q^{11}_{\tau,r}(0)+ 
\sum\limits_{x\in\Z^d}q^{00}_{\tau,r}(x)  V^T(x)\Big]
\nonumber\\
\!&=&\!\fr12 (2\pi)^{-d}
\tr\int\limits_{\T^d}\Big(\hat q^{11}_{\tau,r}(\theta)+ 
\hat q^{00}_{\tau,r}(\theta) \hat V^*(\theta)\Big)\,d\theta=
(2\pi)^{-d}
\tr\int\limits_{\T^d}\hat q^{11}_{\tau,r}(\theta)\,d\theta.
\eeqn
The last equality follows from condition {\bf E2}
 and Remarks \ref{Re3.4} (i).

Similarly, by (\ref{D.0}) we obtain
\beqn
\E_0^\ve [j_k(x',t)]\!\!&=&\!\!\ds\fr12 \sum\limits_{y'\in\Z^d:\,y'_k=0}
\Big(\sum\limits_{m\le0,\,p\ge 1}
\tr\Big[Q^{10}_{\ve,t} (x'+me_k,y'+pe_k)\cdot V^T(x'-y'+(m\!-\!p)e_k)\Big]
\nonumber\\
&&-\sum\limits_{m\ge1,\,p\le 0}
\tr\Big[Q^{10}_{\ve,t}(x'+me_k,y'+pe_k)\cdot V^T(x'-y'+(m\!-\!p)e_k)\Big]\Big).
\nonumber
\eeqn
Therefore, by Theorem \ref{t2.4},
the following limit exists, 
$\lim\limits_{\ve\to0}\E_0^\ve[j_k(x+[r/\ve],\tau/\ve)]={\bf j}_k(\tau,r)$
for any $\tau\not=0$, $r\in\R^d$, $k=1,\dots,d$.
 Here 
\beqn
{\bf j}_k(\tau,r)&=&\ds\fr12\sum\limits_{y'\in\Z^d:\,y'_k=0}
\Big(\sum\limits_{m\le-1,\,p\ge 0}
\tr\Big[q^{10}_{\tau,r}(x'-y'+(m-p)e_k) 
V^T(x'-y'+(m-p)e_k)\Big] \nonumber\\
&&-\sum\limits_{m\ge0,\,p\le -1}
\tr\Big[q^{10}_{\tau,r}(x'-y'+(m-p)e_k) V^T(x'-y'+(m-p)e_k)\Big]\Big).
\nonumber
\eeqn
Write $x'-y'=:z'\in\Z^d$ with $z'_k=0$, $m-p:=s\in\Z^1$ 
and change the order of the summation in the series. Therefore,
\beqn\la{jinfty}
{\bf j}_k(\tau,r)\!\!&=&\!\!-\ds\frac12\sum\limits_{z'\in\Z^d:\,z'_k=0}
\sum\limits_{s\in\Z^1}
\tr [q^{10}_{\tau,r}(z'+s e_k)V^T(z'+se_k)]s
=-\frac12\sum\limits_{z\in\Z^d}
\tr [q^{10}_{\tau,r}(z)z_k V^T(z)]\nonumber\\
\!&=&\!-\ds\fr{i}2(2\pi)^{-d}
\tr\int\limits_{\T^d}\hat q^{10}_{\tau,r}(\theta) \pa_k
\hat V(\theta)\,d\theta,\quad k=1,\dots,d.
\eeqn
Write ${\bf j}(\tau,r)=({\bf j}_1(\tau,r),\dots,{\bf j}_d(\tau,r))$.
Finally, (\ref{etaur}), (\ref{jinfty}) and Corollary \ref{cor2.5} yield 
$$
\pa_\tau {\bf e}(\tau,r)+\nabla_r \cdot {\bf j}(\tau,r)=0,
\quad \tau\in\R,\quad r\in\R^d.
$$
{\bf Remark}.
In Section 2.3, we give the example of the "{\it local equilibrium}" initial
measures $\mu_0^\ve$. Namely, let $q_0^{ij}(z)$ 
be the correlation functions of the 
Gibbs measure $g$ (see Definition \ref{def2.5}) with $\beta=1$, i.e.,
$\hat q_0^{00}(\theta)= \hat V^{-1}(\theta)$,
$\hat q_0^{11}(\theta)=I$,
$\hat q_0^{01}(\theta)= \hat q_0^{10}(\theta)= 0$.
Put $\hat {\bf R}_0(r,\theta)=T(r)\hat q_0(\theta)$,
where the function $T(r)$, $r\in\R^d$, is defined in Section 2.3.
Moreover, let $\mu_0^\ve$, $\ve>0$, be Gaussian measures with 
the correlation functions $Q^{ij}_\ve(z,z')$ defined in (\ref{Qexam}).
Then conditions {\bf V1} and {\bf V2} hold. In this case,
the limit correlation matrices $\hat q^{ij}_{\tau,r}(\theta)$ have a form
$$
\ba{lll}\hat q^{11}_{\tau,r}(\theta)
&=\hat V(\theta)\hat q^{00}_{\tau,r}(\theta)
&=\ds\sum\limits_{\sigma=1}^s {\bf T}_+^{\sigma}(\tau,r;\theta)
\Pi_\sigma(\theta),\nonumber\\
\hat q^{01}_{\tau,r}(\theta)
&=-\hat q^{10}_{\tau,r}(\theta)
&=i\ds\sum\limits_{\sigma=1}^s {\bf T}_-^{\sigma}(\tau,r;\theta)
\omega^{-1}_\sigma(\theta)\Pi_\sigma(\theta),\nonumber
\ea
$$
where, by definition,
 ${\bf T}_\pm^{\sigma}(\tau,r;\theta)
:=(1/2)(T(r+\nabla\omega_\sigma(\theta)\tau)\pm
T(r-\nabla\omega_\sigma(\theta)\tau))$. 
Therefore,
$$
\ba{l}
{\bf e}(\tau,r)=(2\pi)^{-d}\ds
\sum\limits_{\sigma=1}^s\int\limits_{\T^d}{\bf T}_+^{\sigma}(\tau,r;\theta)
\tr \Pi_\sigma(\theta)\,d\theta,\\
{\bf j}(\tau,r)=-(2\pi)^{-d}\ds\sum\limits_{\sigma=1}^s
\int\limits_{\T^d}{\bf T}_-^{\sigma}(\tau,r;\theta)
\nabla\omega_\sigma(\theta)\tr \Pi_\sigma(\theta)\,d\theta.\ea
$$

 In \cite{DPST}, the "{\it locally conserved}" quantities were studied
in the case when $d=1$. Let us consider these quantities 
in many-dimensional case. 
At first, introduce the following matrix-valued functions 
$$
\ba{l}
E(z+h,z;X_0):=\ds\frac12
\Big(v_1(z+h)\otimes v_1(z)+v_0(z+h)\otimes\sum\limits_{z'\in\Z^d}
 V(z-z')v_0(z')\Big),\\
A(z+h,z;X_0):=\ds\frac12
\Big(v_0(z+h)\otimes v_1(z)-v_1(z+h)\otimes v_0(z)\Big),\quad
z,h\in\Z^d,\quad X_0=(v_0,v_1).
\ea
$$
The "locally conserved" quantities $X_h^\ve(\varphi,X_0)$,
$Y_h^\ve(\varphi,X_0)$ are defined as follows. 
For $\varphi\in C_0^1(\R^d)$, we set
\beqn
X_h^\ve(\varphi,X_0)\!&:=&\!\ve^d\sum\limits_{z\in\Z^d}\varphi(\ve z)
E(z+h,z;X_0),\nonumber\\
Y_h^\ve(\varphi,X_0)\!&:=&\!\ve^d\sum\limits_{z\in\Z^d}\varphi(\ve z)
A(z+h,z;X_0),\quad h\in\Z^d,\quad \ve>0.\nonumber
\eeqn
{\bf Theorem D.1}\, 
{\it Let conditions {\bf I1}--{\bf I4}, {\bf V1}, {\bf V2} hold.
Then for any $\varphi\in C_0^1(\R^d)$, $h\in\Z^d$, $\tau\not=0$,
there exist the limits
$$
\lim_{\ve\to0}\E_{\tau/\ve}^\ve[X_h^\ve(\varphi,\cdot)]=
{\bf E}(\varphi;\tau,h),\quad
\lim_{\ve\to0}\E_{\tau/\ve}^\ve[Y_h^\ve(\varphi,\cdot)]=
{\bf A}(\varphi;\tau,h).
$$
The matrices ${\bf E}(\varphi;\tau,h)$ and ${\bf A}(\varphi;\tau,h)$ 
are given by their Fourier transform in the following way.
Write
$$
\hat {\bf E}(\varphi;\tau,\theta):=\sum\limits_{h\in\Z^d}
e^{ih\cdot\theta}{\bf E}(\varphi;\tau,h),\quad
\hat {\bf A}(\varphi;\tau,\theta):=\sum\limits_{h\in\Z^d}
e^{ih\cdot\theta}{\bf A}(\varphi;\tau,h),\quad\theta\in\T^d.
$$ 
Then
\beqn
\hat {\bf E}(\varphi;\tau,\theta)\!\!&=&\!\!\ds\frac12\int\limits_{\R^d}
\varphi(r)\Big(\hat q^{11}_{\tau,r}(\theta)+ 
\hat q^{00}_{\tau,r}(\theta) \hat V(\theta)\Big)\,dr
=\ds\int\limits_{\R^d}\varphi(r)\hat q^{11}_{\tau,r}(\theta)\,dr,
\label{D.5}\\
\hat {\bf A}(\varphi;\tau,\theta)
\!\!&=&\!\!\ds\frac12\int\limits_{\R^d}
\varphi(r)\Big(\hat q^{01}_{\tau,r}(\theta)- 
\hat q^{10}_{\tau,r}(\theta) \Big)\,dr
=\ds\int\limits_{\R^d} \varphi(r)\hat q^{01}_{\tau,r}(\theta)\,dr,
\label{D.6}
\eeqn
where $\hat q^{ij}_{\tau,r}(\theta)$ are defined in (\ref{qt}).}
\bigskip

The proof of Theorem~D.1 is based on Theorem \ref{t2.4} 
and the bound (\ref{D.2}).
\medskip\\
{\bf Remark}\,
It follows from (\ref{D.5}), (\ref{D.6}) and
Corollary~\ref{cor2.5} that $\hat{\bf E}$ and $\hat {\bf A}$
satisfy the equations
$$
\ba{l}
\partial_{\tau}\hat {\bf E}(\varphi;\tau,\theta)
=i\omega_\sigma(\theta) \ds\int\limits_{\R^d}
\nabla\omega_\sigma(\theta)\cdot
\nabla\varphi(r)\hat q^{01}_{\tau,r}(\theta)\,dr
=i\omega_\sigma(\theta)\nabla\omega_\sigma(\theta)\cdot
\hat {\bf A}(\nabla\varphi;\tau,\theta),\\
\partial_{\tau}\hat {\bf A}(\varphi;\tau,\theta)
=-i\omega_\sigma^{-1}(\theta)\ds\int\limits_{\R^d}
\nabla\omega_\sigma(\theta)\cdot
\nabla\varphi(r)\hat q^{11}_{\tau,r}(\theta)\,dr
=-i\omega_\sigma^{-1}(\theta)\nabla\omega_\sigma(\theta)\cdot
\hat {\bf E}(\nabla\varphi;\tau,\theta).
\ea
$$



\begin{thebibliography}{99}

\bibitem{BDS}
Boldrighini, C., Dobrushin, R.L., Suhov, Yu.M., 
One dimensional hard rod caricature
of hydrodynamics, {\em J. Stat. Phys.} {\bf 31} (1983), no.3, 577-615.

\bibitem{BPT} Boldrighini, C., Pellegrinotti, A., and Triolo, L.,
Convergence to stationary states for infinite harmonic systems,
{\it J. Stat. Phys.} {\bf 30} (1983), 123-155. 

\bibitem{DeMasi}
De Masi, A., Ianiro, N., Pellegrinotti, A., Presutti, E., A survey
of the hydrodynamical behavior of many-particle systems.
In: {\it Nonequilibrium phenomena. II. From stochastic to hydrodynamics.}
Lebowitz, J.L., Montroll, E.W. (eds.) pp. 123-294, Amsderdam: 
North-Holland, 1984.

\bibitem{DEL}
De Masi, A., Esposito, R., Lebowitz, J., Incompressible Navier--Stokes 
and Euler limit of the Boltzmann equation,
{\it Commun. Pure Appl. Math.} {\bf 42} (1989), 1189--1214.

\bibitem{DSS} Dobrushin, R.L., Sinai, Ya.G., and Sukhov, Yu.M., 
"Dynamical systems of statistical mechanics and kinetic equations. 
Chapter~10. Dynamical systems of statistical mechanics", 
{\it Dynamical systems}~--~2, Itogi Nauki i Tekhniki. Ser. Sovrem. Probl.
Mat. Fund. Napr., {\bf 2}, VINITI, Moscow, 1985, 235-284
(English transl., "Dynamical systems of statistical mechanics",
{\it Dynamical systems. II. Ergodic theory with applications 
to dynamical systems and statistical mechanics}, 
Encyclopaedia Math. Sci., vol.2, Springer--Verlag,
Berlin 1989, pp.207-278.)

\bibitem{DPST}
Dobrushin, R.L., Pellegrinotti, A., Suhov, Yu.M., and Triolo, L.,
One dimensional harmonic lattice caricature
of hydrodynamics, {\em J. Stat. Phys.} {\bf 43} (1986), 571-607.

\bibitem{DPST2}
Dobrushin, R.L., Pellegrinotti, A., Suhov, Yu.M., and Triolo, L.,
One dimensional harmonic lattice caricature of hydrodynamics:
Second approximation, {\em J. Stat. Phys.} {\bf 52} (1988), 423-439. 

\bibitem{DPS90}
Dobrushin, R.L., Pellegrinotti, A., Suhov, Yu.M., 
One dimensional harmonic lattice caricature of hydrodynamics:
A higher correction, {\em J. Stat. Phys.} {\bf 61} (1990), 387-402. 


\bibitem{DKS} Dudnikova, T.V., Komech, A.I., Spohn, H.,   
On a  two-temperature problem for wave equation,   
 {\it Markov Processes and Related Fields} {\bf 8} (2002), 43-80.   
Preprint at  arXiv:math-ph/0508044.

\bibitem{DK2} Dudnikova, T.V., Komech, A.I.,
On a two-temperature problem  for the Klein--Gordon equation,
 {\it Theory Probab. Appl.} {\bf 50} (2006), no.4,
582-611 (translated from the Russian journal {\it Teoriya Veroyatnostei
i ee Primeneniya} {\bf 50} (2005), no.4, 675-710).

\bibitem{DKS1}  Dudnikova, T.V., Komech, A.I., and Spohn, H., 
 On the convergence to statistical equilibrium for harmonic crystals,
 {\em J. Math. Phys.} {\bf 44} (2003), 2596-2620.
Preprint at arXiv:math-ph/0210039.

\bibitem{DKM} Dudnikova, T.V., Komech, A.I., Mauser, N.,
On two-temperature problem for harmonic crystals,
  {\em J. Stat. Phys.} {\bf 114} (2004), no.3/4, 1035-1083.
Preprint at ArXiv:math-ph/0211017.

\bibitem{DK05} Dudnikova, T.V., Komech, A.I.,
 On the convergence to a statistical equilibrium
 in the crystal coupled to a scalar field,
  {\em Russ. J. Math. Phys.} {\bf 12} (2005), no.3, 301-325.
 Preprint at arXiv:math-ph/0508053.

\bibitem{DS} Dudnikova, T.V., Spohn, H., 
Local stationarity for lattice dynamics in the harmonic approximation, 
{\em Markov Processes and Related Fields}
{\bf 12} (2006), no.4, 645-578. 
Preprint at ArXiv: math-ph/0505031.

\bibitem{D08}  Dudnikova, T.V., On the asymptotical normality 
of statistical solutions for harmonic crystals in half-space,
{\it Russian J. Math. Phys.}  {\bf 15} (2008), no.4, 460-472.
Preprint at ArXiv:0905.3472.

\bibitem{D10} Dudnikova, T.V., Convergence to equilibrium distribution. 
The Klein--Gordon equation coupled to a particle,  
{\it Russian J. Math. Phys.} {\bf 17} (2010), no.1, 77-95.
Preprint at ArXiv: 0711.1091.

\bibitem{D09'} Dudnikova, T.V., 
Lattice dynamics in the half--space. Energy transport equation, 
 {\em J. Math. Phys.} {\bf 51} (2010), 083301.
Preprint at ArXiv: 0905.4806

\bibitem{EMY}
Esposito, R., Marra, R., Yau, H.-T.,
Navier--Stokes equations for stochastic particle systems on the lattice,
{\em Comm. Math. Phys.} {\bf 182} (1996), 395-456.

\bibitem{F}
Fritz, J., {\em An Introduction to the Theory of Hydrodynamic Limits},
Lectures in Mathematical Sciences {\bf 18}. The Graduate School
of Mathematics. The University of Tokyo (2001). 

\bibitem{OVY93}
Olla, S., Varadhan, R.S.R., Yau, H.-T., Hydrodynamical limit for 
a Hamiltonian system with weak noise, {\em Comm. Math. Phys.}
{\bf 155} (1993), 523-560.

\bibitem{QY98}
Quastel, J., Yau, H.-T.,
Lattice gases, large deviations, and the incompressible 
Navier-Stokes equations, {\em Ann. of Math.} {\bf 148} (1998), 51-108. 

\bibitem{Mo}
Morrey, C.B., On the derivation of the equations of hydrodynamics 
from statistical mechanics, {\it Comm. Pure Appl. Math.} {\bf 8} (1955),
279-326.

\bibitem{Sp91} Spohn, H., 
{\em Large Scale Dynamics of Interacting
Particles}, Texts and Monographs in Physics (Springer
Verlag, Heidelberg, 1991).


\bibitem{Tit}
Titchmarsh, E.C., {\it Introduction to the theory of Fourier integrals}, 
Oxford, Clarendon Press (1948), 394 p.

\bibitem{Fed} Fedoryuk, M.V., The
stationary phase method and pseudodifferential operators,
{\em Russ. Math. Surveys} {\bf 26} (1971),
no.1, 65--115.

\bibitem{Halmos} Halmos, P.R., {\it Measure Theory}, New York (1950).

\end{thebibliography}
\end{document}